\documentclass[12pt]{article}
\usepackage{fullpage}
\usepackage{natbib,epsf,color}
\usepackage{graphicx}
\usepackage{subfigure}
\usepackage{multirow}
\usepackage{booktabs}
\usepackage{amssymb}
\usepackage{amsthm}
\usepackage{amsmath}
\usepackage{multirow,setspace}
\usepackage[figuresright]{rotating}

\begin{document}
\newtheorem{theorem}{Theorem}
\newtheorem{lemma}{Lemma}
\newtheorem{proposition}{Proposition}
\newtheorem{corollary}{Corollary}
\theoremstyle{remark}
\newtheorem{remark}{Remark}
\title{Bayesian Quantile Regression for Single-Index Models }
\author{Yuao Hu$^{a}$, Robert B.~Gramacy$^{b}$ and Heng Lian$^{a}$\\
\begin{tabular}{l}
 {\small\it $^{a}$Division of Mathematical Sciences, SPMS, Nanyang Technological University, Singapore}\\
 {\small\it $^{b}$Booth School of Business, University of Chicago, Chicago, USA}\\
\end{tabular}
}


\maketitle

\begin{abstract}
Using an asymmetric Laplace distribution, which provides a mechanism for Bayesian inference of quantile regression models, we develop a fully Bayesian approach to fitting single-index models in conditional quantile regression. In this work, we use a Gaussian process prior for the unknown nonparametric link function and a Laplace distribution on the index vector, with the latter motivated by the recent popularity of the Bayesian lasso idea. We design a Markov chain Monte Carlo algorithm for posterior inference. Careful consideration of the singularity of the kernel matrix, and tractability of some of the full conditional distributions leads to a partially collapsed approach where the nonparametric link function is integrated out in some of the sampling steps. Our simulations demonstrate the superior performance of the Bayesian method versus the frequentist approach. The method is further illustrated by an application to the hurricane data. 

\end{abstract}
\section{Introduction}
Single-index models (SIM) provide an efficient way of coping with high-dimensional nonparametric estimation problems (\cite{hardle1993optimal,yu2002penalized}) and avoid the ``curse of dimensionality" in nonparametric problems by assuming that the response is only related to a single linear combination of the covariates. 
Compared to fully nonparametric regression, it offers a nice tradeoff between simplicity and modelling power.
The fitting of single-index models, commonly based on splines or kernel methods, has found wide application in the literature. For example, \cite{hardle1993optimal} used SIM to study the dependence of the severity of side impacts on the velocity and acceleration of the automobile in an accident, and \cite{xia2004goodness} demonstrated that SIM provides a good fit in a study trying to identify causal factors associated with the prevalence and incidence of depression. However, efficient and stable estimation of SIMs is still a challenging problem and has inspired many recent works in this area (\cite{wang2010estimation,liang2010estimation}). 

Although frequentist estimation of SIMs has a long history, the Bayesian approach to fitting these models has only appeared quite recently. \cite{antoniadis2004bayesian} and \cite{wang2009bayesian} proposed a Bayesian approach using polynomial splines to model the nonparametric link function, while \cite{choi2011gaussian} and \cite{gramacy2010gaussian} use a Gaussian process (GP) prior. As noted in \cite{gramacy2010gaussian}, one advantage of using GPs as the prior for the link function is that the index vector does not have to be normalized to have unit norm, which makes the choice of prior easier, and subsequently the sampling algorithm simplifies too. 

However, the restriction of these works to mean regression, that is on estimating the conditional mean regression function, may be a limitation. As a useful supplement to mean regression, quantile regression produces a more complete description of the conditional response distribution. In particular, it can uncover different structural relationships between covariates and responses at the upper or lower tails, which is sometimes of significant interest in econometrics applications. Furthermore, compared to mean regression, median regression (which is a special case of quantile regression) is more robust to outliers or heavy-tailed random errors.

In this article, we consider a single-index quantile regression model. For a given quantile level $\tau\in(0,1)$ and i.i.d.~pairs $(\boldsymbol{x}_i, y_i)$, it is given by
\begin{equation*}\label{model}
Q_{y_{i}|\boldsymbol{x}_i}(\tau)=\eta(\boldsymbol{x_{i}}^{T}\boldsymbol{\beta}),\;\;\; i=1,2,\cdots,n.
\end{equation*}
Here $y_{i}$ is the response, $\boldsymbol{x_{i}}=(x_{i1},\cdots,x_{ip})^{T}$ is the $p$-dimensional predictor vector, $Q_{y_{i}|\boldsymbol{x}_i}(.)=F^{-1}_{y_{i}|\boldsymbol{x}_i}(.)$ is the inverse cumulative distribution function of the response given the predictors, $\eta(.)$ is the unknown univariate link function, and $\boldsymbol{\beta}=(\beta_{1},\beta_{2},\cdots,\beta_{p})^{T}$ is the index which implicitly depends on the desired quantile $\tau$. Dimension reduction is achieved specifically by the index vector so that $\eta$ is a univariate function instead of $p$-variate one, as in fully nonparametric regression. 

In this paper, we propose a Bayesian treatment of the single-index quantile regression model. Recently, \cite{wu2010single} has considered a similar model using kernel regression, which will serve as a frequentist benchmark for our methods. 
We establish a hierarchical Bayesian model by adopting the asymmetric Laplace distribution, which is one common approach among a few alternatives in the Bayesian quantile regression literature \citep{yu2001bayesian,kottas2009bayesian,reich2010flexible,lancaster2010bayesian,tokdar2011simultaneous}. Following \cite{gramacy2010gaussian}, we assign a Gaussian process prior on the link function, to obtain a flexible nonparametric quantile regression model. The posterior inference of all parameters is performed via Markov chain Monte Carlo computations which automatically incorporates all sources of uncertainty. 

The remainder of the paper proceeds as follows.  In Section 2, we describe the structure of our hierarchical Bayesian single-index quantile regression model and discuss our prior choices. We also consider the posterior sampling algorithm focusing on a more efficient partially collapsed sampler, where the link function is integrated out when drawing samples of the index vector. This is explained in detail in the Appendix. Then, numerical illustrations including simulation studies and a real data example are presented in Section 3. We conclude the paper with a discussion in Section 4.

\section{Hierarchical Bayesian Modelling}
At the $\tau$-th quantile, we model the residual errors by the asymmetric Laplace distribution (ALD, \cite{yu2001bayesian,geraci2007quantile,youxi2010bayesian}). More specifically, the probability distribution of $y$ given $\mu=\eta(\boldsymbol{x}^T\boldsymbol{\beta})$ is assumed to be
\begin{equation*}
\pi(y|\mu, \sigma)=\frac{\tau(1-\tau)}{\sigma}\exp\left\{-\frac{1}{\sigma}\rho_{\tau}(y-\mu)\right\},
\end{equation*}
where $\rho_{\tau}(u)=u(\tau-I(u\leq 0))$ is the so-called {\em check function}, the quantile level $\tau$ is the skewness parameter in the distribution, $\mu$ is the location parameter, and $\sigma$ is the scale parameter. In our context, with the setting $\mu_{i}=\eta(\boldsymbol{x}_{i}^{T}\boldsymbol{\beta})$, and $\boldsymbol{y}=(y_{1},y_{2},\cdots, y_{n})^{T}$, the conditional distribution for the observations is 
\begin{equation}
\label{quantile}
\pi(\boldsymbol{y}|\boldsymbol{\beta}, \eta, \sigma)=\frac{\tau^{n}(1-\tau)^{n}}{\sigma^{n}}\exp\left\{-\frac{1}{\sigma}\mathop{\sum}\limits_{i=1}\limits^{n}\rho_{\tau}(y_{i}-\eta(\boldsymbol{x_{i}}^{T}\boldsymbol{\beta}))\right\}.
\end{equation}

Quantile regression is typically based on minimization of the check loss function. However, direct use of the likelihood above is rather inconvenient for Bayesian inference. A location-scale mixture representation of the ALD (\cite{kozumi2009gibbs}) is helpful here. We can write the observations satisfying (\ref{quantile}) alternatively as
\begin{equation*}
y_{i}=\eta(\boldsymbol{x}_{i}^{T}\boldsymbol{\beta})+k_{1}e_{i}+\sqrt{k_{2}\sigma e_{i}}z_{i},
\end{equation*}
where $e_{i}\sim\exp(1/\sigma)$ is an exponential random variable with mean $\sigma$, $z_{i}$ is a standard normal random variable and is independent of $e_i$, $k_{1}=\frac{1-2\tau}{\tau(1-\tau)}$, and $k_{2}=\frac{2}{\tau(1-\tau)}$. This suggests treating the $e_i$ as latent variables, where the conditional distribution of $\boldsymbol{y}$ is rewritten as
\begin{equation*}\label{yDistribution}
\begin{split}
\pi(\boldsymbol{y}|\boldsymbol{\beta}, \eta, \sigma, \boldsymbol{e}_{n})&=\mathop{\prod}\limits_{i=1}\limits^{n}(2\pi k_{2}\sigma e_{i})^{-1/2}\exp\left\{-\frac{1}{2k_{2}\sigma e_{i}}(y_{i}-\eta(\boldsymbol{x}_{i}^{T}\boldsymbol{\beta})-k_{1}e_{i})^{2}\right\}\\
&\propto\exp\left\{-\frac{(\boldsymbol{y}-\boldsymbol{\eta}_{n}-k_{1}\boldsymbol{e}_{n})^{T}\boldsymbol{E}^{-1}(\boldsymbol{y}-\boldsymbol{\eta}_{n}-k_{1}\boldsymbol{e}_{n})}{2}\right\}(\det[\boldsymbol{E}])^{- 1/2}.
\end{split}
\end{equation*}
Here $\boldsymbol{e}_{n}=(e_{1}, e_{2},\cdots e_{n})^{T}$, $\boldsymbol{E}=k_{2}\sigma \rm{diag}$$(e_{1}, e_{2},\cdots e_{n})$, and
 $\boldsymbol{\eta}_{n}=(\eta_{1},\cdots,\eta_{n})^{T}=$\\$(\eta(\boldsymbol{x}_{1}^{T}\boldsymbol{\beta}),\cdots, \eta(\boldsymbol{x}_{n}^{T}\boldsymbol{\beta}))^{T}$.

As in \cite{choi2011gaussian} and \cite{gramacy2010gaussian}, we model the link function by a Gaussian process prior distribution. More specifically, $\eta$ is a Gaussian process a priori, with zero mean and a squared-exponential covariance function, 
\begin{equation}
\label{eqn:d}
\eta\sim \text{GP}(\boldsymbol{0,\; C(\cdot, \cdot)}), \;\;\;\boldsymbol{C}(x,x')=\gamma\exp\{-(x-x')^2/d\},   
\end{equation}
where $\gamma$ and $d$ are hyperparameters. Writing this out in the single-index model framework using the observed covariates $\boldsymbol{x_i}$, we have 
\begin{equation*}\label{EtaPrior}
\pi(\boldsymbol{\eta}_{n}|\boldsymbol{\beta},  \gamma)\propto \det[\boldsymbol{C}_{n}]^{-1/2}\exp\left\{-\frac{\boldsymbol{\eta}_{n}^{T}\boldsymbol{C}_{n}^{-1}\boldsymbol{\eta}_{n}}{2}\right\},
\end{equation*}
where $\boldsymbol{C}_n$ is an $n\times n$ matrix with entries $\boldsymbol{C}(\boldsymbol{x}_{i}, \boldsymbol{x}_{j})=\gamma\exp\{-(\boldsymbol{x}_{i}^{T}\boldsymbol{\beta}-\boldsymbol{x}_{j}^{T}\boldsymbol{\beta})^{2}/d\}$.

In the literature of single-index models, it is well-known that $\eta$ and $\boldsymbol{\beta}$ are unidentifiable since $\eta(\boldsymbol{x}^T\boldsymbol{\beta})=\eta_c(\boldsymbol{x}^T(c\boldsymbol{\beta}))$, $c\neq 0$, where $\eta_c(.)=\eta(./c)$ and thus $\boldsymbol{\beta}$ is only identifiable up to a constant scale. It is typically assumed that $\|\boldsymbol{\beta}\|=1$ so that $\boldsymbol{\beta}$ is identifiable up to sign ($\boldsymbol{\beta}$ and $-\boldsymbol{\beta}$ leads to exactly the same model fit). Accordingly, \cite{choi2011gaussian} also suppose the support of the prior distribution for $\boldsymbol{\beta}$ is on the unit sphere. On the other hand, \cite{gramacy2010gaussian} noted that without the constraint $\|\boldsymbol{\beta}\|=1$, only $\boldsymbol{\beta}/\sqrt{d}$ is identifiable when using the Gaussian process prior, and thus one can remove the range parameter $d$ and also remove the unit norm constraint on $\boldsymbol{\beta}$, which is mathematically equivalent to imposing $\|\boldsymbol{\beta}\|=1$ and keeping the range parameter $d$. With the latter approach, we have the advantage that the prior on $\boldsymbol{\beta}$ is more easily specified, and one fewer parameter ($d$) to worry about. Note that the model leaves the sign of $\boldsymbol{\beta}$ unidentified under either specification. In some cases when $\boldsymbol{\beta}$ is not of direct interest, it does not matter at all. When inference for $\boldsymbol{\beta}$ is a primary goal, some simple heuristics for reconciling the signs in \cite{gramacy2010gaussian} can be used.

We therefore adopt the approach in \cite{gramacy2010gaussian}, so that the entries of $\boldsymbol{C}_n$ are 
 $\boldsymbol{C}(\boldsymbol{x}_{i}, \boldsymbol{x}_{j})=\gamma\exp\{-(\boldsymbol{x}_{i}^{T}\boldsymbol{\beta}-\boldsymbol{x}_{j}^{T}\boldsymbol{\beta})^{2}\}.$ 
Since $\boldsymbol{\beta}$ is not constrained to have unit norm, we are free to choose any prior for $\boldsymbol{\beta}$.  A typical choice is to put an independent Gaussian prior on each component, which is sometimes called a {\em ridge prior}. One can also consider the popular g-prior \citep{zellner86,george2000calibration}. A generalization of the ridge prior is the so-called 
Bayesian lasso, which has been of much interest in the recent literature
(\cite{park2008bayesian,hans2009bayesian}).
Under this prior, $\beta_{j}, \, j=1,\ldots,p $ are independent and identically Laplace, 
\begin{equation*}
\label{BetaPrior}
\pi(\boldsymbol{\beta}|\sigma,\lambda)=\mathop{\prod}\limits_{j=1}\limits^{p}\frac{\lambda}{2\sigma}e^{-\lambda|\beta_{j}|/\sigma},\; \lambda>0.
\end{equation*}
There are a suite of similar priors which share attractive properties and yet further generalize the lasso.  Examples include the normal--gamma prior (\cite{griffin:brown:2010}), the Bayesian elastic set (e.g., \cite{li2010bayesian}) and the horseshoe (e.g., \cite{carvalho2010horseshoe}).  
Our reasons for calling these ``generalizations'' have to do with the form of their hierarchical latent variable representations, and corresponding data augmentation Gibbs samplers.  We focus on the Bayesian lasso as a representative case in this paper.  Simplifications (to the ridge) and further generalizations to the others are straightforward. Recognizing that estimators for $\boldsymbol{\beta}$ are not equivariant under such priors (e.g., \cite{park2008bayesian}), we take the common pre-processing step of scaling the inputs $\boldsymbol{x}_i$ to have a unit $L_2$-norm.  This also simplifies the choice of default priors for $\lambda$ and $\sigma$.


To summarize, our Bayesian hierarchical formulation is provided below.
\begin{equation*}
\begin{split}
\boldsymbol{y}|\boldsymbol{\eta}_n,\boldsymbol{e}_n,\sigma &\sim N(\boldsymbol{\eta}_n+k_1\boldsymbol{e}_n,\boldsymbol{E}),\\
\boldsymbol{\eta}_n|\boldsymbol{x}_i, \gamma,\boldsymbol{\beta}&\sim N(\boldsymbol{0},\boldsymbol{C}_n),\\
\boldsymbol{\beta}&\sim\pi(\boldsymbol{\beta}|\sigma,\lambda),\;\;e_{i}\stackrel{i.i.d.}{\sim} \exp(1/\sigma),\\
\sigma&\sim \pi(\sigma),\,\lambda\sim \pi(\lambda),\,\gamma\sim \pi(\gamma).\\
\end{split}
\end{equation*}
The hyperpriors for $\sigma, \lambda, \gamma$ are set to be $IG(a_{\sigma}, b_{\sigma})$, $Ga(a_{\lambda}, b_{\lambda})$ and $IG(a_{\gamma}, b_{\gamma})$, where $IG$ denotes the inverse Gamma distribution and $Ga$ denotes the Gamma distribution. All of the hyperparameters $a_{\sigma}, b_{\sigma}, a_{\lambda}, b_{\lambda}, a_{\gamma}, b_{\gamma}$ are set to be 0.5 in all our numerical experiments. Sensitivity analyses reveal that our results are not sensitive to these choices. 

The posterior distribution of various variables and parameters is found via MCMC, using a partially collapsed version integrating out $\boldsymbol{\eta}_n$. The details are left to the Appendix.

\section{Numerical Illustrations}
We present three simulation examples and a real data application to illustrate the proposed method, the Bayesian quantile single-index regression model, which is denoted by BQSIM for short in the rest of the article. The MCMC algorithm is implemented in {\sf R}, and available upon request.

\subsection{Simulations}
We illustrate the performance of the proposed method by comparing it with a non-Bayesian single-index quantile regression approach described by \cite{wu2010single}, based on kernel estimation. This frequentist method is denoted by QSIM in the following. 
Since the frequentist approach requires the identifiability constraint $\|\boldsymbol{\beta}\|=1$, we also normalized the Bayesian estimate of the index vector to have unit norm and furthermore require the first component of the index vector to be positive to resolve the sign indeterminacy. The following three simulation examples are directly taken from \cite{wu2010single}.\\

\subsubsection*{Example 1}
\begin{figure}[ht!]
\begin{center}
\begin{minipage}{5in}
\includegraphics[width=5in,trim=3 20 0 0,clip=TRUE]{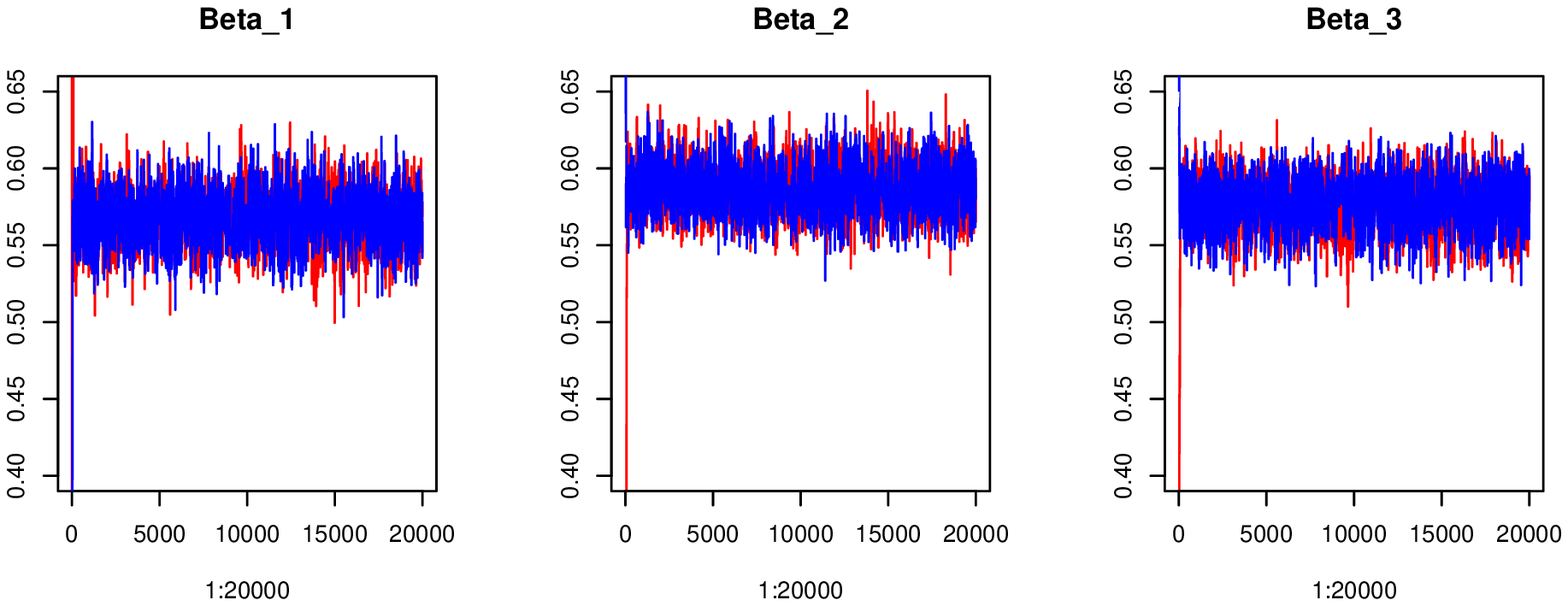}
\end{minipage}

\begin{minipage}{5in}
\includegraphics[width=5in,trim=0 20 0 10,clip=TRUE]{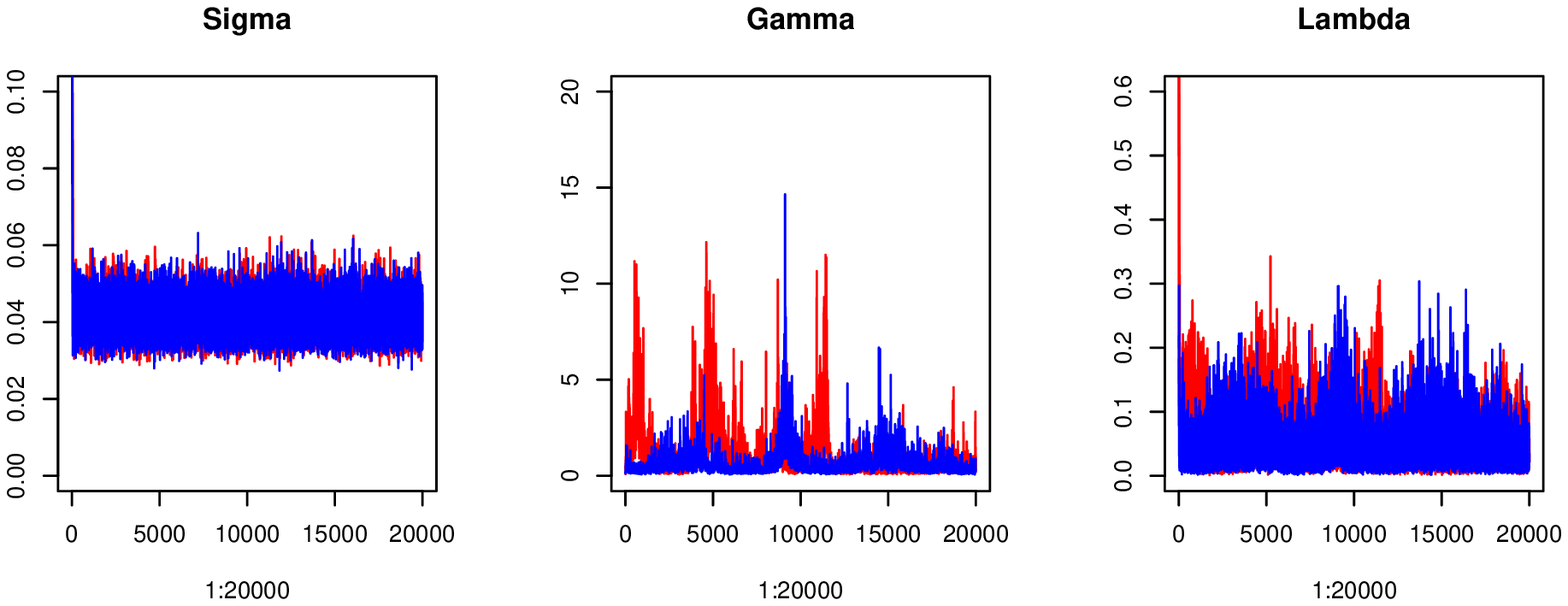}
\end{minipage}
\end{center}
\caption{Trace plots of $\beta_{1}$, $\beta_{2}$, $\beta_{3}$, $\sigma$, $\gamma$ and $\lambda$ at quantile 0.5 for simulation example 1, when $n=100$. Two chains with different starting values are illustrated.} 
\label{traceplot_sim1}
\end{figure}

Consider data generated from the following single-index model with homoscedastic errors,
\begin{equation*}\label{SIM1}
y=\eta(\boldsymbol{x}^{T}\boldsymbol{\beta})+0.1 \mathcal{Z},\; \eta(t)=\rm{sin}\left(\frac{\pi(t-A)}{C-A}\right),
\end{equation*}
with
$\boldsymbol{\beta}=(\beta_{1},\beta_{2},\beta_{3})^{T}=\frac{1}{\sqrt{3}}(1,1,1)^{T}$, $A=\frac{\sqrt{3}}{2}-\frac{1.645}{\sqrt{12}}$, $C=\frac{\sqrt{3}}{2}+\frac{1.645}{\sqrt{12}}$, $\mathcal{Z}\sim N(0,1)$. The predictors $\boldsymbol{x}=(x_1,x_2,x_3)^T$ are uniform in $[0, 1]^{3}$. We consider sample sizes $n=100$ and $n=200$. For each sample size, we fit the models  at seven different quantiles $\tau=0.1,0.25,0.5,0.75,0.9,0.95,0.99$. 
In each case, the MCMC algorithm is run for 20000 iterations with a burn-in of 10000. For convergence diagnosis, we present trace plots of $\boldsymbol{\beta}$, $\sigma$, $\lambda$ and $\gamma$ in Figure \ref{traceplot_sim1} with two different sets of initial values using $\tau=0.5$. The plots suggest that the constructed chains mix quickly.


For a Bayesian point estimator we consider both the posterior mean and posterior median based on the sampled values after burn-in.  
The resulting estimates are summarized in Tables \ref{result_sim1_1} and \ref{result_sim1_1cont}, together with the sample standard deviation (S.D.) and 2.5\% and 97.5\% quantiles of the sampled values after burn-in. It is seen that both posterior mean and posterior median estimators work well and give similar estimates. Thus we only focus on posterior mean as our point estimate in the following.

\begin{figure}[ht!]
\begin{center}
\begin{minipage}{0.75in}
$\tau = 0.1$ 
\end{minipage}
\begin{minipage}{4.5in}
\includegraphics[width=3.9in,trim=0 40 0 0,clip=TRUE]{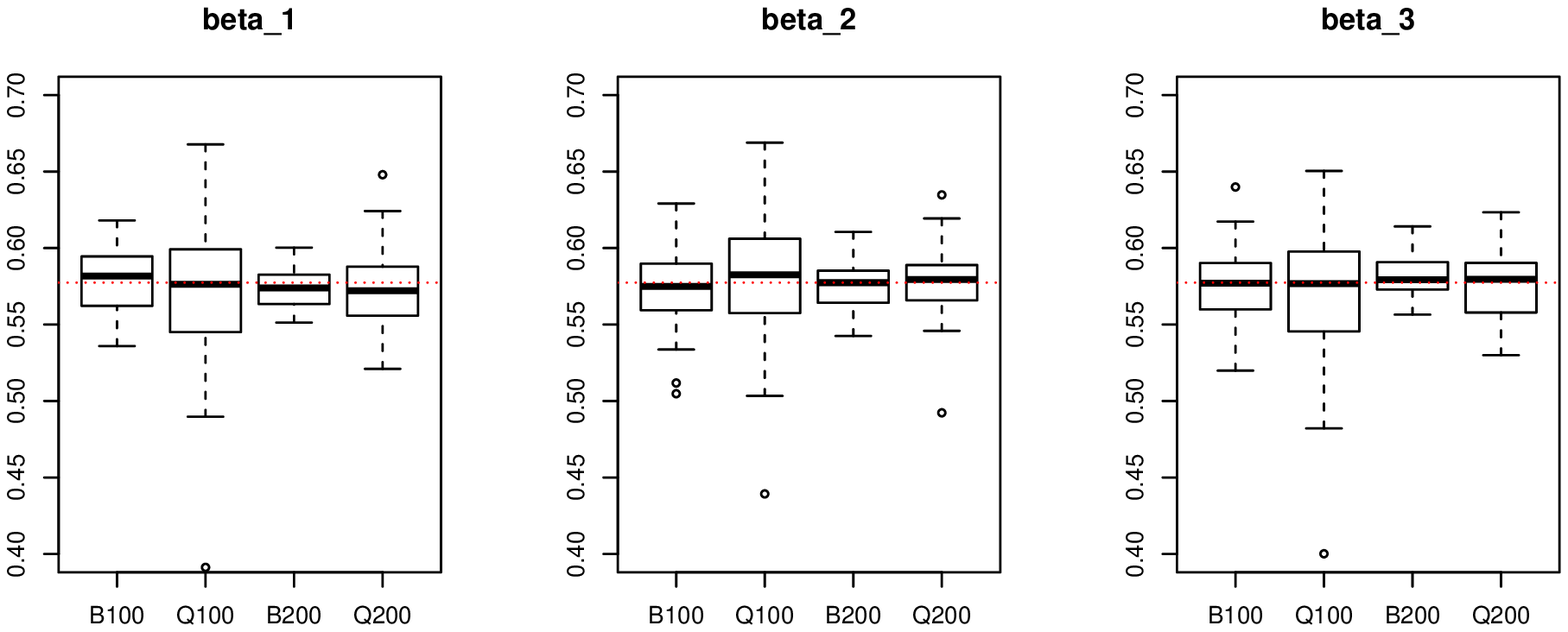}
\end{minipage}

\begin{minipage}{0.75in}
$\tau = 0.25$ 
\end{minipage}
\begin{minipage}{4.5in}
\includegraphics[width=3.9in,trim=0 40 0 30,clip=TRUE]{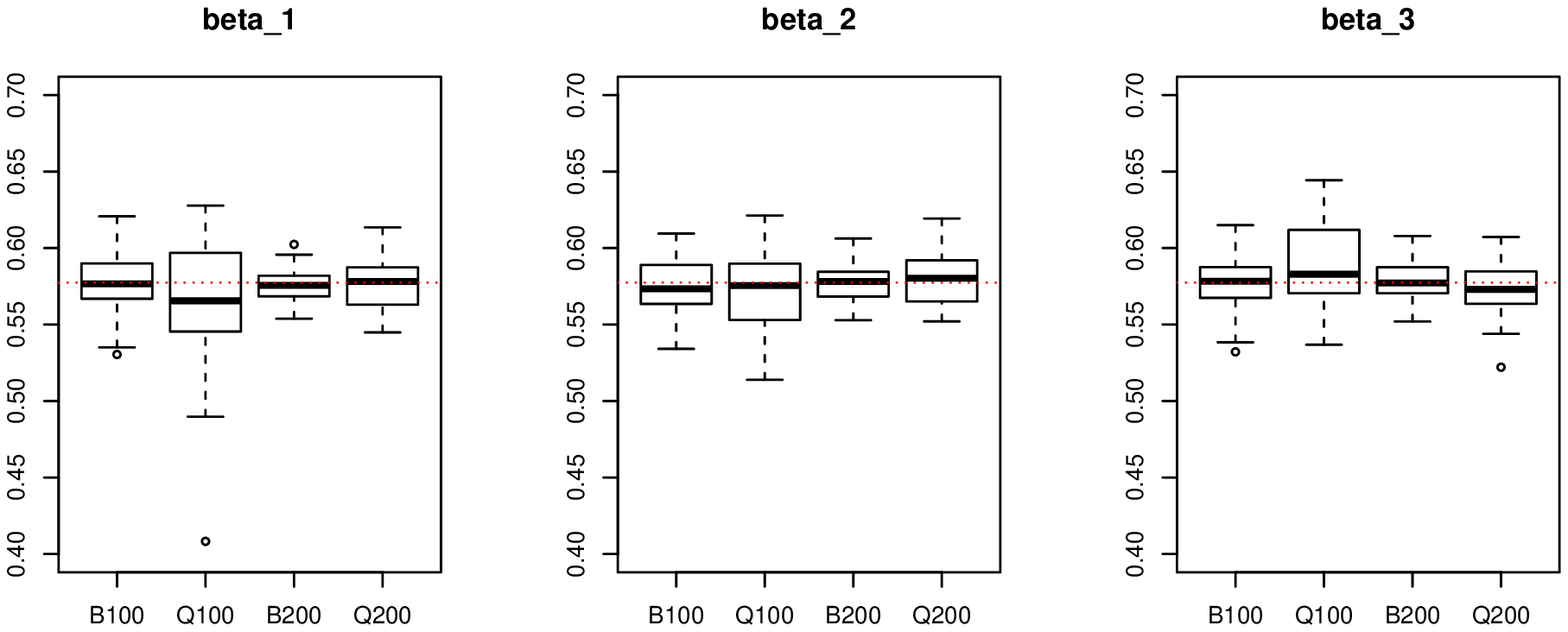}
\end{minipage}

\begin{minipage}{0.75in}
$\tau = 0.5$ 
\end{minipage}
\begin{minipage}{4.5in}
\includegraphics[width=3.9in,trim=0 40 0 30,clip=TRUE]{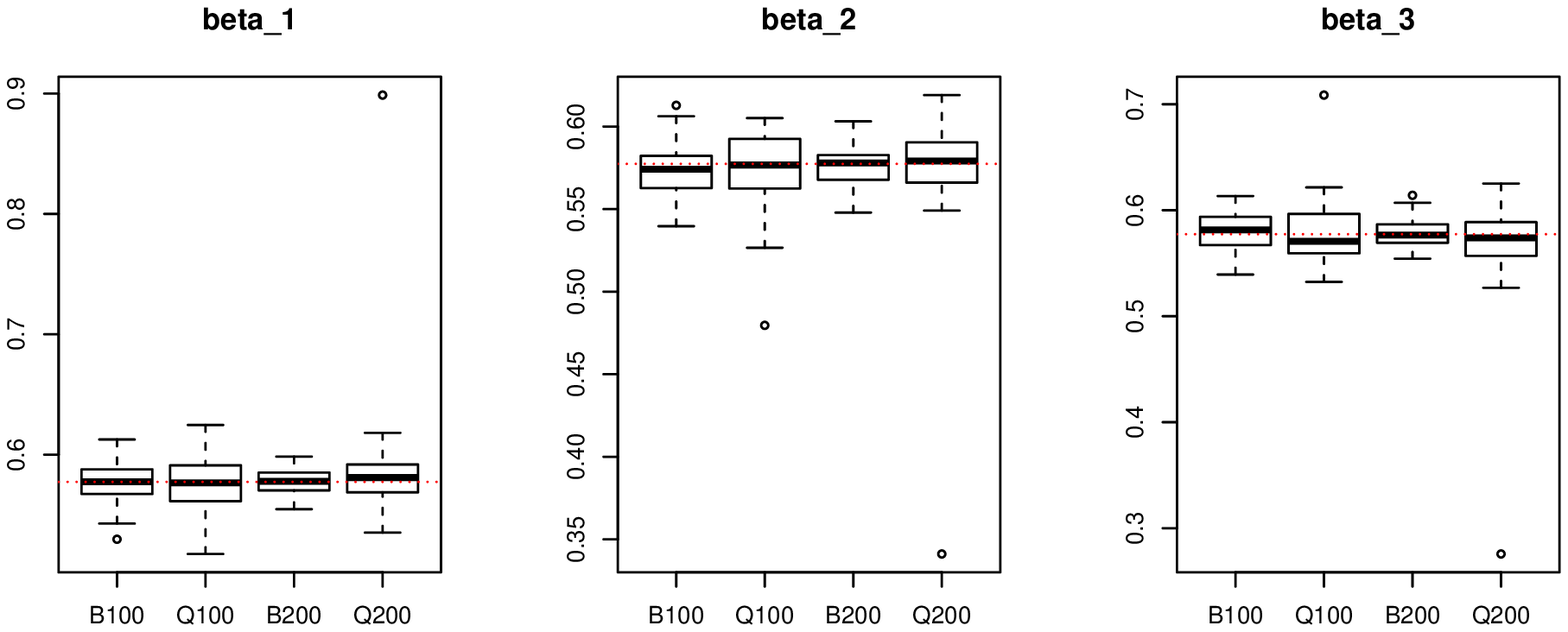}
\end{minipage}

\begin{minipage}{0.75in}
$\tau = 0.75$ 
\end{minipage}
\begin{minipage}{4.5in}
\includegraphics[width=3.9in,trim=0 40 0 30,clip=TRUE]{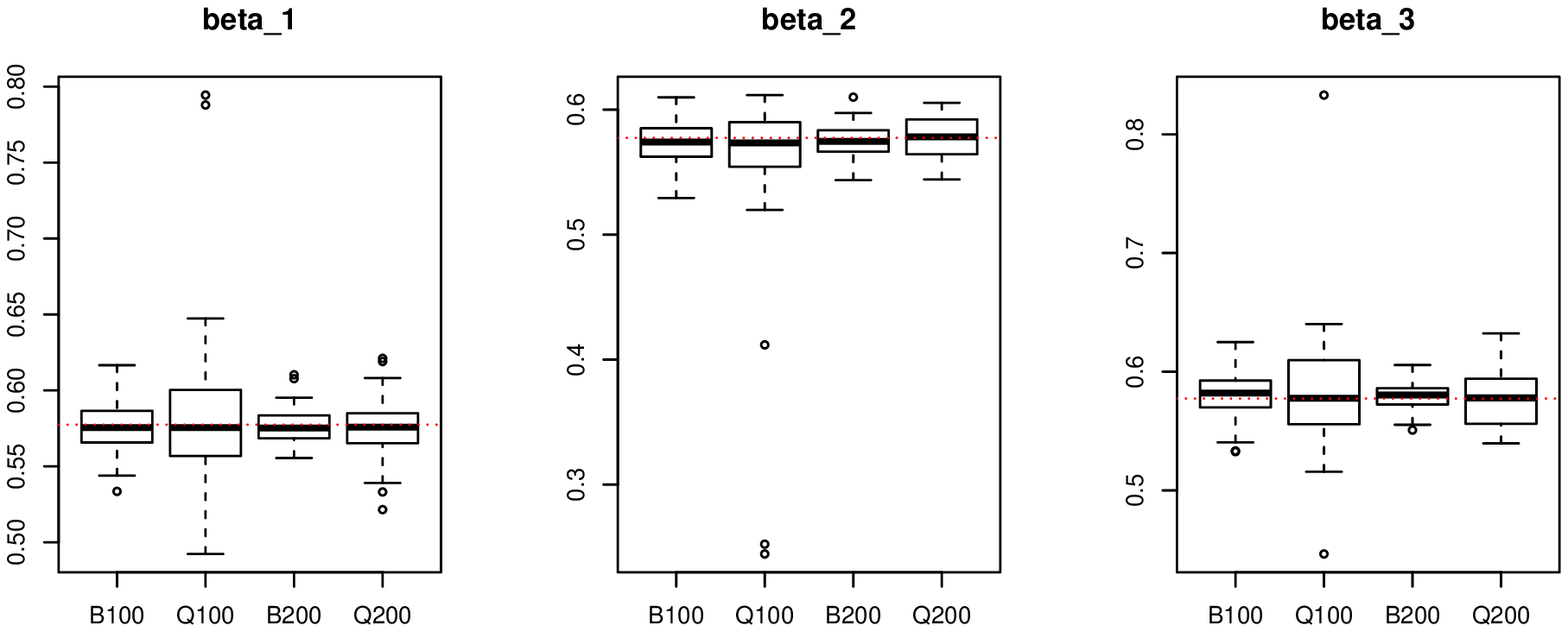}
\end{minipage}

\begin{minipage}{0.75in}
$\tau = 0.9$ 
\end{minipage}
\begin{minipage}{4.5in}
\includegraphics[width=3.9in,trim=0 15 0 30,clip=TRUE]{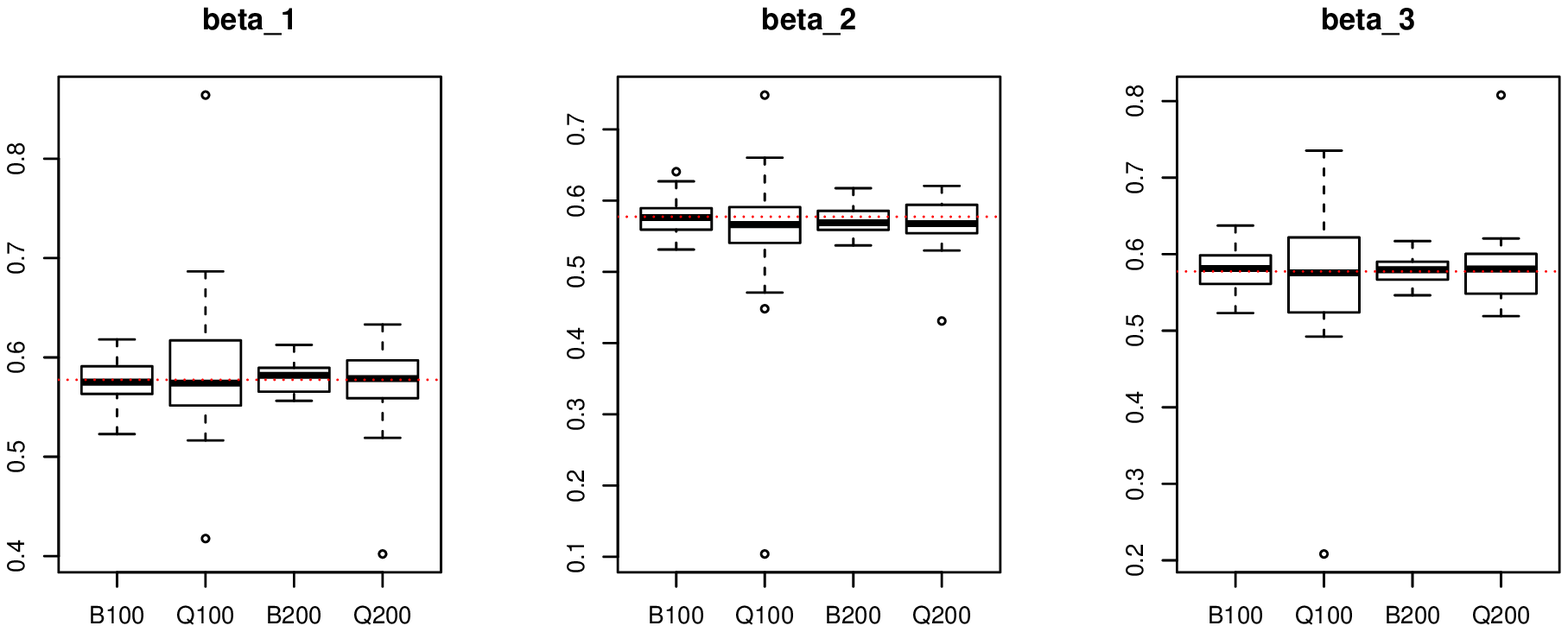}
\end{minipage}
\caption{Summarizing estimators of $\beta$ for $n=100,\,200$ in example 1. `B100' (`Q100') denotes BQSIM (QSIM) with $n=100$, for example.}
\label{boxplot_sim1_5}
\end{center}
\end{figure}

Figure \ref{boxplot_sim1_5} displays the boxplots of the estimated index vectors, comparing BQSIM and QSIM, with $\tau\in \{0.1,0.25,0.5,0.75,0.9\}$. To save space, cases with $\tau=0.95$ and $0.99$ are not presented. The plots are based on 100 independently generated datasets in each case and show that the Bayesian estimates have smaller bias and lower variance. These plots generally give the impression that BQSIM produces more precise and stable estimates than QSIM. Mean squared errors (MSE) of the estimates based on these 100 replications in each case are shown in Table \ref{result_sim1_2} for both sample sizes and all seven quantile levels. 

\begin{figure}[ht!]
\begin{center}
\includegraphics[width=5in,trim=0 50 0 0]{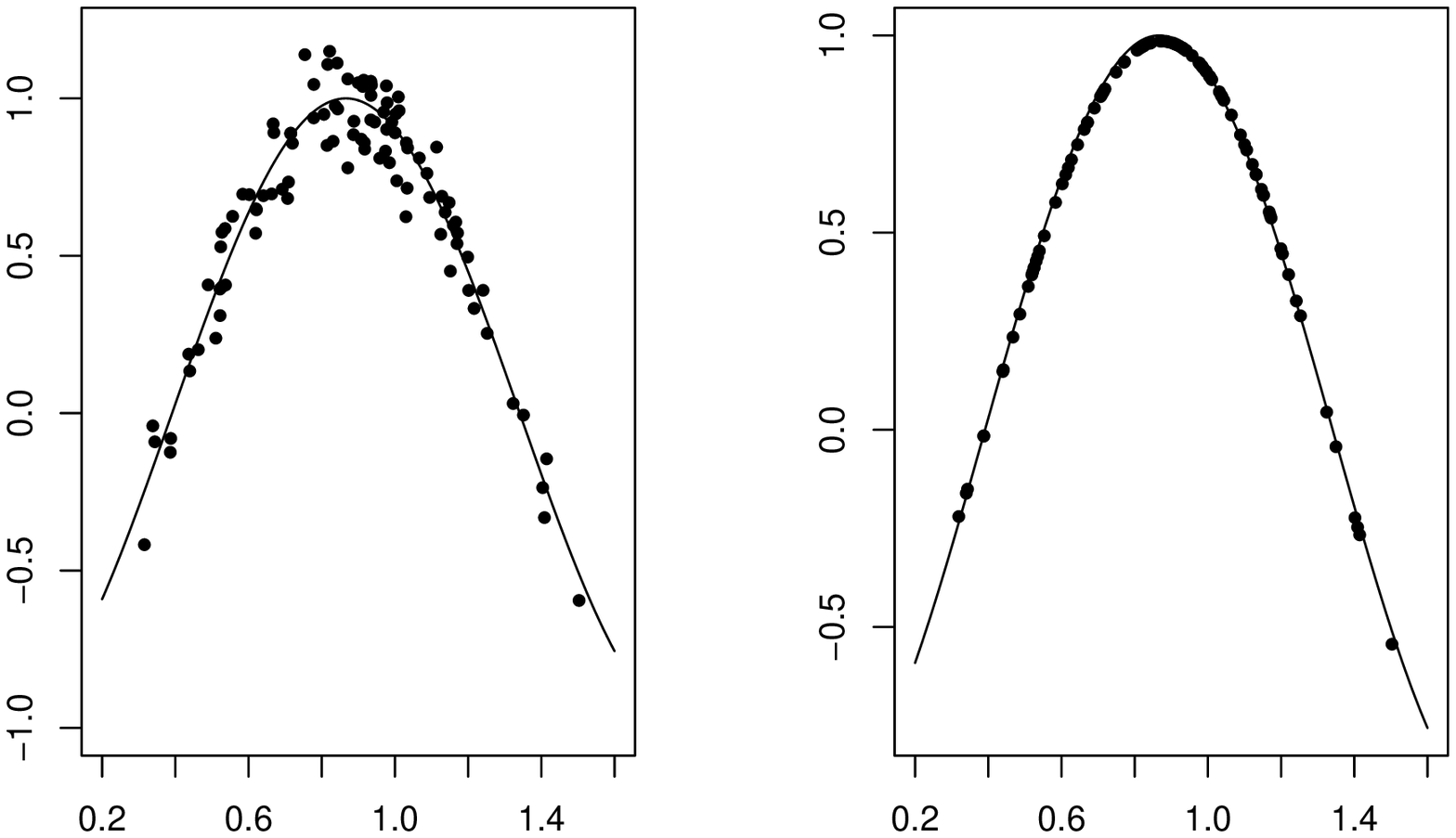}
\end{center}
\caption{The estimated link functions. On the left panel, the values of estimated $\boldsymbol{\eta}_n$ are plotted against the  true index, while for the right panel the values of estimated $\boldsymbol{\eta}_n$ are plotted against the estimated index. }
\label{linkfunction_sim1}
\end{figure}

Figure \ref{linkfunction_sim1} shows the fitted $\boldsymbol{\eta}_n$ values (posterior mean) by BQSIM at $\tau=0.5$ on a typical run. In the left panel, the fitted $\boldsymbol{\eta}_n$ are plotted against the true index $\boldsymbol{x}_i^T\boldsymbol{\beta}$ where $\boldsymbol{\beta}$ is the true value in the model. The true link function is also shown on the same figure. In the right panel, the fitted $\boldsymbol{\eta}_n$ are plotted against the fitted index $\boldsymbol{x}_i^T\hat{\boldsymbol{\beta}}$, where $\hat{\boldsymbol{\beta}}$ is the posterior mean estimate, leading to visually smoother fitted values.

\begin{figure}[ht!]
\begin{center}
\includegraphics[width=5in]{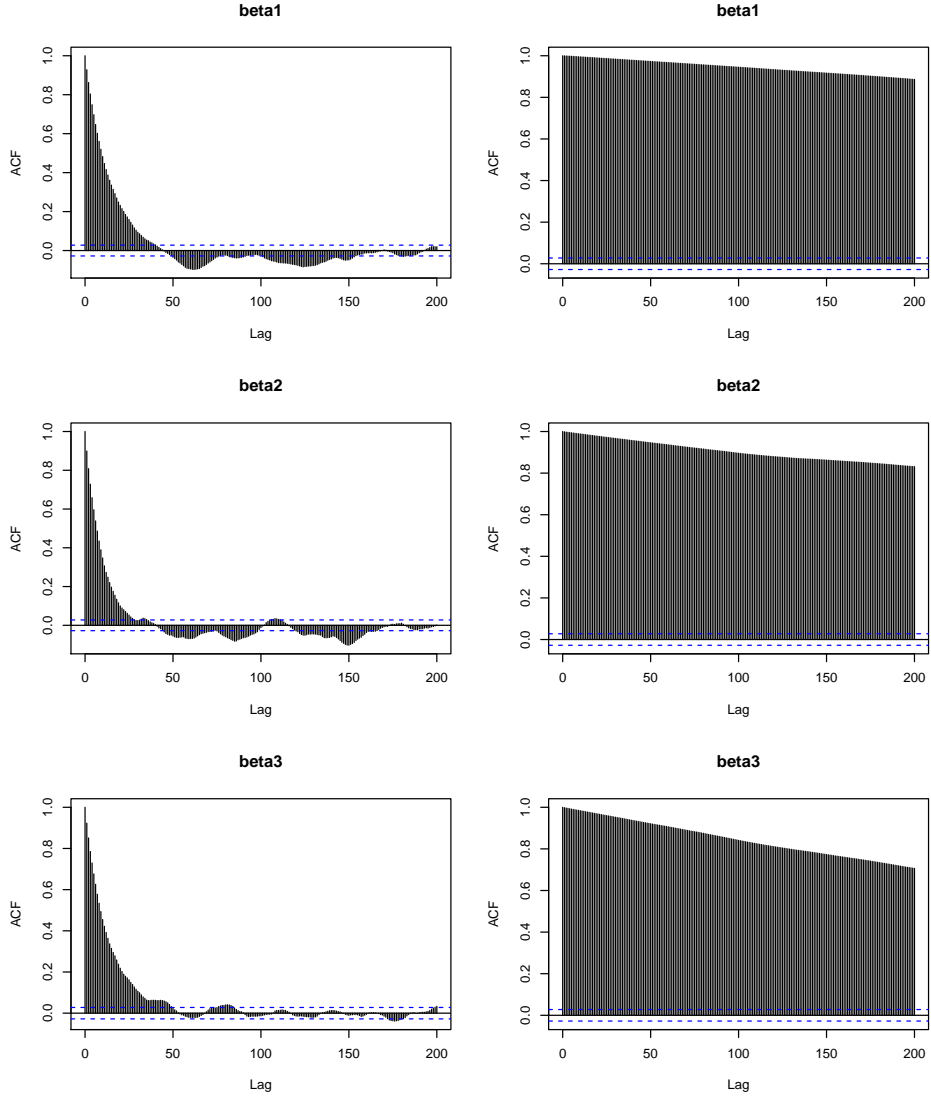}
\end{center}
\caption{The autocorrelation plots for $\beta_1,\beta_2,\beta_3$. The left column represents the series produced by our partially collapsed sampler and the right column represents the uncollapsed sampler.  }
\label{fig:auto}
\end{figure}

Finally, to demonstrate that partial collapsing can significantly improve mixing, we show the autocorrelation plots of the $\boldsymbol{\beta}$ series in Figure \ref{fig:auto}. Observe that the autocorrelation of the series produced by the uncollapsed chain is much higher, implying the samples are much ``stickier". As discussed in the Appendix, $\boldsymbol{C}_n$ is nearly singular which caused numerical problems when using the uncollapsed Gibbs sampler. To avoid this numerical problem, a small artificial nugget effect is added to the matrix (that is we use $\boldsymbol{C}_n+\epsilon I$ in place of $\boldsymbol{C}_n$ in evaluating the full conditional density, with $\epsilon=10^{-5}$).

\begin{table}[ht!]
\footnotesize
\caption{Results of BQSIM for simulation example 1, $n=100$.}
\label{result_sim1_1}
\vspace{0.1in}
\centering
\begin{tabular}{cccccccc}

\toprule
\multicolumn{2}{c}{$n=100$} &{True}&{Mean} &{Median}&{S.D.}&{2.5\%}&{97.5\%}\\
\midrule
\multirow{3}{*} {$\tau$=0.10} & $\mathbf{\beta}_{1}$&0.5774&0.5666&0.5817&0.0157&0.5354&0.5970\\
                              & $\mathbf{\beta}_{2}$&0.5774&0.5739&0.5749&0.0159&0.5420&0.6044\\
                              & $\mathbf{\beta}_{3}$&0.5774&0.5741&0.5770&0.0164&0.5410&0.6052                                   
                                   \\\midrule
\multirow{3}{*} {$\tau$=0.25} & $\mathbf{\beta}_{1}$&0.5774&0.5785&0.5766&0.0154&0.5481&0.6084\\
                              & $\mathbf{\beta}_{2}$&0.5774&0.5748&0.5734&0.0153&0.5442&0.6047\\
                              & $\mathbf{\beta}_{3}$&0.5774&0.5773&0.5783&0.0155&0.5466&0.6071                                  
                                   \\\midrule
\multirow{3}{*} {$\tau$=0.50} & $\mathbf{\beta}_{1}$&0.5774&0.5773&0.5775&0.0157&0.5465&0.6080\\
                              & $\mathbf{\beta}_{2}$&0.5774&0.5732&0.5741&0.0158&0.5426&0.6043\\
                              & $\mathbf{\beta}_{3}$&0.5774&0.5802&0.5815&0.0159&0.5486&0.6118                                              
                                                                     \\\midrule                                   
\multirow{3}{*} {$\tau$=0.75} & $\mathbf{\beta}_{1}$&0.5774&0.5756&0.5755&0.0156&0.5455&0.6067\\
                              & $\mathbf{\beta}_{2}$&0.5774&0.5738&0.5740&0.0156&0.5429&0.6046\\
                              & $\mathbf{\beta}_{3}$&0.5774&0.5812&0.5802&0.0157&0.5503&0.6118  
                                    \\\midrule                           
\multirow{3}{*} {$\tau$=0.90} & $\mathbf{\beta}_{1}$&0.5774&0.5760&0.5751&0.0157&0.5453&0.6071\\
                              & $\mathbf{\beta}_{2}$&0.5774&0.5751&0.5762&0.0152&0.5458&0.6053\\
                              & $\mathbf{\beta}_{3}$&0.5774&0.5790&0.5811&0.0155&0.5490&0.6102
                                                                                                  \\\midrule                                   
\multirow{3}{*} {$\tau$=0.95} & $\mathbf{\beta}_{1}$&0.5774&0.5819&0.5823&0.0159&0.5511&0.6127\\
                              & $\mathbf{\beta}_{2}$&0.5774&0.5745&0.5754&0.0154&0.5443&0.6047\\
                              & $\mathbf{\beta}_{3}$&0.5774&0.5730&0.5735&0.0157&0.5432&0.6045  
                                    \\\midrule                           
\multirow{3}{*} {$\tau$=0.99} & $\mathbf{\beta}_{1}$&0.5774&0.5771&0.5786&0.0212&0.5338&0.6142\\
                              & $\mathbf{\beta}_{2}$&0.5774&0.5710&0.5668&0.0184&0.5362&0.6077\\
                              & $\mathbf{\beta}_{3}$&0.5774&0.5729&0.5802&0.0225&0.5304&0.6124                                        
                                                                     \\\bottomrule                                    
\end{tabular}
\end{table}

\begin{table}[ht!]
\footnotesize
\caption{Results of BQSIM for simulation example 1, $n=200$.}
\label{result_sim1_1cont}
\vspace{0.1in}
\centering
\begin{tabular}{cccccccc}

\toprule
\multicolumn{2}{c}{$n=200$} &{True}&{Mean} &{Median}&{S.D.}&{2.5\%}&{97.5\%}\\
\midrule
\multirow{3}{*} {$\tau$=0.10} & $\mathbf{\beta}_{1}$&0.5774&0.5742&0.5745&0.0104&0.5543&0.5947\\
                              & $\mathbf{\beta}_{2}$&0.5774&0.5749&0.5768&0.0010&0.5567&0.5959\\
                              & $\mathbf{\beta}_{3}$&0.5774&0.5821&0.5789&0.0104&0.5592&0.6002                                   
                                   \\\midrule
\multirow{3}{*} {$\tau$=0.25} & $\mathbf{\beta}_{1}$&0.5774&0.5761&0.5756&0.0106&0.5552&0.5968\\
                              & $\mathbf{\beta}_{2}$&0.5774&0.5764&0.5782&0.0105&0.5562&0.5972\\
                              & $\mathbf{\beta}_{3}$&0.5774&0.5788&0.5771&0.0106&0.5575&0.5991                                  
                                   \\\midrule
\multirow{3}{*} {$\tau$=0.50} & $\mathbf{\beta}_{1}$&0.5774&0.5771&0.5779&0.0106&0.5561&0.5971\\
                              & $\mathbf{\beta}_{2}$&0.5774&0.5760&0.5779&0.0106&0.5561&0.5976\\
                              & $\mathbf{\beta}_{3}$&0.5774&0.5783&0.5766&0.0104&0.5571&0.5979                                              
                                                                     \\\midrule                                   
\multirow{3}{*} {$\tau$=0.75} & $\mathbf{\beta}_{1}$&0.5774&0.5769&0.5752&0.0107&0.5567&0.5987\\
                              & $\mathbf{\beta}_{2}$&0.5774&0.5754&0.5746&0.0109&0.5548&0.5973\\
                              & $\mathbf{\beta}_{3}$&0.5774&0.5790&0.5803&0.0109&0.5568&0.5992  
                                    \\\midrule                           
\multirow{3}{*} {$\tau$=0.90} & $\mathbf{\beta}_{1}$&0.5774&0.5817&0.5802&0.0108&0.5600&0.6027\\
                              & $\mathbf{\beta}_{2}$&0.5774&0.5816&0.5802&0.0098&0.5630&0.6017\\
                              & $\mathbf{\beta}_{3}$&0.5774&0.5675&0.5644&0.0106&0.5471&0.5881
                                                                                                  \\\midrule                                   
\multirow{3}{*} {$\tau$=0.95} & $\mathbf{\beta}_{1}$&0.5774&0.5754&0.5766&0.0109&0.5546&0.5959\\
                              & $\mathbf{\beta}_{2}$&0.5774&0.5810&0.5809&0.0114&0.5599&0.6037\\
                              & $\mathbf{\beta}_{3}$&0.5774&0.5743&0.5738&0.0105&0.5543&0.5951  
                                    \\\midrule                           
\multirow{3}{*} {$\tau$=0.99} & $\mathbf{\beta}_{1}$&0.5774&0.5590&0.5749&0.0241&0.5244&0.6029\\
                              & $\mathbf{\beta}_{2}$&0.5774&0.5700&0.5780&0.0155&0.5425&0.5983\\
                              & $\mathbf{\beta}_{3}$&0.5774&0.5601&0.5722&0.0218&0.5277&0.5984                                        
                                                                     \\\bottomrule                                    
\end{tabular}
\end{table}

\begin{table}[ht!]
\footnotesize
\caption{Comparison of MSE for BQSIM (using posterior mean as the estimator) and QSIM based on 100 replications in each case for simulation example 1.}
\label{result_sim1_2}
\vspace{0.1in}
\centering 
\begin{tabular}{llcccccc}
\toprule \multicolumn{2}{c}{}&\multicolumn{3}{c}{MSE ($n=100$)}&\multicolumn{3}{c}{MSE ($n=200$)}\\
\cmidrule(r){3-5} \cmidrule(r){6-8}
\multicolumn{2}{c}{} &{$\beta_{1}$} &{$\beta_{2}$}&{$\beta_{3}$}&{$\beta_{1}$}&{$\beta_{2}$}&{$\beta_{3}$}\\
\midrule
\multirow{2}{*} {$\tau$=0.10} & $\rm{QSIM}$ &0.00239&0.00447&0.00886&0.00346&0.00473&0.00052\\
                              & $\rm{BQSIM}$&0.00041&0.00051&0.00045&0.00017&0.00023&0.00020 
                              \\\midrule
\multirow{2}{*} {$\tau$=0.25} & $\rm{QSIM}$ &0.00169&0.00312&0.00303&0.00276&0.00138&0.00310\\
                              & $\rm{BQSIM}$&0.00029&0.00029&0.00030&0.00012&0.00014&0.00019
                              \\\midrule
\multirow{2}{*} {$\tau$=0.50} & $\rm{QSIM}$ &0.00269&0.00154&0.00254&0.00058&0.00053&0.00089\\
                              & $\rm{BQSIM}$&0.00023&0.00025&0.00029&0.00013&0.00013&0.00018
                              \\\midrule
\multirow{2}{*} {$\tau$=0.75} & $\rm{QSIM}$ &0.00340&0.00601&0.00291&0.00039&0.00028&0.00048\\
                              & $\rm{BQSIM}$&0.00026&0.00029&0.00038&0.00017&0.00016&0.00016
                              \\\midrule
\multirow{2}{*} {$\tau$=0.90} & $\rm{QSIM}$ &0.00424&0.00794&0.00641&0.00139&0.00100&0.00192\\
                              & $\rm{BQSIM}$&0.00040&0.00049&0.00059&0.00023&0.00028&0.00030
                              \\\midrule
\multirow{2}{*} {$\tau$=0.95} & $\rm{QSIM}$ &0.00857&0.00739&0.00951&0.00157&0.00183&0.00245\\
                              & $\rm{BQSIM}$&0.00051&0.00075&0.00094&0.00042&0.00029&0.00041
                              \\\midrule
\multirow{2}{*} {$\tau$=0.99} & $\rm{QSIM}$ &0.05034&0.04259&0.06166&0.00494&0.09708&0.07046\\
                              & $\rm{BQSIM}$&0.00083&0.00098&0.00125&0.00062&0.00192&0.00099\\
                                                          
\bottomrule
\end{tabular}
\end{table}

\subsubsection*{Example 2}
\begin{figure}[ht!]
\begin{center}
\begin{minipage}{0.75in}
$\tau = 0.1$ 
\end{minipage}
\begin{minipage}{4.5in}
\includegraphics[width=3.5in,trim=0 70 0 20,clip=TRUE]{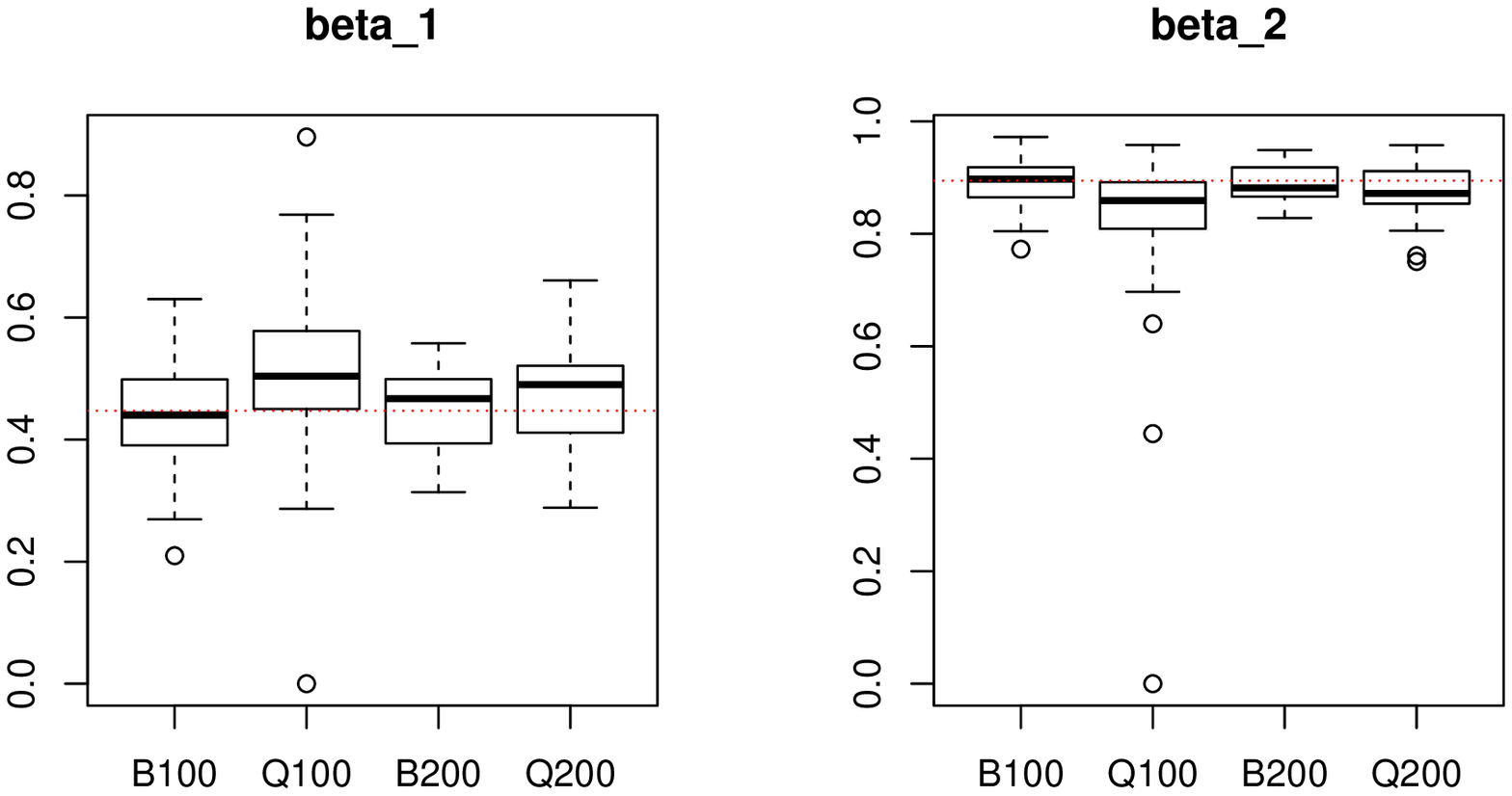}
\end{minipage}

\begin{minipage}{0.75in}
$\tau = 0.25$ 
\end{minipage}
\begin{minipage}{4.5in}
\includegraphics[width=3.5in,trim=0 70 0 50,clip=TRUE]{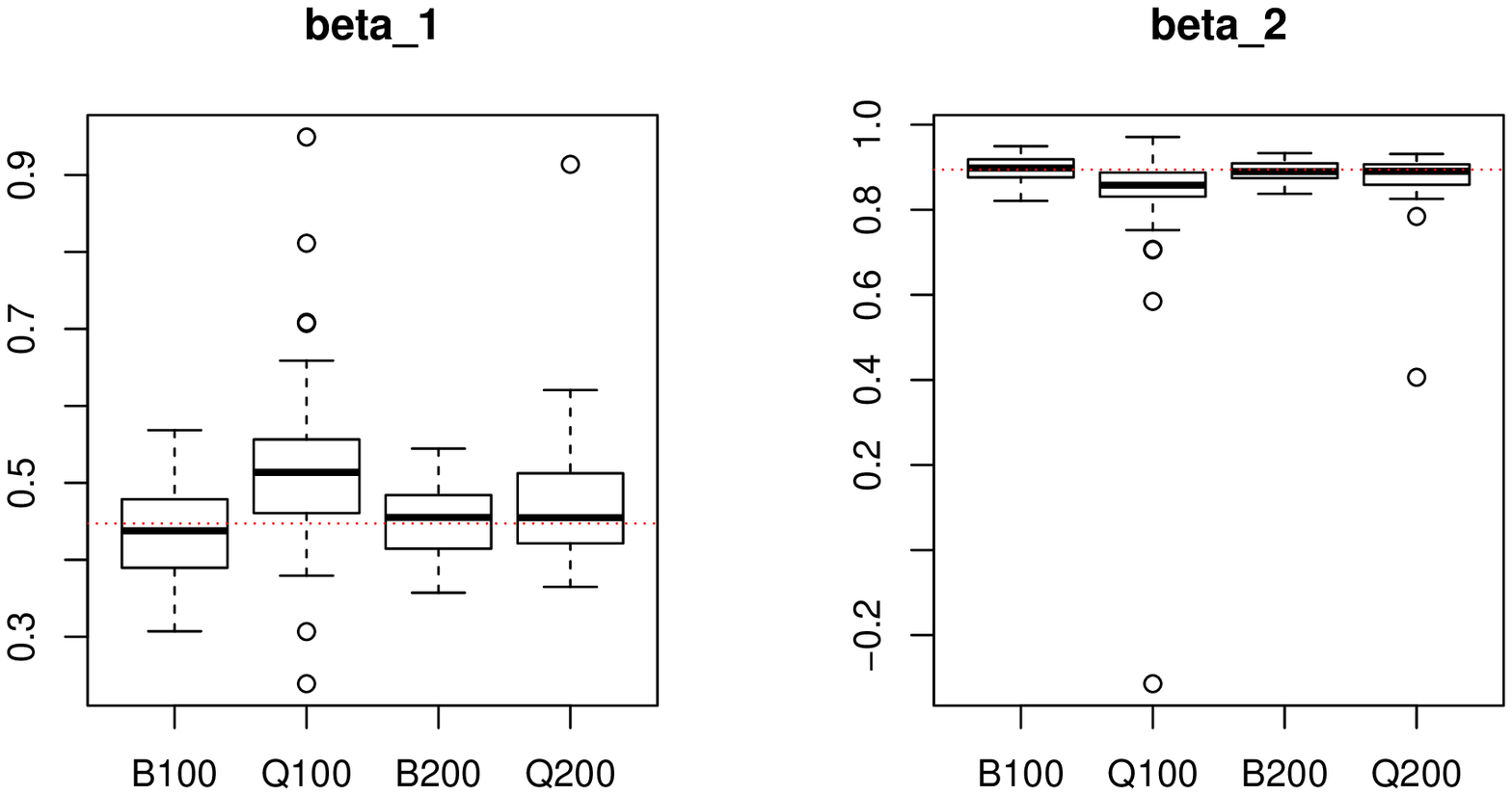}
\end{minipage}

\begin{minipage}{0.75in}
$\tau = 0.5$ 
\end{minipage}
\begin{minipage}{4.5in}
\includegraphics[width=3.5in,trim=0 70 0 50,clip=TRUE]{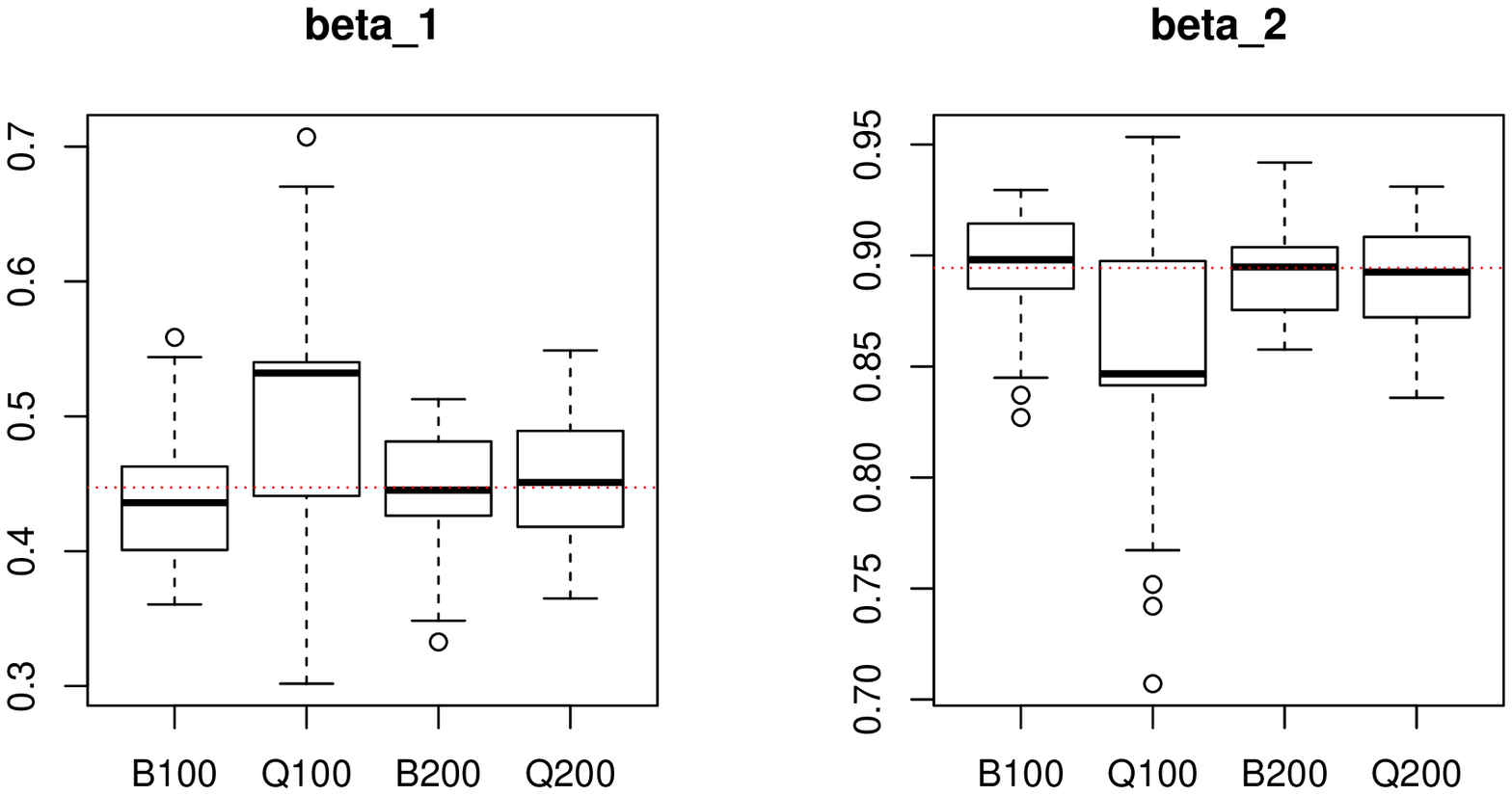}
\end{minipage}

\begin{minipage}{0.75in}
$\tau = 0.75$ 
\end{minipage}
\begin{minipage}{4.5in}
\includegraphics[width=3.5in,trim=0 70 0 50,clip=TRUE]{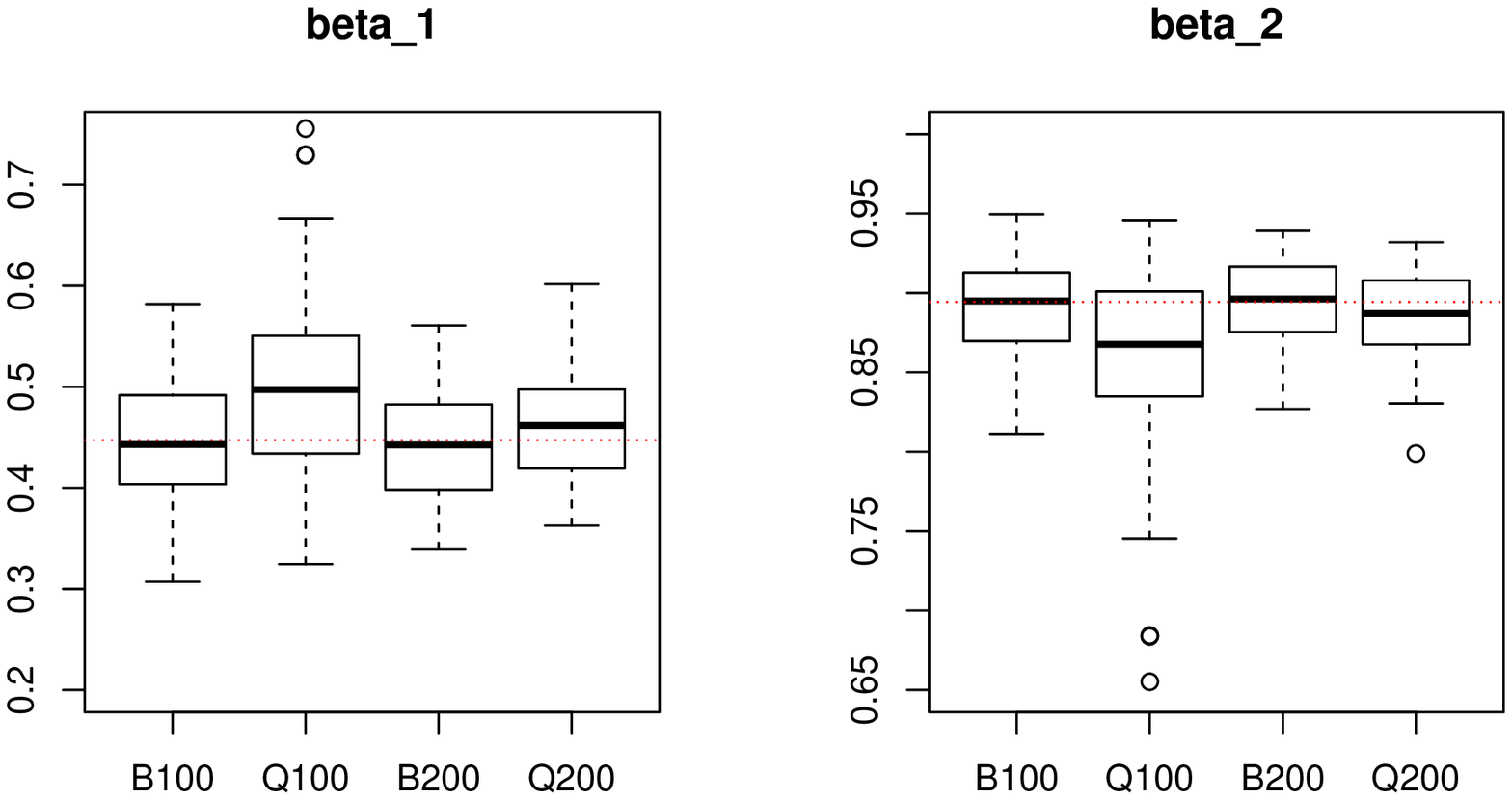}
\end{minipage}

\begin{minipage}{0.75in}
$\tau = 0.9$ 
\end{minipage}
\begin{minipage}{4.5in}
\includegraphics[width=3.5in,trim=0 40 0 50,clip=TRUE]{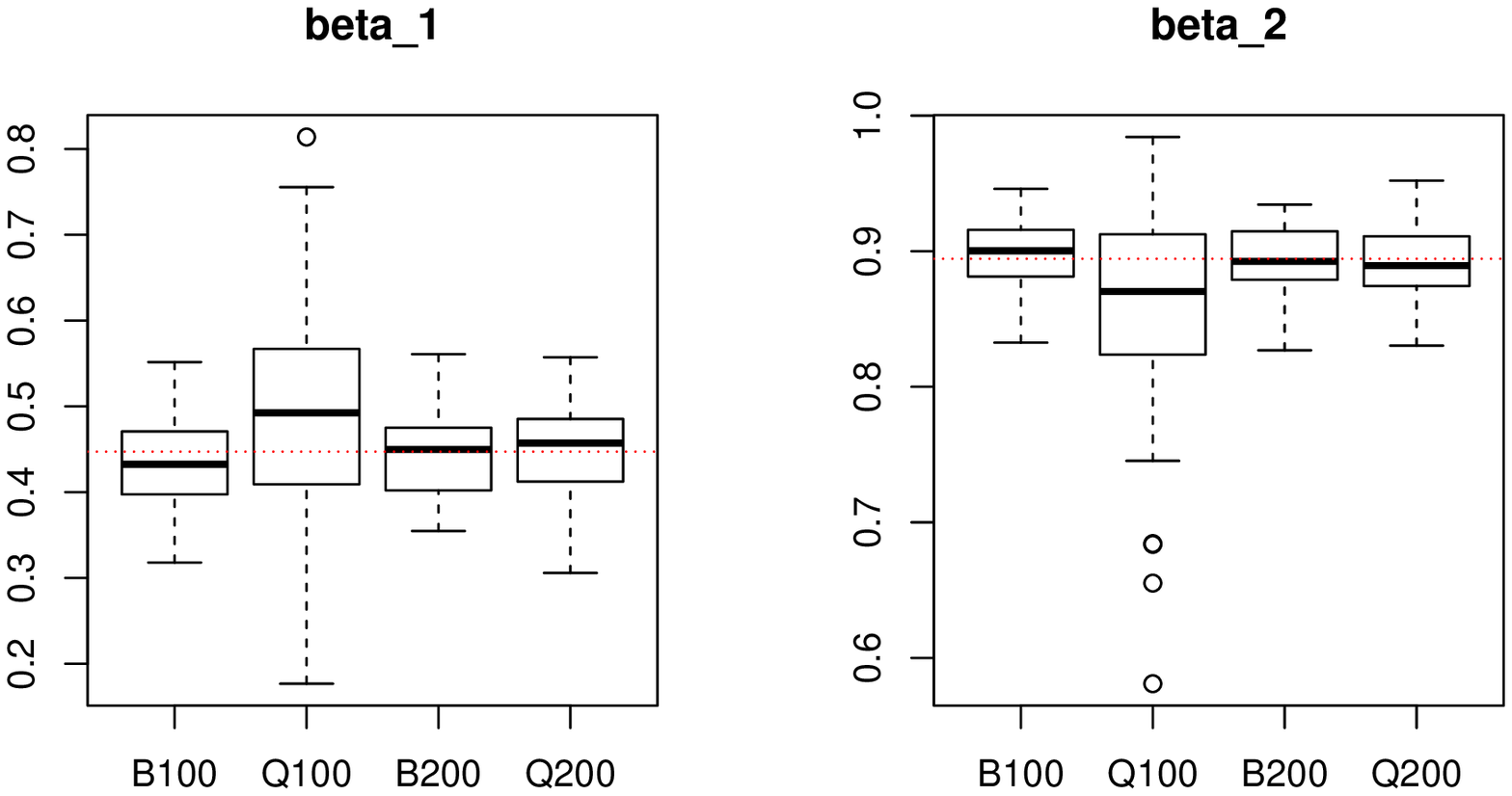}
\end{minipage}
\caption{Summarizing estimators of $\beta$ for $n=100,\,200$ in example 2.}
\label{boxplot_sim2_5}
\end{center}
\end{figure}

Now consider data generated as follows:
\begin{equation*}\label{SIM2}
 y=\eta(\boldsymbol{x}^{T}\boldsymbol{\beta})+\sqrt{(\rm{sin}(\boldsymbol{x}^{T}\boldsymbol{\beta})+1)}\mathcal{Z},\;\mbox{ where }\; \eta(t)=10\sin(0.75t),
\end{equation*}
and $\boldsymbol{\beta}=(\beta_{1},\beta_{2})^{T}=\frac{1}{\sqrt{5}}(1,2)^{T}$, $\boldsymbol{x}=(x_{1}, x_{2})^{T}$, and $\mathcal{Z}$ is a standard normal random variable. The $x_{j}$s, $(j=1,2)$ are drawn identically and independently from a normal distribution with mean 0 and variance $0.25^{2}$.  We conduct simulations at $\tau\in\{0.1,0.25,0.5,0.75,0.9,0.95,0.99\}$ with $n=100$ and $n=200$, each with 100 replications. Figure \ref{boxplot_sim2_5} shows the boxplots for the estimated index vector and Table \ref{result_sim2_2} reports the mean squared errors.

\begin{table}[ht!]
\footnotesize
\caption{Comparison of MSE for BQSIM (using posterior mean as the estimator) and QSIM based on 100 replications in each case for simulation example 2.}
\label{result_sim2_2}
\vspace{0.1in}
\centering 
\begin{tabular}{llcccc}
\toprule \multicolumn{2}{c}{}&\multicolumn{2}{c}{MSE ($n=100$)}&\multicolumn{2}{c}{MSE ($n=200$)}\\
\cmidrule(r){3-4} \cmidrule(r){5-6}
\multicolumn{2}{c}{} &{$\beta_{1}$} &{$\beta_{2}$}&{$\beta_{1}$}&{$\beta_{2}$}\\
\midrule
\multirow{2}{*} {$\tau$=0.10} & $\rm{QSIM}$ &0.02510& 0.02991&0.00744& 0.00238\\                            
                              & $\rm{BQSIM}$ &0.00841& 0.00197&0.00451& 0.00115
                              \\\midrule
\multirow{2}{*} {$\tau$=0.25} & $\rm{QSIM}$ &0.01997& 0.00396&0.00823& 0.00645 \\                            
                              & $\rm{BQSIM}$ &0.00401& 0.00095&0.00220& 0.00059
                              \\\midrule
\multirow{2}{*} {$\tau$=0.50} & $\rm{QSIM}$ &0.01098& 0.00409&0.00214& 0.00058\\                            
                              & $\rm{BQSIM}$ &0.00225& 0.00056&0.00179& 0.00041
                              \\\midrule
\multirow{2}{*} {$\tau$=0.75} & $\rm{QSIM}$ &0.02025 & 0.00868&0.00345& 0.00089\\                            
                              & $\rm{BQSIM}$ &0.00222&0.00059&0.00021& 0.00056 
                              \\\midrule
\multirow{2}{*} {$\tau$=0.90} & $\rm{QSIM}$ &0.01983& 0.01227&0.00584& 0.00196\\                            
                              & $\rm{BQSIM}$ &0.00393& 0.00104&0.00393& 0.00104
                                 \\\midrule
\multirow{2}{*} {$\tau$=0.95} & $\rm{QSIM}$ &0.02048 & 0.00974&0.01044& 0.01279\\                            
                              & $\rm{BQSIM}$ &0.00647& 0.00150&0.00488& 0.00126 
                              \\\midrule
\multirow{2}{*} {$\tau$=0.99} & $\rm{QSIM}$ &0.06640& 0.04652&0.05911& 0.03972\\                            
                              & $\rm{BQSIM}$ &0.00953& 0.00226&0.00997& 0.00210\\                                                
                              \bottomrule
\end{tabular}
\end{table}

\subsubsection*{Example 3}
\begin{figure}[ht!]
\begin{center}
\begin{minipage}{0.75in}
$\tau = 0.1$ 
\end{minipage}
\begin{minipage}{4.5in}
\includegraphics[width=3.5in,trim=0 70 0 20,clip=TRUE]{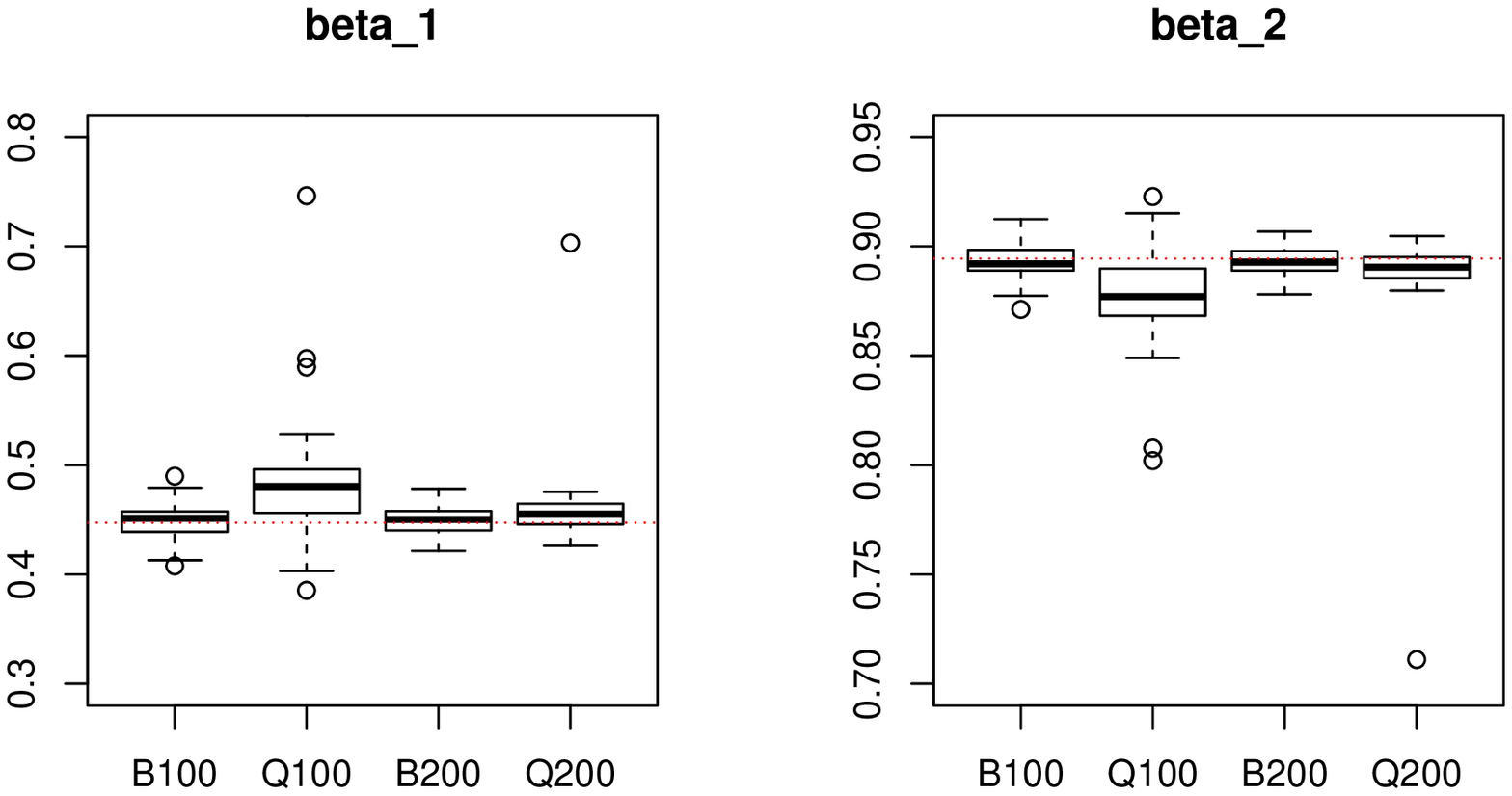}
\end{minipage}

\begin{minipage}{0.75in}
$\tau = 0.25$ 
\end{minipage}
\begin{minipage}{4.5in}
\includegraphics[width=3.5in,trim=0 70 0 50,clip=TRUE]{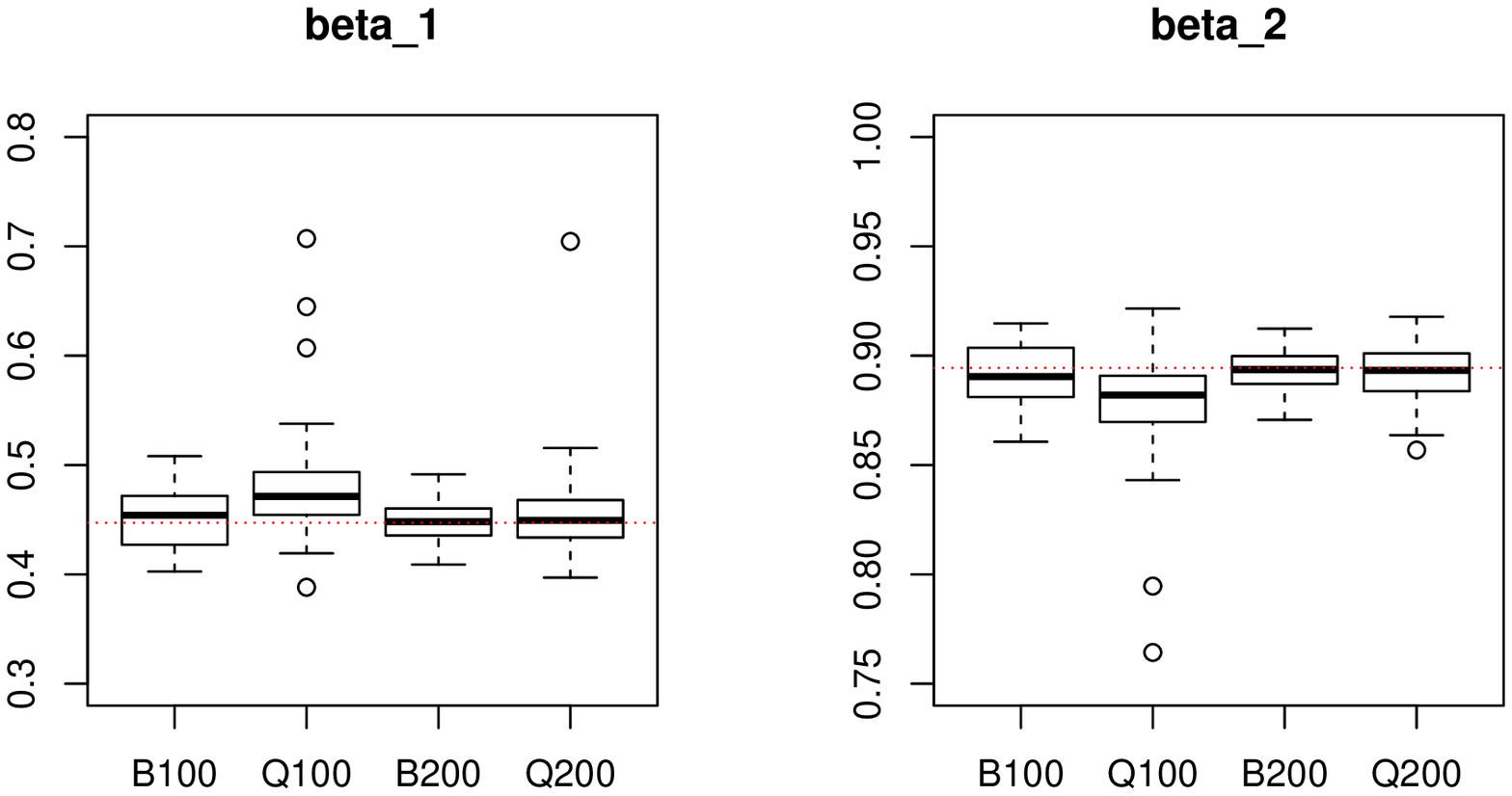}
\end{minipage}

\begin{minipage}{0.75in}
$\tau = 0.5$ 
\end{minipage}
\begin{minipage}{4.5in}
\includegraphics[width=3.5in,trim=0 70 0 50,clip=TRUE]{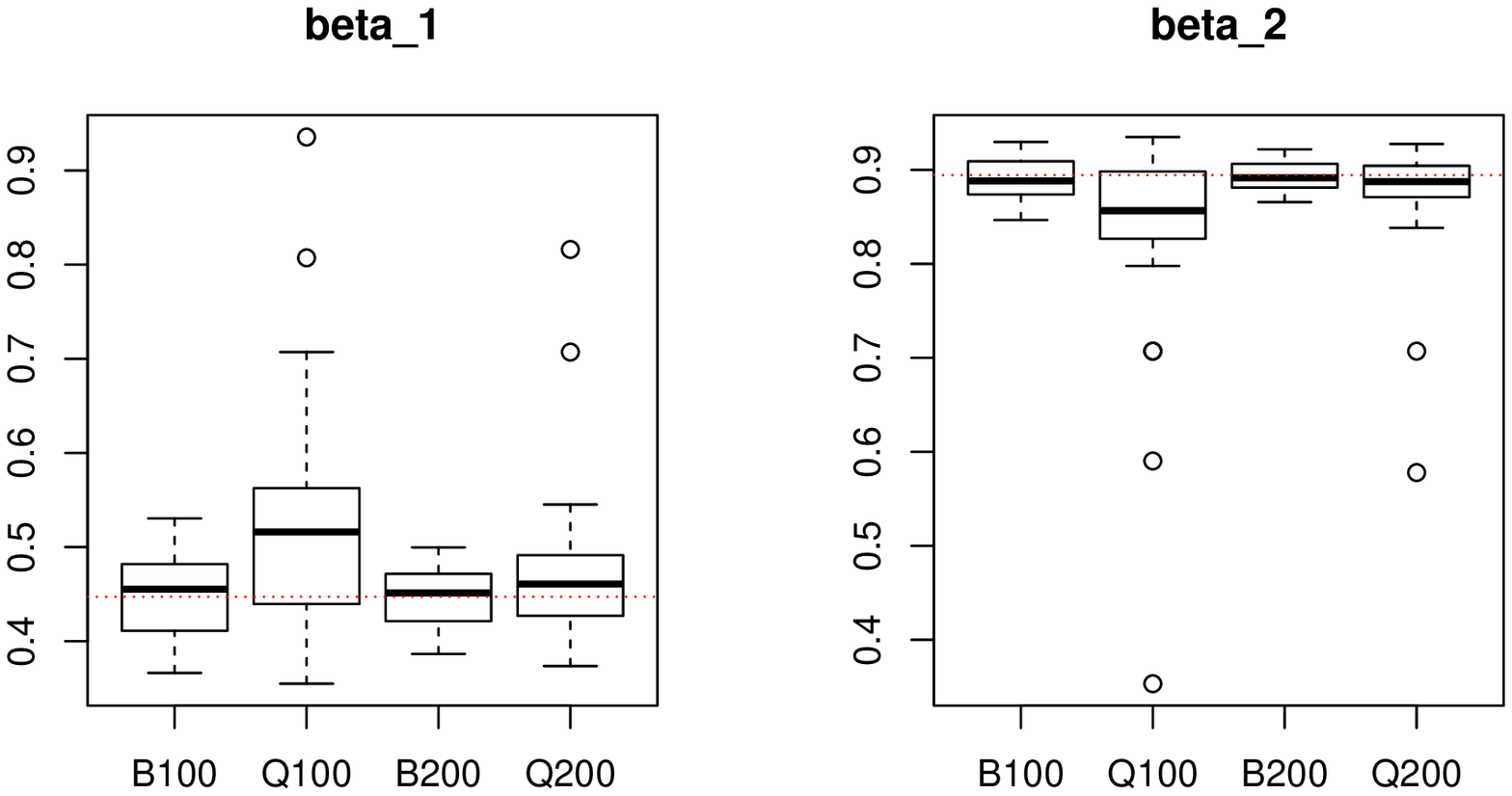}
\end{minipage}

\begin{minipage}{0.75in}
$\tau = 0.75$ 
\end{minipage}
\begin{minipage}{4.5in}
\includegraphics[width=3.5in,trim=0 70 0 50,clip=TRUE]{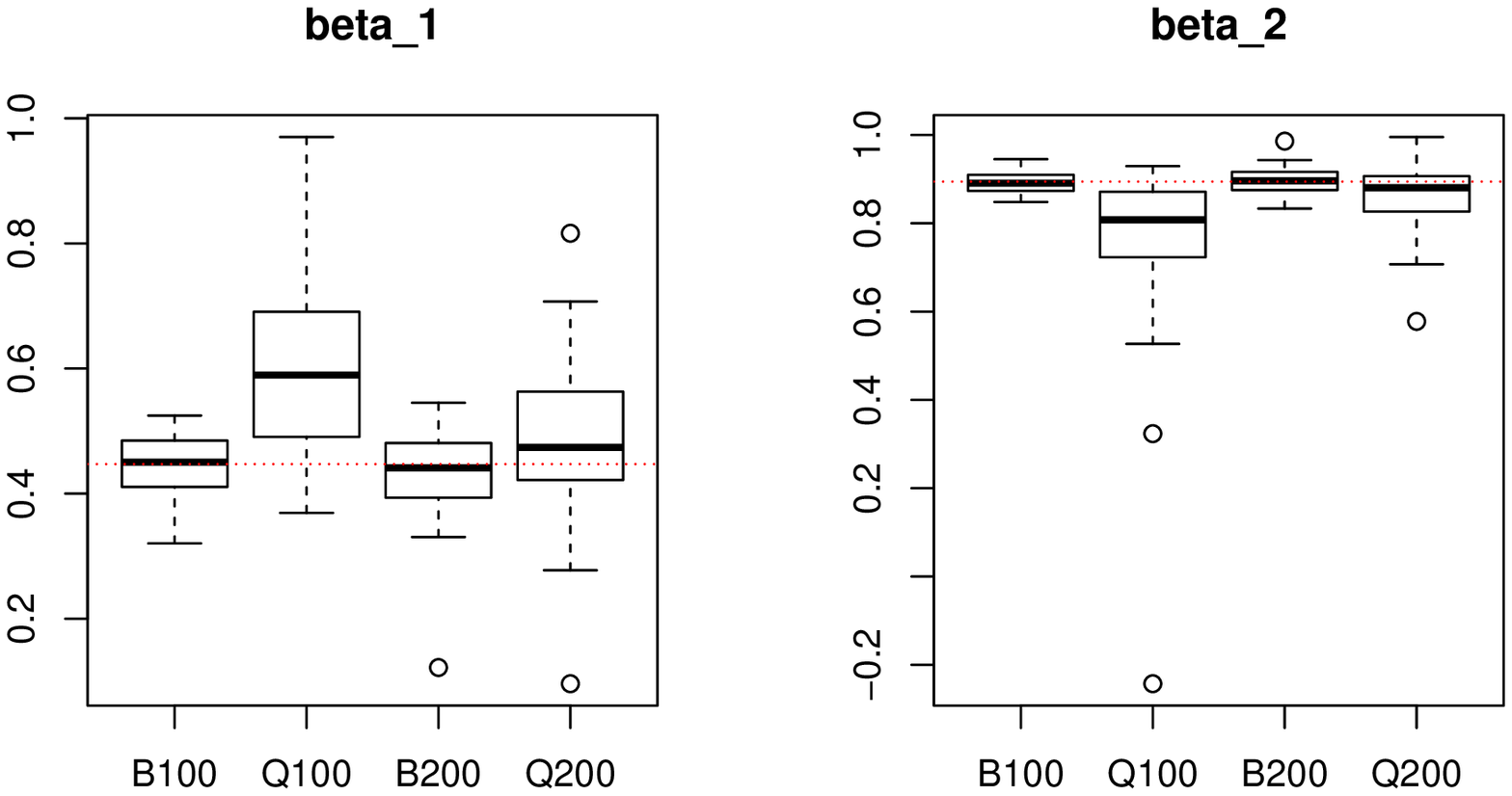}
\end{minipage}

\begin{minipage}{0.75in}
$\tau = 0.9$ 
\end{minipage}
\begin{minipage}{4.5in}
\includegraphics[width=3.5in,trim=0 40 0 50,clip=TRUE]{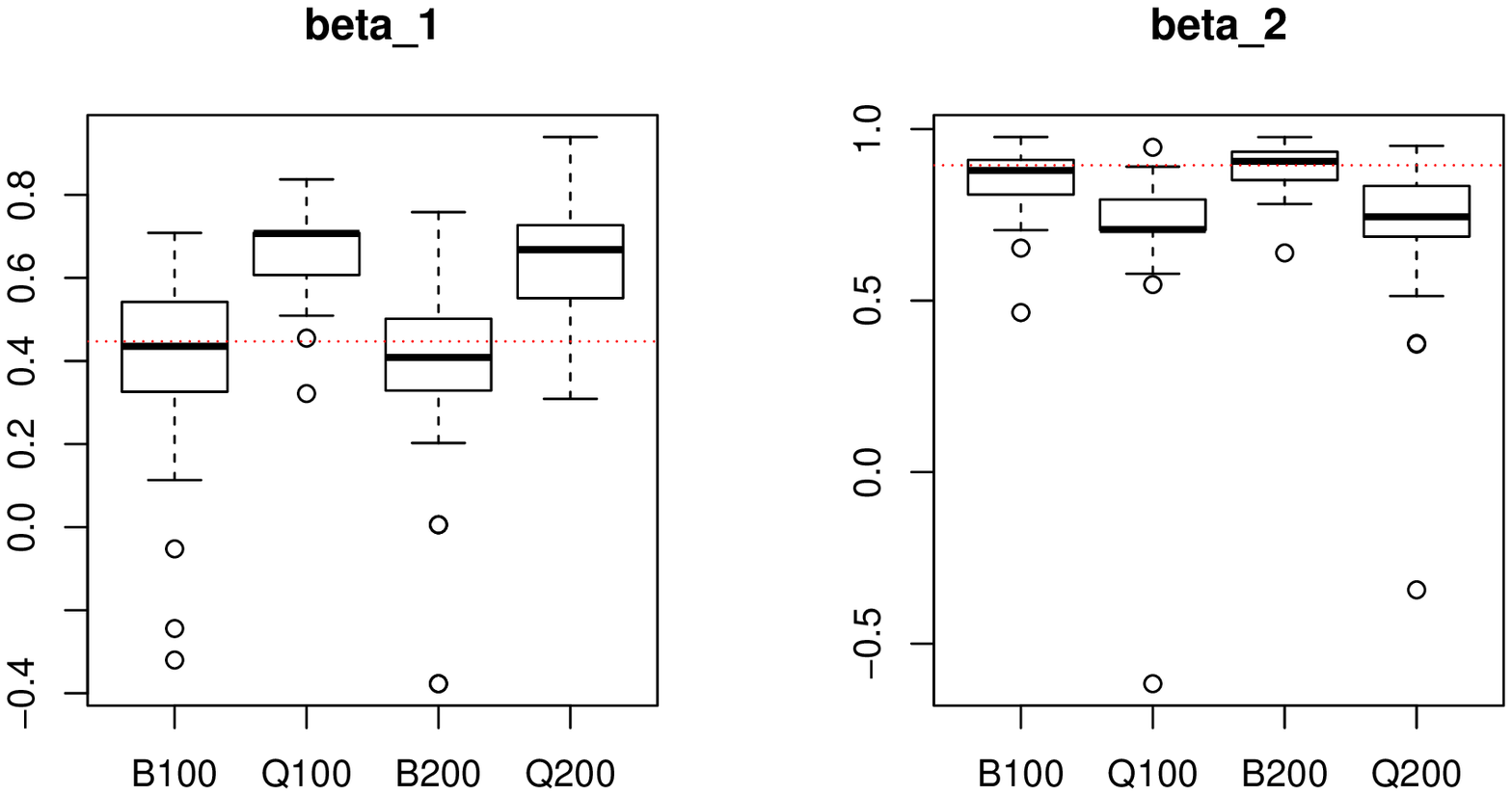}
\end{minipage}
\caption{Summarizing estimators of $\beta$ for $n=100,\,200$ in example 3.}
\label{boxplot_sim3_5}
\end{center}
\end{figure}

Next, we consider a regression model with exponentially distributed errors,
\begin{equation*}
\label{SIM3}
y=\eta(\boldsymbol{x}^{T}\boldsymbol{\beta})+\mathcal{E},\; \eta(t)=5\rm{cos}(t)+\exp(-t^{2}),
\end{equation*}
where $\boldsymbol{\beta}=(\beta_{1},\beta_{2})^{T}=\frac{1}{\sqrt{5}}(1,2)^{T}$,$\boldsymbol{x}=(x_{1}, x_{2})^{T}$. $x_{j}\stackrel{i.i.d.}{\sim} N(0,1)$, $j=1,2$, and $\mathcal{E}\sim\exp (1/2)$. Using the same sample sizes and quantile levels as for the previous two examples, the results are presented in Figure \ref{boxplot_sim3_5} and Table \ref{result_sim3_2}, which again demonstrate the superiority of BQSIM.

\begin{table}[ht!]
\footnotesize 
\caption{Comparison of MSE for BQSIM (using posterior mean as the estimator) and QSIM based on 100 replications in each case for simulation example 3.}
\label{result_sim3_2}
\centering 
\vspace{0.1in}
\begin{tabular}{llcccc}
\toprule \multicolumn{2}{c}{}&\multicolumn{2}{c}{MSE ($n=100$)}&\multicolumn{2}{c}{MSE ($n=200$)}\\
\cmidrule(r){3-4} \cmidrule(r){5-6}
\multicolumn{2}{c}{} &{$\beta_{1}$} &{$\beta_{2}$}&{$\beta_{1}$}&{$\beta_{2}$}\\
\midrule
\multirow{2}{*} {$\tau$=0.10} & $\rm{QSIM}$ &0.00767& 0.00447&0.00174 & 0.00114 \\                            
                              & $\rm{BQSIM}$ &0.00033&0.00008&0.00018 & 0.00005
                              \\\midrule
\multirow{2}{*} {$\tau$=0.25} & $\rm{QSIM}$ &0.00421& 0.00175&0.00647& 0.00384 \\                            
                              & $\rm{BQSIM}$ &0.00074& 0.00020 &0.00035& 0.00009
                              \\\midrule
\multirow{2}{*} {$\tau$=0.50} & $\rm{QSIM}$ &0.00843 & 0.00306&0.00631& 0.00351\\                            
                              & $\rm{BQSIM}$ &0.00173 & 0.00044&0.00090& 0.00022
                              \\\midrule
\multirow{2}{*} {$\tau$=0.75} & $\rm{QSIM}$ &0.03992& 0.05156&0.01429& 0.00690\\                            
                              & $\rm{BQSIM}$ &0.00221 & 0.00050&0.00573 & 0.00098
                              \\\midrule
\multirow{2}{*} {$\tau$=0.90} & $\rm{QSIM}$ &0.06118 & 0.08449 &0.06311& 0.07486\\                            
                              & $\rm{BQSIM}$ &0.04856& 0.01129 &0.04571 & 0.00219 
                                  \\\midrule
\multirow{2}{*} {$\tau$=0.95} & $\rm{QSIM}$ &0.07744& 0.05669&0.09499& 0.17752\\                            
                              & $\rm{BQSIM}$ &0.07094& 0.05106&0.08097 & 0.05106
                              \\\midrule
\multirow{2}{*} {$\tau$=0.99} & $\rm{QSIM}$ &0.10262& 0.10510 &0.12993& 0.13203\\                            
                              & $\rm{BQSIM}$ &0.10064& 0.05996&0.12071 & 0.06277\\                                                         
                              \bottomrule
\end{tabular}
\end{table}

\subsubsection*{Example 4}
Here we follow a similar setup as in Example 1, except that the additive errors follow the ALD with $\sigma=0.05$. That is we generate data sets from the model (\ref{quantile}), independently for each value of $\tau$. This example mainly serves as an illustration that when the errors indeed follow ALD, so that estimation of $\sigma$ becomes meaningful, we can indeed estimate its value satisfactorily.
These results are presented in Tables \ref{result_ALD_100} and \ref{result_ALD_200} for $n=100$ and $n=200$ respectively. We emphasize that the estimated $\sigma$ is meaningful only if the true ALD is used in estimation. As an illustration of this point, we consider data generated from (\ref{quantile}) using $\tau=0.5$ and fitted using BQSIM at quantile level $\tau=0.75$. In this case, the index vector $\boldsymbol{\beta}$ can still be estimated very close to the true values, while the estimated $\sigma$ is about $0.01$. In general, BQSIM cannot be used to estimate the error distribution directly. This limitation is discussed further in Section \ref{sec:dis}. In this example, the performance at $\tau=0.99$ is less satisfactory than in previous examples. 

\begin{table}[ht!]
\footnotesize
\caption{Results of BQSIM for simulation example 4, $n=100$.}
\label{result_ALD_100}
\vspace{0.1in}
\centering
\begin{tabular}{ccccccccc}

\toprule
\multicolumn{2}{c}{$n=100$} &{True}&{Mean} &{Median}&{S.D.}&{2.5\%}&{97.5\%}&{MSE}\\
\midrule
\multirow{4}{*} {$\tau$=0.10} & $\mathbf{\beta}_{1}$&0.5774&0.5805&0.5818&0.0353&0.5084&0.6476&0.00117\\
                               & $\mathbf{\beta}_{2}$&0.5774&0.5814&0.5771&0.0338&0.5123&0.6463&0.00069\\
                               & $\mathbf{\beta}_{3}$&0.5774&0.5641&0.5615&0.0350&0.4923&0.6304&0.00132\\                                                             & $\mathbf{\sigma}$&0.0500&0.0521&0.0527&0.0054&0.0430&0.0637&0.00002
                                   \\\midrule
\multirow{4}{*} {$\tau$=0.25} & $\mathbf{\beta}_{1}$&0.5774&0.5809&0.5823&0.0215&0.5377&0.6227&0.00055\\
                               & $\mathbf{\beta}_{2}$&0.5774&0.5762&0.5760&0.0212&0.5341&0.6171&0.00043\\
                               & $\mathbf{\beta}_{3}$&0.5774&0.5723&0.5704&0.0218&0.5290&0.6145&0.00063\\                                                  & $\mathbf{\sigma}$&0.0500&0.0520&0.0520&0.0054&0.0427&0.0638&0.00003
                                   \\\midrule
\multirow{4}{*} {$\tau$=0.50} & $\mathbf{\beta}_{1}$&0.5774&0.5787&0.5786&0.0176&0.5432&0.6129&0.00030\\
                               & $\mathbf{\beta}_{2}$&0.5774&0.5740&0.5759&0.0176&0.5394&0.6084&0.00026\\
                               & $\mathbf{\beta}_{3}$&0.5774&0.5778&0.5783&0.0179&0.5425&0.6127&0.00030\\                                                             & $\mathbf{\sigma}$&0.0500&0.0514&0.0515&0.0053&0.0421&0.0631&0.00003
                                   \\\midrule
\multirow{4}{*} {$\tau$=0.75} & $\mathbf{\beta}_{1}$&0.5774&0.5750&0.5765&0.0208&0.5346&0.6165&0.00041\\
                               & $\mathbf{\beta}_{2}$&0.5774&0.5745&0.5750&0.0210&0.5330&0.6155&0.00047\\
                               & $\mathbf{\beta}_{3}$&0.5774&0.5802&0.5780&0.0212&0.5384&0.6219&0.00043\\                                                             & $\mathbf{\sigma}$&0.0500&0.0519&0.0514&0.0053&0.0423&0.0634&0.00004
                                   \\\midrule
\multirow{4}{*} {$\tau$=0.90} & $\mathbf{\beta}_{1}$&0.5774&0.5734&0.5682&0.0342&0.5055&0.6401&0.00083\\
                               & $\mathbf{\beta}_{2}$&0.5774&0.5685&0.5672&0.0336&0.5020&0.6343&0.00094\\
                               & $\mathbf{\beta}_{3}$&0.5774&0.5842&0.5829&0.0342&0.5178&0.6521&0.00131\\                                                             & $\mathbf{\sigma}$&0.0500&0.0514&0.0517&0.0054&0.0422&0.0633&0.00003
                                   \\\midrule
\multirow{4}{*} {$\tau$=0.95} & $\mathbf{\beta}_{1}$&0.5774&0.5455&0.5652&0.0851&0.3827&0.6887&0.01151\\
                               & $\mathbf{\beta}_{2}$&0.5774&0.5611&0.5658&0.0611&0.4412&0.6818&0.00352\\
                               & $\mathbf{\beta}_{3}$&0.5774&0.5676&0.5912&0.0714&0.4175&0.6979&0.01261\\                                                             & $\mathbf{\sigma}$&0.0500&0.0518&0.0518&0.0053&0.0422&0.0640&0.00003
                                   \\\midrule
\multirow{4}{*} {$\tau$=0.99} & $\mathbf{\beta}_{1}$&0.5774&0.1115&0.1589&0.4957&-0.7874&0.8895&0.28021\\
                               & $\mathbf{\beta}_{2}$&0.5774&0.4797&0.4677&0.2595&0.0817&0.9325&0.02603\\
                               & $\mathbf{\beta}_{3}$&0.5774&0.1051&0.1513&0.4771&-0.7418&0.8867&0.27783\\                                                             & $\mathbf{\sigma}$&0.0500&0.0529&0.0530&0.0054&0.0433&0.0646&0.00003
   
                                                                    \\\bottomrule                                    
\end{tabular}
\end{table}

\begin{table}[ht!]
\footnotesize
\caption{Results of BQSIM for simulation example 4, $n=200$.}
\label{result_ALD_200}
\vspace{0.1in}
\centering
\begin{tabular}{ccccccccc}

\toprule
\multicolumn{2}{c}{$n=200$} &{True}&{Mean} &{Median}&{S.D.}&{2.5\%}&{97.5\%}&{MSE}\\
\midrule
\multirow{4}{*} {$\tau$=0.10} & $\mathbf{\beta}_{1}$&0.5774&0.5813&0.5797&0.0209&0.5389&0.6213&0.00042\\
                               & $\mathbf{\beta}_{2}$&0.5774&0.5742&0.5736&0.0221&0.5299&0.6167&0.00041\\
                               & $\mathbf{\beta}_{3}$&0.5774&0.5742&0.5753&0.0217&0.5301&0.6159&0.00041\\                                                             & $\mathbf{\sigma}$&0.0500&0.0521&0.0519&0.0037&0.0430&0.0637&0.00002
                                   \\\midrule
\multirow{4}{*} {$\tau$=0.25} & $\mathbf{\beta}_{1}$&0.5774&0.5799&0.5798&0.0135&0.5534&0.6057&0.00020\\
                               & $\mathbf{\beta}_{2}$&0.5774&0.5764&0.5806&0.0138&0.5493&0.6028&0.00025\\
                               & $\mathbf{\beta}_{3}$&0.5774&0.5747&0.5754&0.0139&0.5476&0.6018&0.00025\\                                                  & $\mathbf{\sigma}$&0.0500&0.0515&0.0528&0.0037&0.0427&0.0638&0.00002
                                   \\\midrule
\multirow{4}{*} {$\tau$=0.50} & $\mathbf{\beta}_{1}$&0.5774&0.5816&0.5830&0.0119&0.5584&0.6045&0.00030\\
                               & $\mathbf{\beta}_{2}$&0.5774&0.5748&0.5738&0.0118&0.5516&0.5980&0.00026\\
                               & $\mathbf{\beta}_{3}$&0.5774&0.5749&0.5752&0.0118&0.5517&0.5978&0.00030\\                                                             & $\mathbf{\sigma}$&0.0500&0.0522&0.0517&0.0038&0.0447&0.0592&0.00001
                                   \\\midrule
\multirow{4}{*} {$\tau$=0.75} & $\mathbf{\beta}_{1}$&0.5774&0.5771&0.5789&0.0136&0.5511&0.6035&0.00027\\
                               & $\mathbf{\beta}_{2}$&0.5774&0.5769&0.5774&0.0138&0.5503&0.6039&0.00023\\
                               & $\mathbf{\beta}_{3}$&0.5774&0.5770&0.5751&0.0135&0.5505&0.6029&0.00022\\                                                             & $\mathbf{\sigma}$&0.0500&0.0512&0.0512&0.0037&0.0448&0.0594&0.00002
                                   \\\midrule
\multirow{4}{*} {$\tau$=0.90} & $\mathbf{\beta}_{1}$&0.5774&0.5774&0.5760&0.0213&0.5363&0.6201&0.00057\\
                               & $\mathbf{\beta}_{2}$&0.5774&0.5800&0.5788&0.0217&0.5370&0.6224&0.00044\\
                               & $\mathbf{\beta}_{3}$&0.5774&0.5720&0.5697&0.0217&0.5295&0.6152&0.00047\\                                                             & $\mathbf{\sigma}$&0.0500&0.0511&0.0518&0.0037&0.0447&0.0593&0.00001
                                   \\\midrule
\multirow{4}{*} {$\tau$=0.95} & $\mathbf{\beta}_{1}$&0.5774&0.5766&0.5739&0.0374&0.5009&0.6447&0.00197\\
                               & $\mathbf{\beta}_{2}$&0.5774&0.5795&0.5841&0.0337&0.5128&0.6471&0.00099\\
                               & $\mathbf{\beta}_{3}$&0.5774&0.5679&0.5688&0.0354&0.5003&0.6385&0.00145\\                                                             & $\mathbf{\sigma}$&0.0500&0.0518&0.0529&0.0038&0.0450&0.0649&0.00002
                                   \\\midrule
\multirow{4}{*} {$\tau$=0.99} & $\mathbf{\beta}_{1}$&0.5774&0.3256&0.3507&0.3545&-0.3724&0.8083&0.12130\\
                               & $\mathbf{\beta}_{2}$&0.5774&0.5350&0.5428&0.1791&0.2028&0.8372&0.02907\\
                               & $\mathbf{\beta}_{3}$&0.5774&0.3318&0.3344&0.2977&-0.2720&0.7914&0.12535\\                                                             & $\mathbf{\sigma}$&0.0500&0.0524&0.0520&0.0038&0.0433&0.0646&0.00002
   
                                                                    \\\bottomrule                                    
\end{tabular}
\end{table}

\subsubsection*{Example 5}
Finally, we increase the dimension in Example 1 to $p=10$ to demonstrate the performance in higher dimensions. The only difference in the setup from Example 1 is that we now set $\boldsymbol{\beta}=(1,1,1,0,\ldots,0)^T$ and $\boldsymbol{\beta}=\frac{1}{\sqrt{10}}(1,1,1,1,\ldots,1)^T$, and only consider $n=100$. For these two cases, the estimation results are shown in Tables \ref{result_sim5_1} and \ref{result_sim5_2} respectively. To save space, boxplots comparing BQSIM and QSIM are not shown now that we are estimating $10$ coefficients. Our methods still perform much better than QSIM in these cases with higher dimensions.

\begin{table}[ht!]
\footnotesize
\caption{Comparison of MSE for BQSIM and QSIM based on 100 replications in each case for simulation example 5, where $\boldsymbol{\beta}=(1,1,1,0,\ldots,0)^T$.}
\label{result_sim5_1}
\vspace{0.1in}
\centering
\begin{tabular}{llcccccc}

\toprule
\multicolumn{2}{c}{$n=100$} &{$\beta_1$} &{$\beta_2$}&{$\beta_3$}&{$\beta_4$}&{$\beta_5$}\\
\midrule
\multirow{2}{*} {$\tau$=0.10} & $\rm{QSIM}$&0.03051&0.06458&0.09402&0.03894&0.03484\\
                                     & $\rm{BQSIM}$&0.01000&0.00251&0.03926&0.00068&0.00460                             
                                   \\\midrule
\multirow{2}{*} {$\tau$=0.25} & $\rm{QSIM}$&0.00822&0.01876&0.04851&0.01342&0.00825\\
                                     & $\rm{BQSIM}$&0.00054&0.00046&0.00064&0.00045&0.00038                           
                                   \\\midrule
\multirow{2}{*} {$\tau$=0.50} & $\rm{QSIM}$&0.01385&0.00483&0.00308&0.00617&0.00417\\
                                   & $\rm{BQSIM}$&0.00031&0.00039&0.00043&0.00033&0.00039                                     
                                                                     \\\midrule                                   
\multirow{2}{*} {$\tau$=0.75} & $\rm{QSIM}$&0.00106&0.03489&0.03508&0.00102& 0.00396\\                                
                                  &$\rm{BQSIM}$&0.00046&0.00039&0.00058&0.00063&0.00068  
                                    \\\midrule                           
\multirow{2}{*} {$\tau$=0.90} & $\rm{QSIM}$&0.00532&0.03687&0.03949&0.01752&0.00940\\                                
                                   &$\rm{BQSIM}$ &0.00148 & 0.00106&0.00116&0.00156& 0.00151 
                                   \\\midrule
\multirow{2}{*} {$\tau$=0.95} & $\rm{QSIM}$&0.03913& 0.06571& 0.10765& 0.03061 & 0.030221\\                                
                                   &$\rm{BQSIM}$ &0.00698& 0.00368& 0.00401& 0.00178& 0.00287
                                   \\\midrule
\multirow{2}{*} {$\tau$=0.99} & $\rm{QSIM}$&0.04228& 0.05799& 0.06567 & 0.06315 & 0.06315\\                                
                                   &$\rm{BQSIM}$ &0.02259& 0.01966& 0.03773 & 0.02252 & 0.01228                                            
                                                                     \\\bottomrule                                    
\toprule
\multicolumn{2}{c}{$n=100$} &{$\beta_6$} &{$\beta_7$}&{$\beta_8$}&{$\beta_9$}&{$\beta_{10}$}\\
\midrule
\multirow{2}{*} {$\tau$=0.10} & $\rm{QSIM}$&0.02318&0.02520&0.02052&0.03534&0.04236\\
                                     & $\rm{BQSIM}$&0.01572&0.00628&0.00214&0.00194&0.00432                             
                                   \\\midrule
\multirow{2}{*} {$\tau$=0.25} & $\rm{QSIM}$&0.01096&0.01086&0.00622&0.00749&0.01302\\
                                     & $\rm{BQSIM}$&0.00056&0.00034&0.00038&0.00054&0.00036                           
                                   \\\midrule
\multirow{2}{*} {$\tau$=0.50} & $\rm{QSIM}$&0.00717&0.01045&0.00452&0.00622&0.00974\\
                                   & $\rm{BQSIM}$&0.00043&0.00047&0.00029&0.00049&0.00032                                     
                                                                     \\\midrule                                   
\multirow{2}{*} {$\tau$=0.75} & $\rm{QSIM}$&0.00117& 0.00386& 0.00157& 0.01653& 0.00803\\                                
                                  &$\rm{BQSIM}$&0.00035&0.00055&0.00049&0.00064&0.00046  
                                    \\\midrule                           
\multirow{2}{*} {$\tau$=0.90} & $\rm{QSIM}$&0.00766&0.04293&0.01118&0.01324&0.02452\\                                
                                   &$\rm{BQSIM}$ &0.00118&0.00167& 0.00065& 0.00133& 0.00181
                                   \\\midrule 
\multirow{2}{*} {$\tau$=0.95} & $\rm{QSIM}$&0.02806& 0.03451& 0.02941& 0.03739& 0.02923\\                                
                                   &$\rm{BQSIM}$ &0.00288& 0.00200& 0.00451& 0.00305& 0.00263
                                   \\\midrule                                      
\multirow{2}{*} {$\tau$=0.99} & $\rm{QSIM}$&0.08010& 0.06853 & 0.05461 & 0.06722 & 0.05835\\                                
                                   &$\rm{BQSIM}$ &0.01346 & 0.01363& 0.01514 & 0.01219 & 0.01140                        
                                   \\\bottomrule

\end{tabular}
\end{table}

\begin{table}[ht!]
\footnotesize
\caption{Comparison of MSE for BQSIM and QSIM based on 100 replications in each case for simulation example 5, where $\boldsymbol{\beta}=\frac{1}{\sqrt{10}}(1,1,1,1,\ldots,1)^T$.}
\label{result_sim5_2}
\vspace{0.1in}
\centering
\begin{tabular}{llcccccc}

\toprule
\multicolumn{2}{c}{$n=100$} &{$\beta_1$} &{$\beta_2$}&{$\beta_3$}&{$\beta_4$}&{$\beta_5$}\\
\midrule
\multirow{2}{*} {$\tau$=0.10} & $\rm{QSIM}$&0.00124&0.00149&0.00114&0.00133&0.00140\\
                                     & $\rm{BQSIM}$&0.00066&0.00085&0.00067&0.00079&0.00078                             
                                   \\\midrule
\multirow{2}{*} {$\tau$=0.25} & $\rm{QSIM}$&0.00101&0.00077&0.00072&0.00089&0.00050\\
                                     & $\rm{BQSIM}$&0.00047&0.00035&0.00040&0.00039&0.00038                        
                                   \\\midrule
\multirow{2}{*} {$\tau$=0.50} & $\rm{QSIM}$&0.00086&0.00062&0.00367&0.00061&0.00567\\
                                   & $\rm{BQSIM}$&0.00048&0.00032&0.00026&0.00048&0.00055                                     
                                                                     \\\midrule                                   
\multirow{2}{*} {$\tau$=0.75} & $\rm{QSIM}$&0.00102&0.01325&0.00135&0.00122& 0.00173\\                                
                                  &$\rm{BQSIM}$&0.00064&0.00048&0.00045&0.00053&0.00048  
                                    \\\midrule                           
\multirow{2}{*} {$\tau$=0.90} & $\rm{QSIM}$&0.00308&0.00256&0.01116&0.00666&0.00280\\                                
                                   &$\rm{BQSIM}$ &0.00095 & 0.00082&0.00058&0.00081& 0.00061 
                                   \\\midrule
\multirow{2}{*} {$\tau$=0.95} & $\rm{QSIM}$&0.01535& 0.01194& 0.03278& 0.02291& 0.02617\\                                
                                   &$\rm{BQSIM}$ &0.00129& 0.00144& 0.00125& 0.00162& 0.00129
                                   \\\midrule
\multirow{2}{*} {$\tau$=0.99} & $\rm{QSIM}$&0.01629& 0.05267& 0.02109 & 0.04479 & 0.04052\\                                
                                   &$\rm{BQSIM}$ &0.01493& 0.00967& 0.01255 & 0.01236 & 0.01534                                            
                                                                     \\\bottomrule                                    
\toprule
\multicolumn{2}{c}{$n=100$} &{$\beta_6$} &{$\beta_7$}&{$\beta_8$}&{$\beta_9$}&{$\beta_{10}$}\\
\midrule
\multirow{2}{*} {$\tau$=0.10} & $\rm{QSIM}$&0.00123&0.00160&0.00141&0.00151&0.00100\\
                                     & $\rm{BQSIM}$&0.00096&0.00086&0.0065&0.00107&0.00066                             
                                   \\\midrule
\multirow{2}{*} {$\tau$=0.25} & $\rm{QSIM}$&0.00074&0.00101&0.00069&0.00067&0.00084\\
                                     & $\rm{BQSIM}$&0.00054&0.00048&0.00052&0.00043&0.00033                           
                                   \\\midrule
\multirow{2}{*} {$\tau$=0.50} & $\rm{QSIM}$&0.00098&0.00058&0.00180&0.00178&0.00064\\
                                   & $\rm{BQSIM}$&0.00037&0.00048&0.00033&0.00037&0.00029                                     
                                                                     \\\midrule                                   
\multirow{2}{*} {$\tau$=0.75} & $\rm{QSIM}$&0.00283& 0.00123& 0.00428& 0.00093& 0.00123\\                                
                                  &$\rm{BQSIM}$&0.00057&0.00049&0.00057&0.00041&0.00041 
                                    \\\midrule                           
\multirow{2}{*} {$\tau$=0.90} & $\rm{QSIM}$&0.00385&0.00258&0.01294&0.00964&0.00341\\                                
                                   &$\rm{BQSIM}$ &0.00070&0.00071& 0.00079& 0.00105& 0.00064
                                   \\\midrule 
\multirow{2}{*} {$\tau$=0.95} & $\rm{QSIM}$&0.02083& 0.02902& 0.04447& 0.02839& 0.05003\\                                
                                   &$\rm{BQSIM}$ &0.00132& 0.00170& 0.00077& 0.00142& 0.00104
                                   \\\midrule                                      
\multirow{2}{*} {$\tau$=0.99} & $\rm{QSIM}$&0.04999& 0.04875 & 0.07079 & 0.06024 & 0.05711\\                                
                                   &$\rm{BQSIM}$ &0.01191 & 0.01088& 0.01134 & 0.02032 & 0.01921                        
                                   \\\bottomrule

\end{tabular}
\end{table}

\subsection{Real data analysis}
Finally, we apply the proposed method to the tropical cyclone (TC) data. Coastal tropical cyclones pose a serious threat to social and economic institutions. It is necessary and useful to provide a statistical way to analyze the TC data and evaluate the risk of the next catastrophic cyclone. Here we consider a dataset consisting of a sample of 422 TCs occurring near the US coastline over a 108-year period (1899-2006). Following \cite{jagger2009modeling} and \cite{bondell2010noncrossing}, we model the wind speeds from  TCs with four climate variables: the North Atlantic Oscillation Index (NAO), the Southern Oscillation Index (SOI), the Atlantic sea-surface temperature (SST) and the average sun spot number (SSN). The values of SOI, SST, and SSN are averaged values over the peak season of August through October and the values of NAO are averaged over the preseason and early season months of May and June. Both \cite{jagger2009modeling} and \cite{bondell2010noncrossing} used linear quantile regression to analyze this data. Here we apply our BQSIM to analyze how these climate variables influence the wind speeds of TCs. We also fitted QSIM described in \cite{wu2010single} for comparison.

The particular focus for this type of data is on the upper quantiles, as these extreme hurricane-strength storms are of considerable importance. We consider three different quantile levels $\tau=(0.5, 0.75, 0.9, 0.95, 0.99)$. All covariates are scaled to have mean zero and standard deviation one. 

\begin{figure}[ht!]
\begin{center}
\begin{minipage}{0.75in}
$\tau=0.5$
\end{minipage}
\begin{minipage}{2in}
\includegraphics[width=1.7in,trim=0 30 0 30,clip=TRUE]{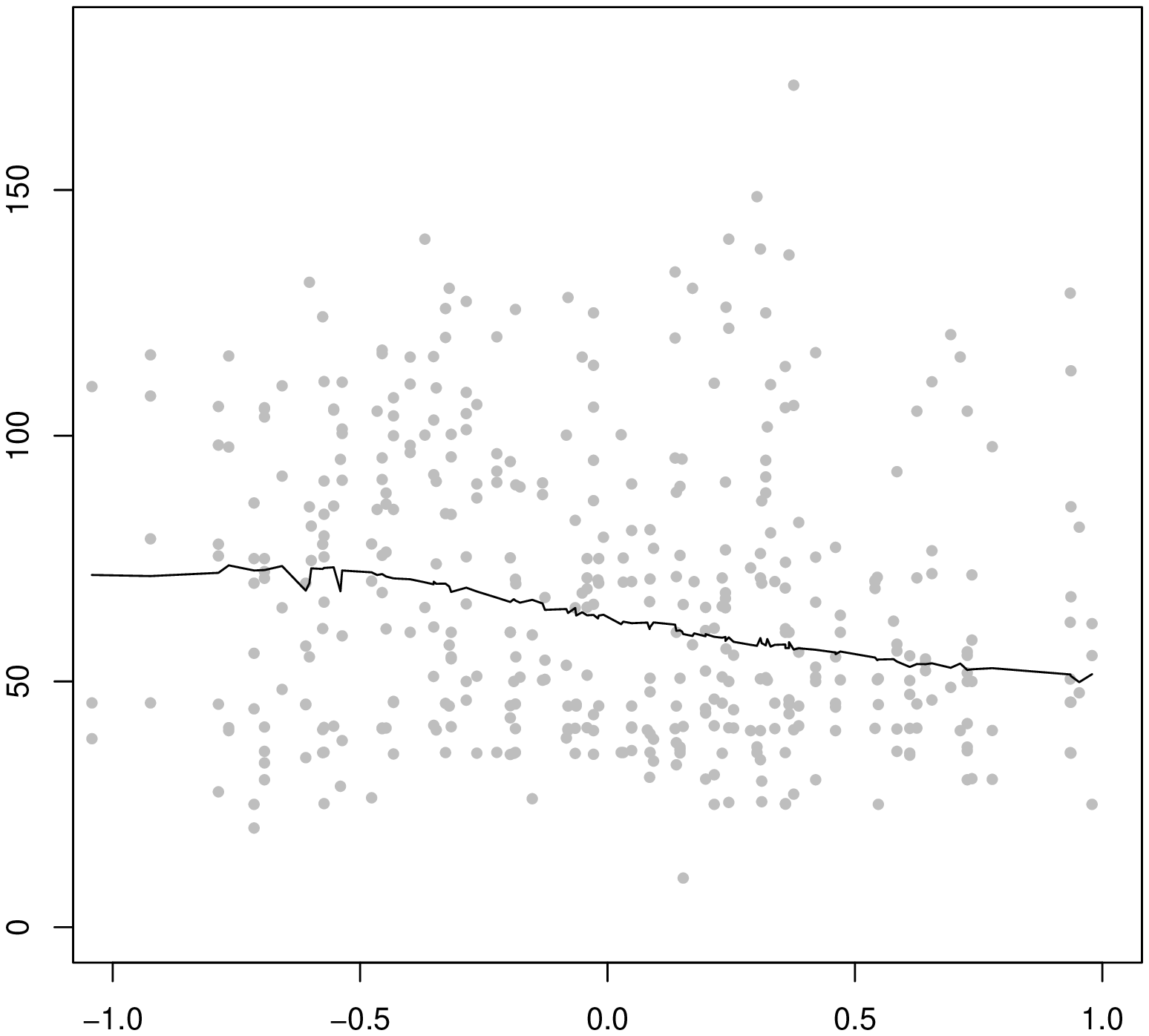}
\end{minipage}
\begin{minipage}{2in}
\includegraphics[width=1.7in,trim=0 30 0 30,clip=TRUE]{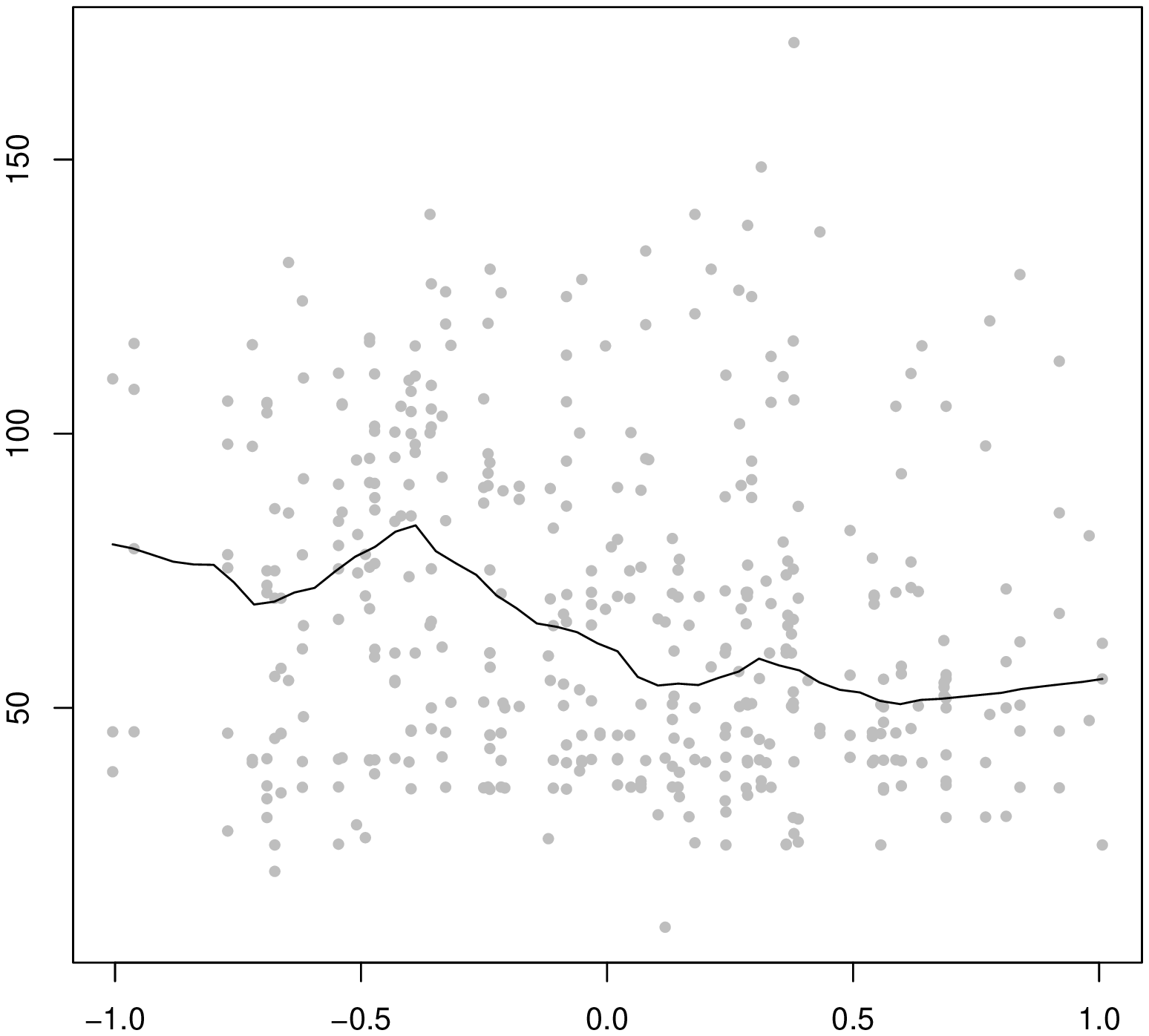}
\end{minipage}

\begin{minipage}{0.75in}
$\tau=0.75$
\end{minipage}
\begin{minipage}{2in}
\includegraphics[width=1.7in,trim=0 30 0 50,clip=TRUE]{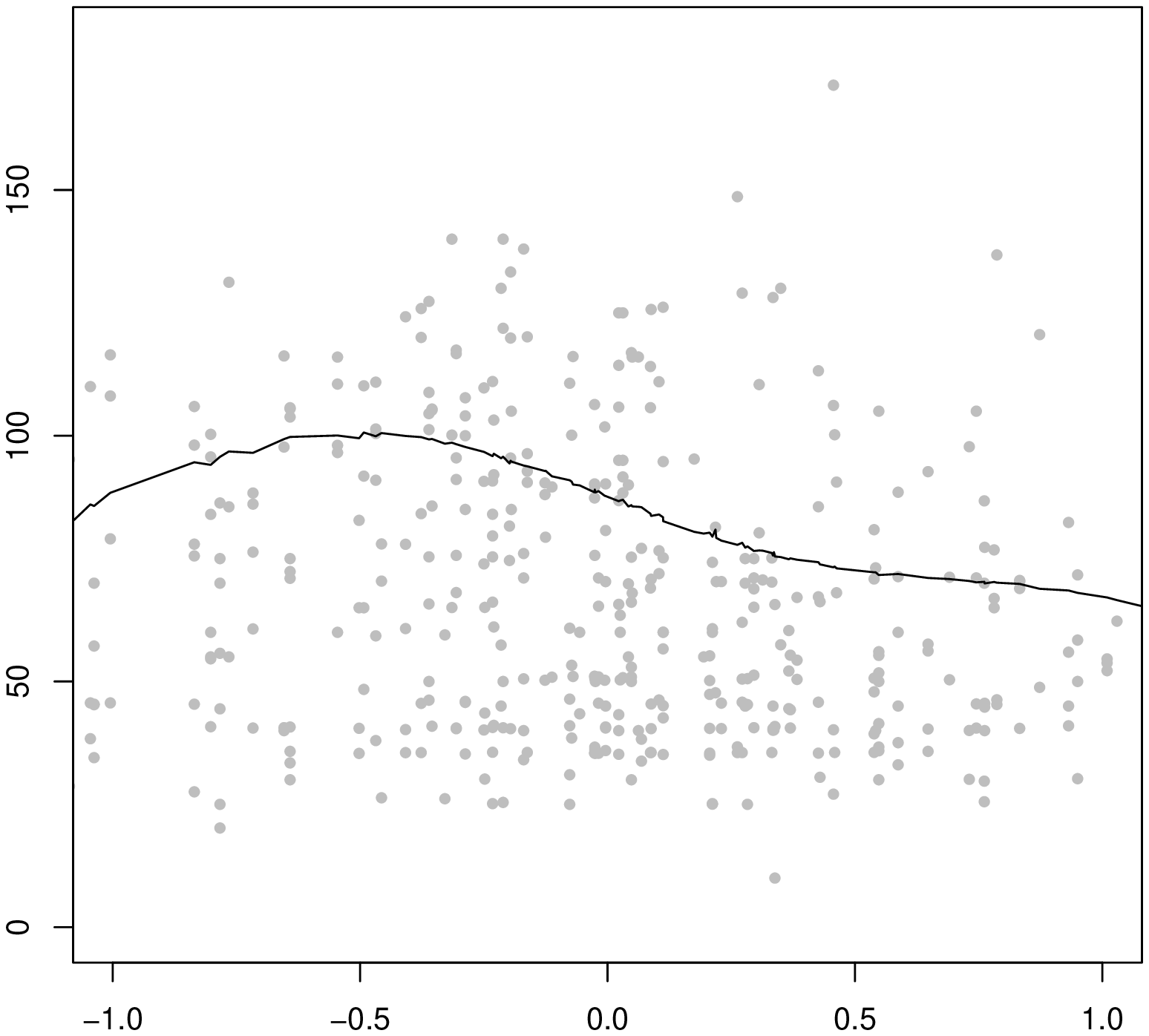}
\end{minipage}
\begin{minipage}{2in}
\includegraphics[width=1.7in,trim=0 30 0 50,clip=TRUE]{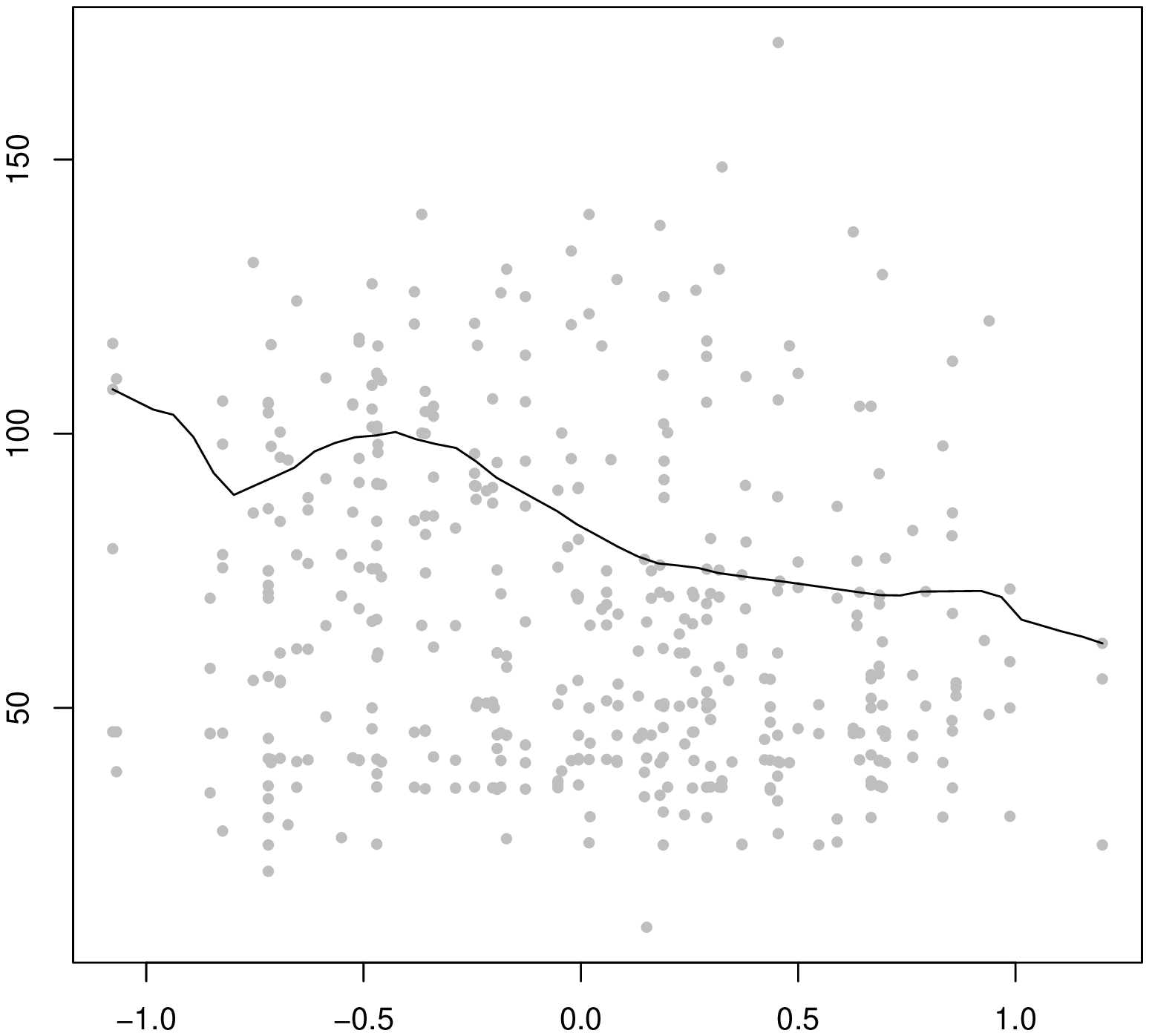}
\end{minipage}

\begin{minipage}{0.75in}
$\tau=0.9$
\end{minipage}
\begin{minipage}{2in}
\includegraphics[width=1.7in,trim=0 30 0 50,clip=TRUE]{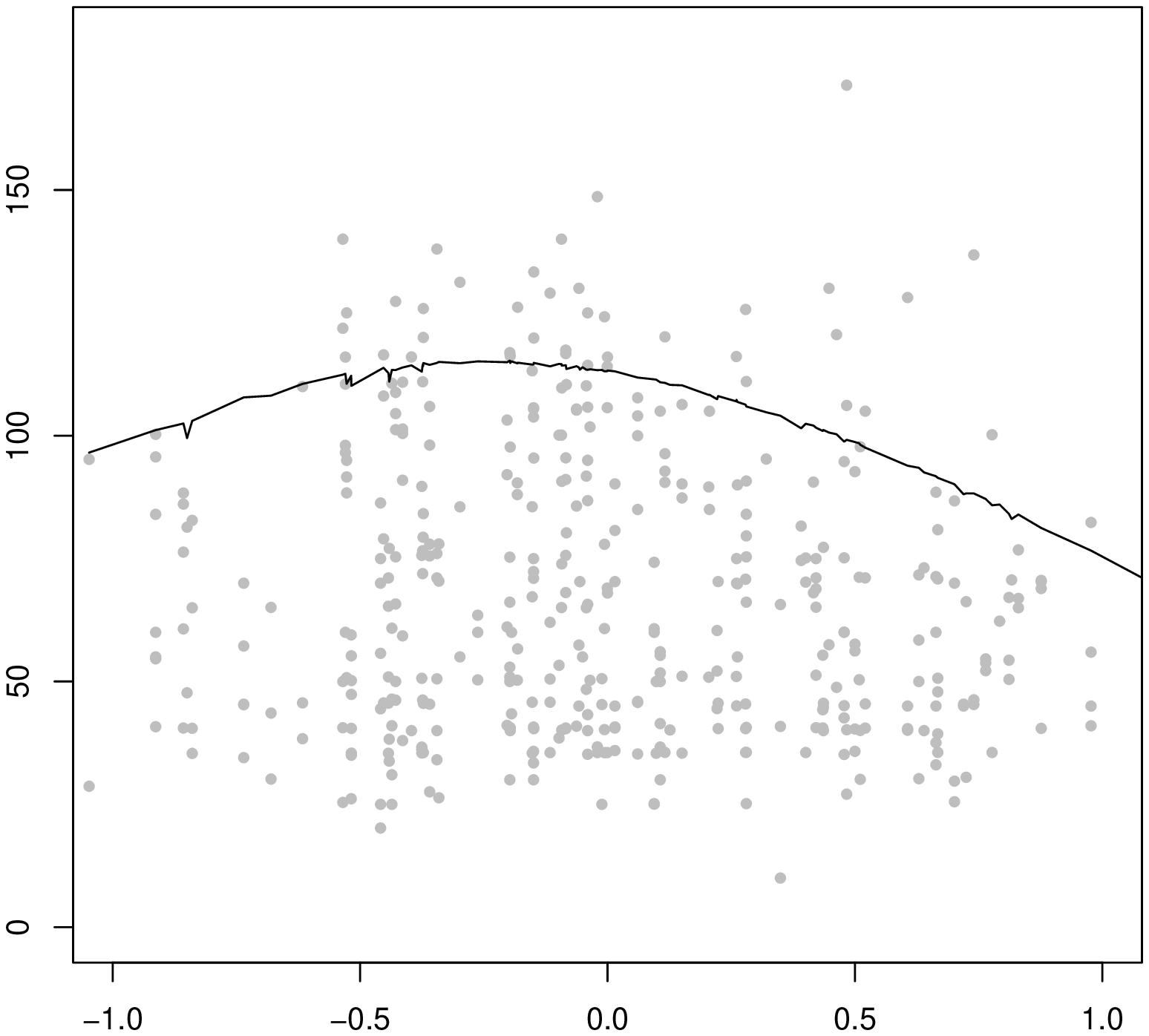}
\end{minipage}
\begin{minipage}{2in}
\includegraphics[width=1.7in,trim=0 30 0 50,clip=TRUE]{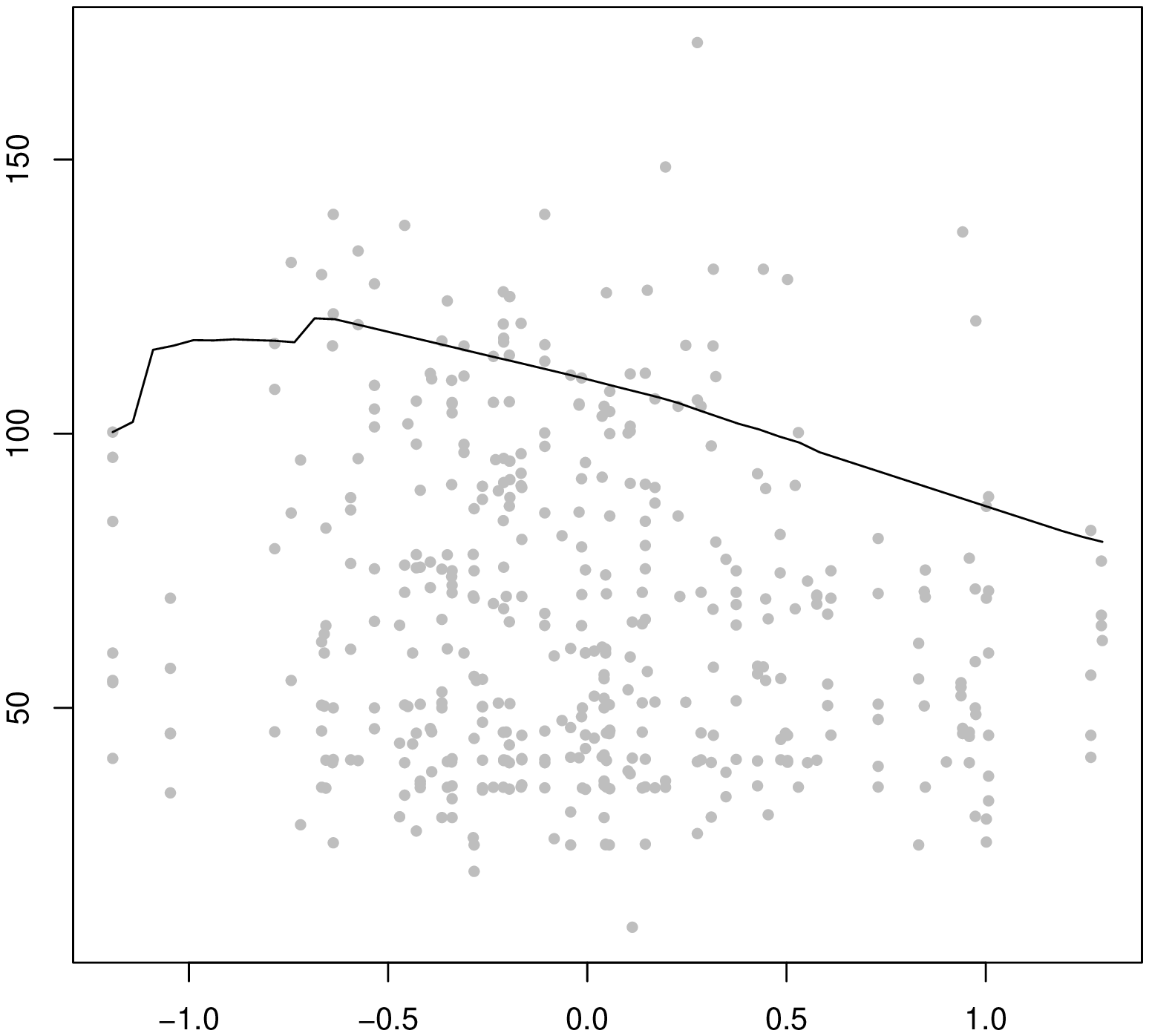}
\end{minipage}

\begin{minipage}{0.75in}
$\tau=0.95$
\end{minipage}\begin{minipage}{2in}
\includegraphics[width=1.7in,trim=0 30 0 50,clip=TRUE]{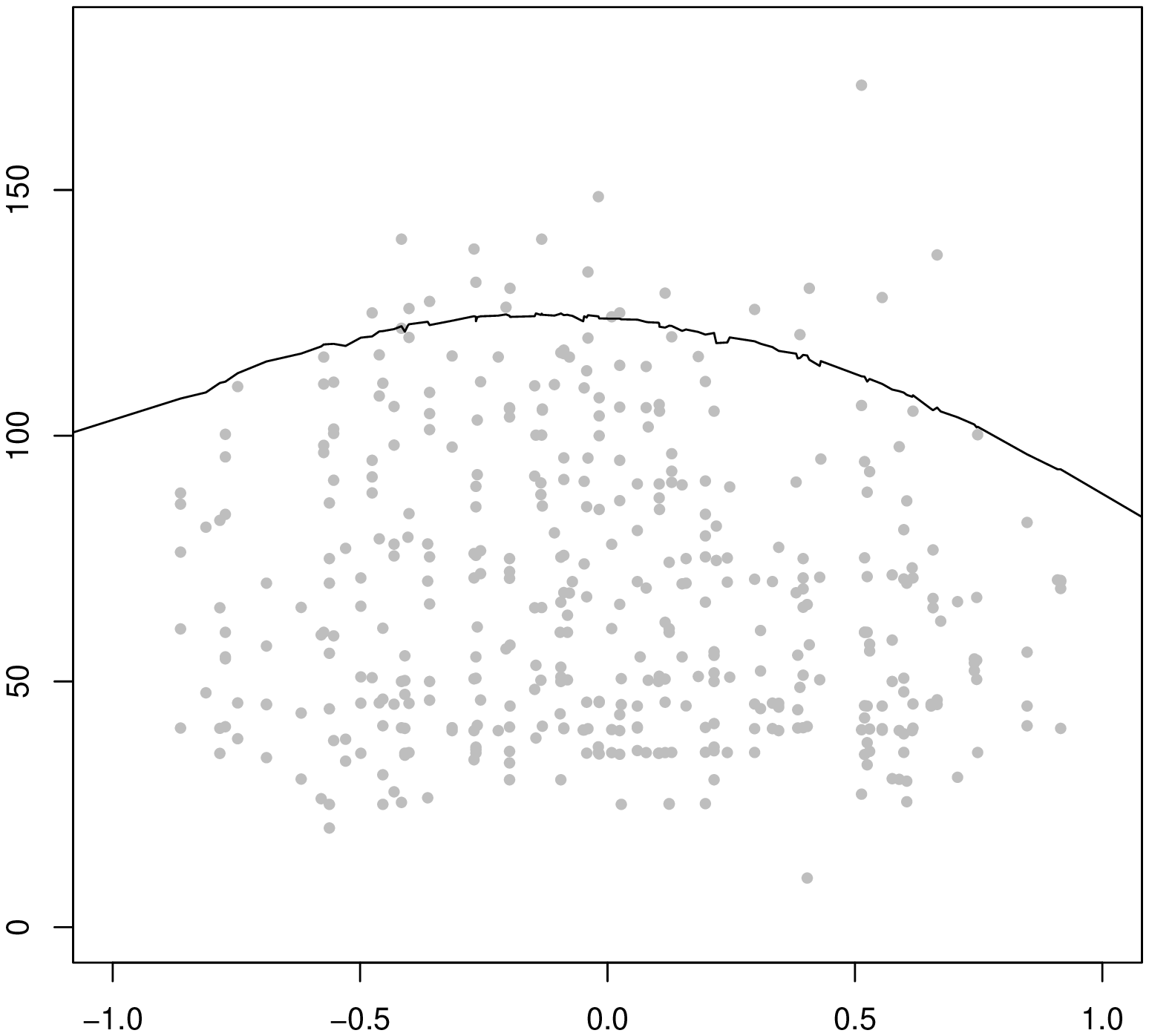}
\end{minipage}
\begin{minipage}{2in}
\includegraphics[width=1.7in,trim=0 30 0 50,clip=TRUE]{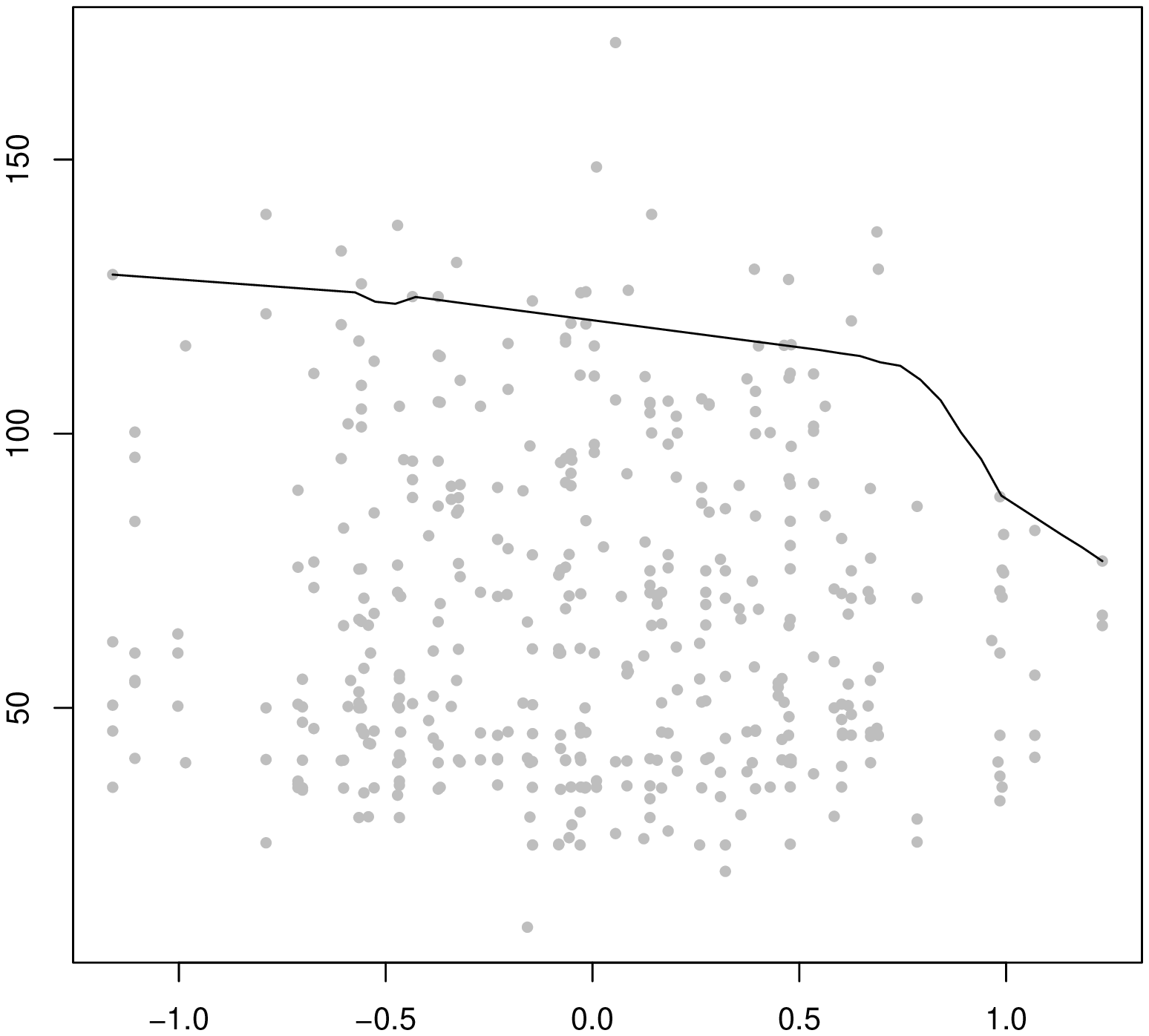}
\end{minipage}

\begin{minipage}{0.75in}
$\tau=0.99$
\end{minipage}
\begin{minipage}{2in}
\includegraphics[width=1.7in,trim=0 30 0 50,clip=TRUE]{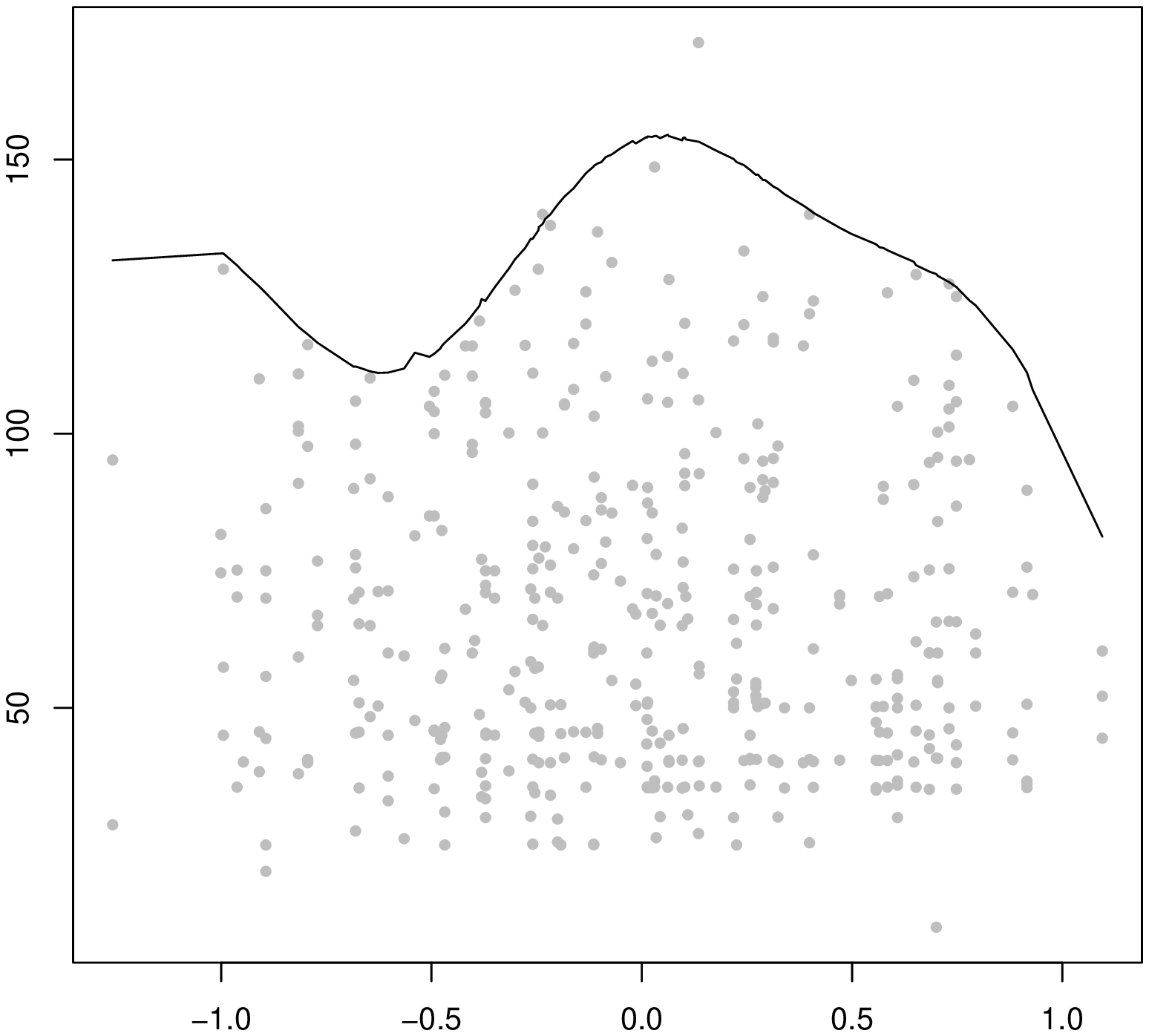}
\end{minipage}
\begin{minipage}{2in}
\includegraphics[width=1.7in,trim=0 30 0 50,clip=TRUE]{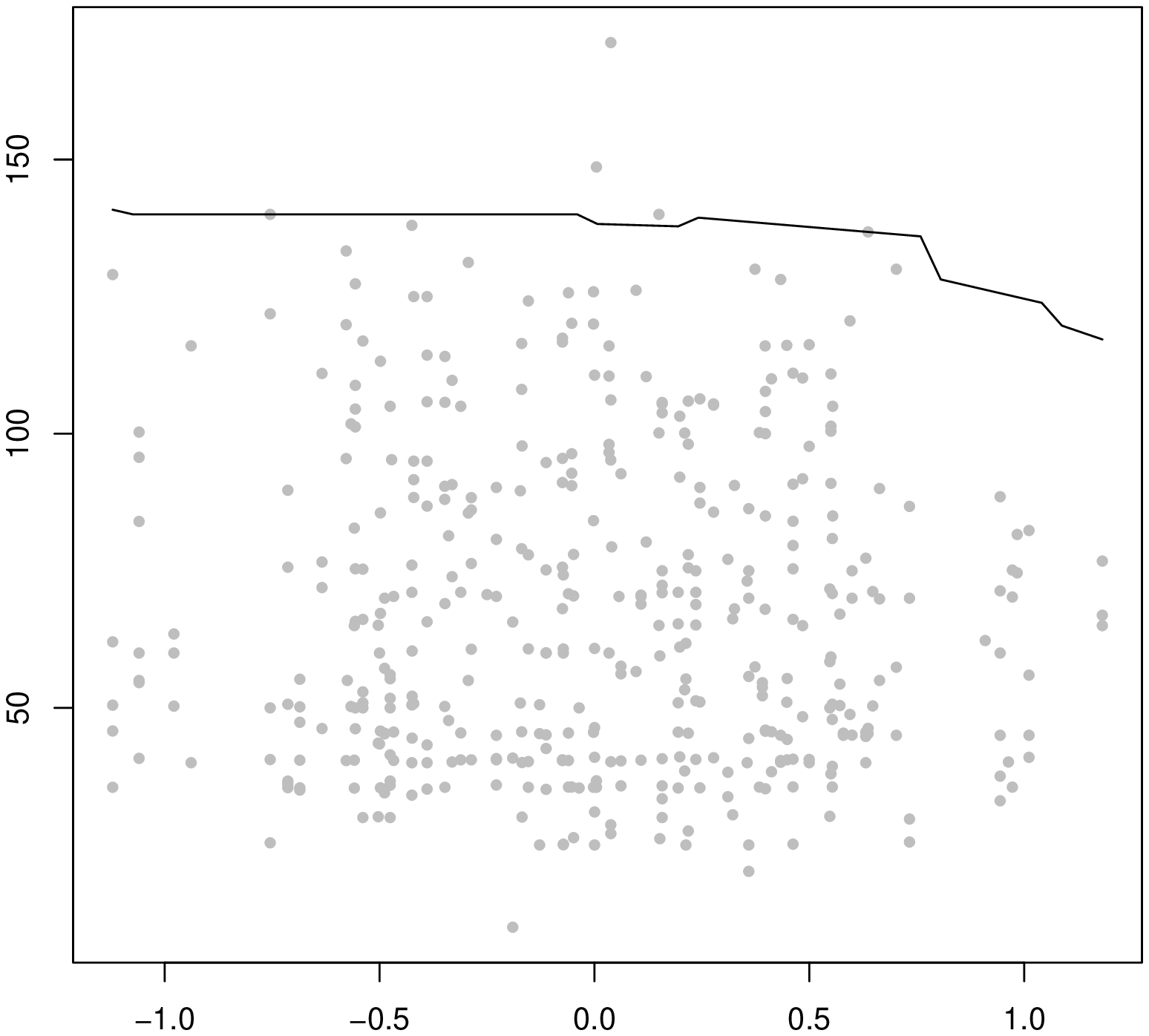}
\end{minipage}
\end{center}
\caption{The estimated link functions by BQSIM (left column) and QSIM (right column), for TC intensity at different quantiles.} 
\label{quantileplot_tc}
\end{figure}

\begin{figure}
\begin{center}
\begin{minipage}{0.75in}
$\tau=0.5$
\end{minipage}
\begin{minipage}{2in}
\includegraphics[width=1.7in,trim=0 30 0 50,clip=TRUE]{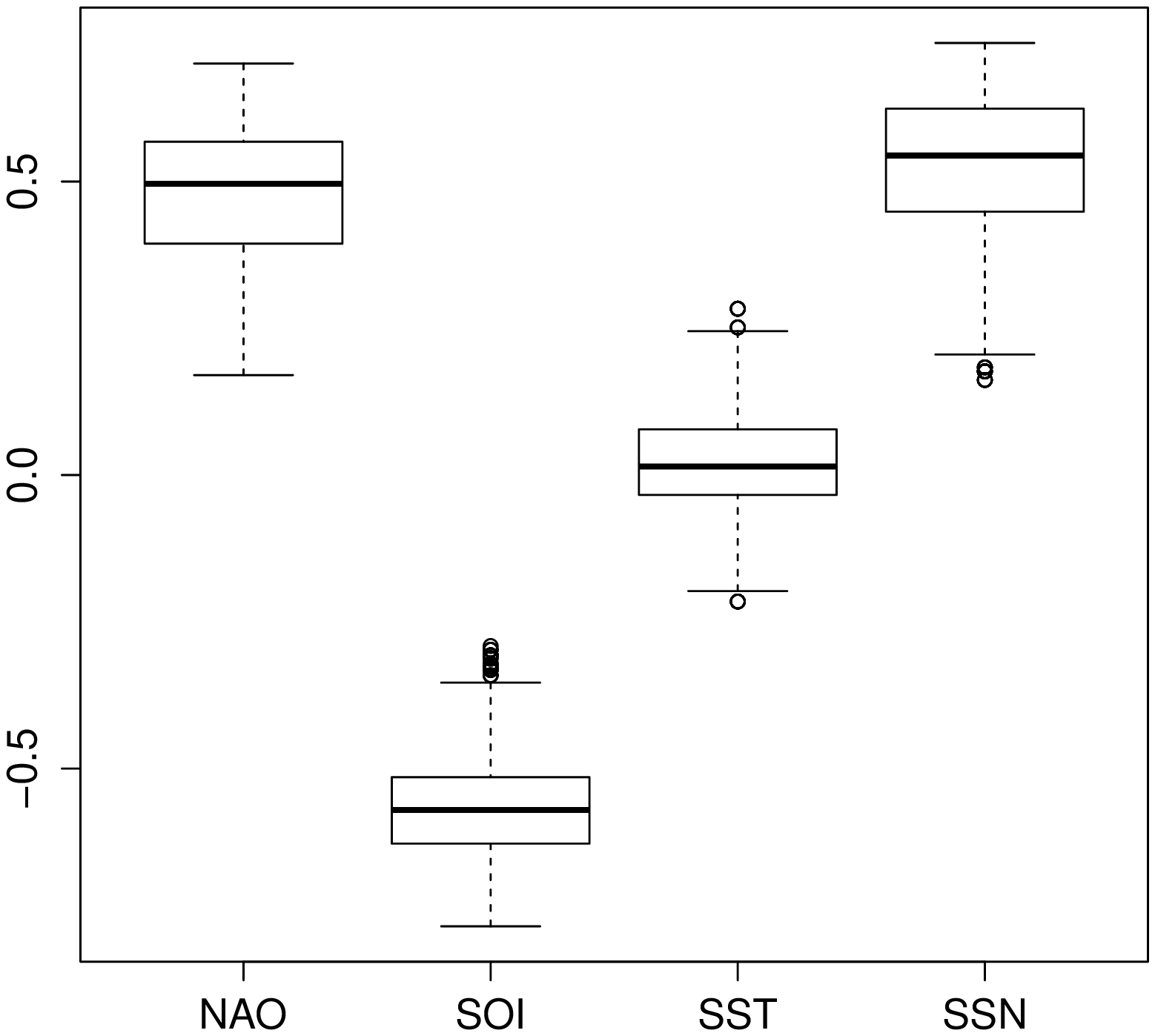}
\end{minipage}
\begin{minipage}{2in}
\includegraphics[width=1.7in,trim=0 30 0 50,clip=TRUE]{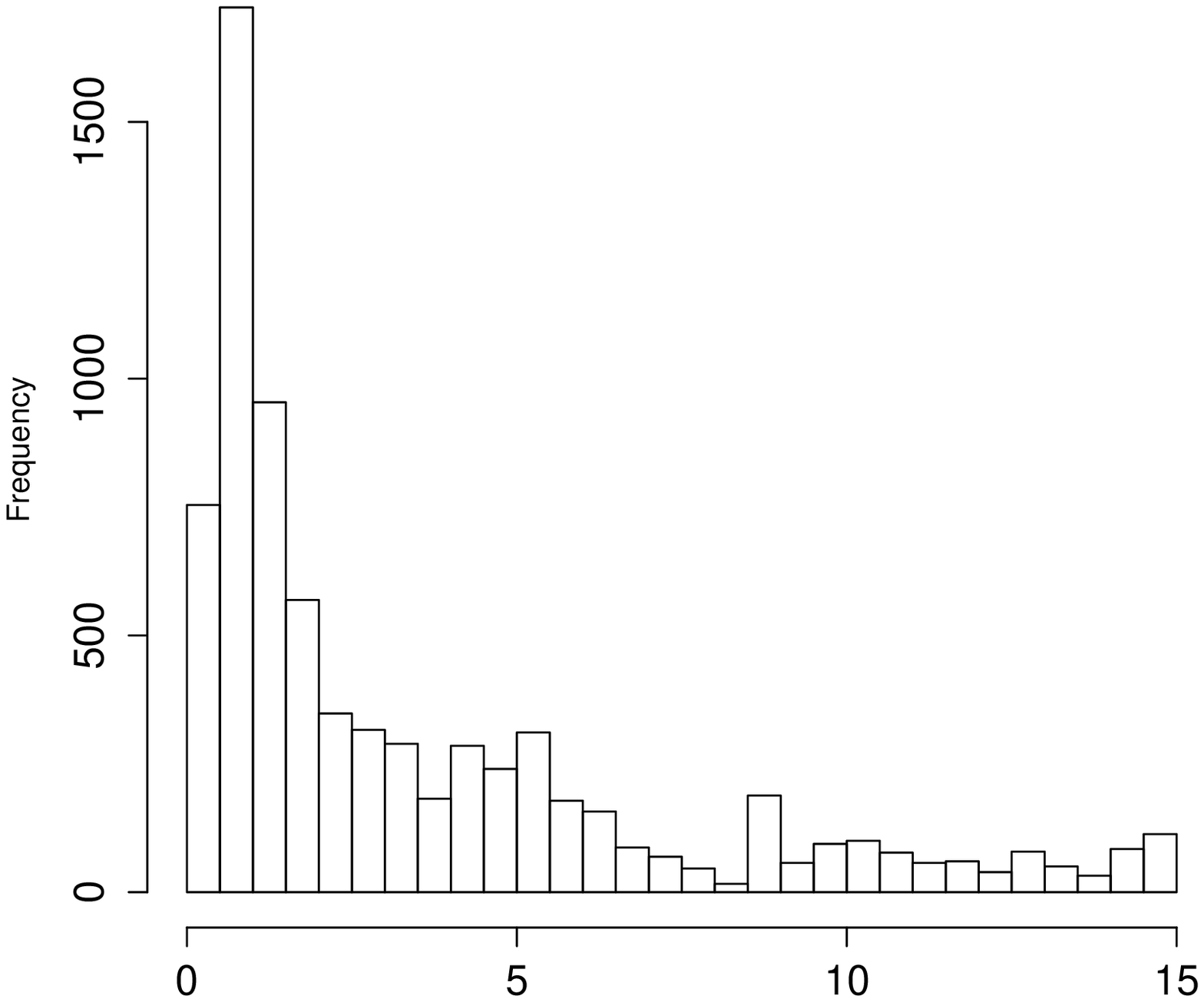}
\end{minipage}

\begin{minipage}{0.75in}
$\tau=0.75$
\end{minipage}
\begin{minipage}{2in}
\includegraphics[width=1.7in,trim=0 30 0 50,clip=TRUE]{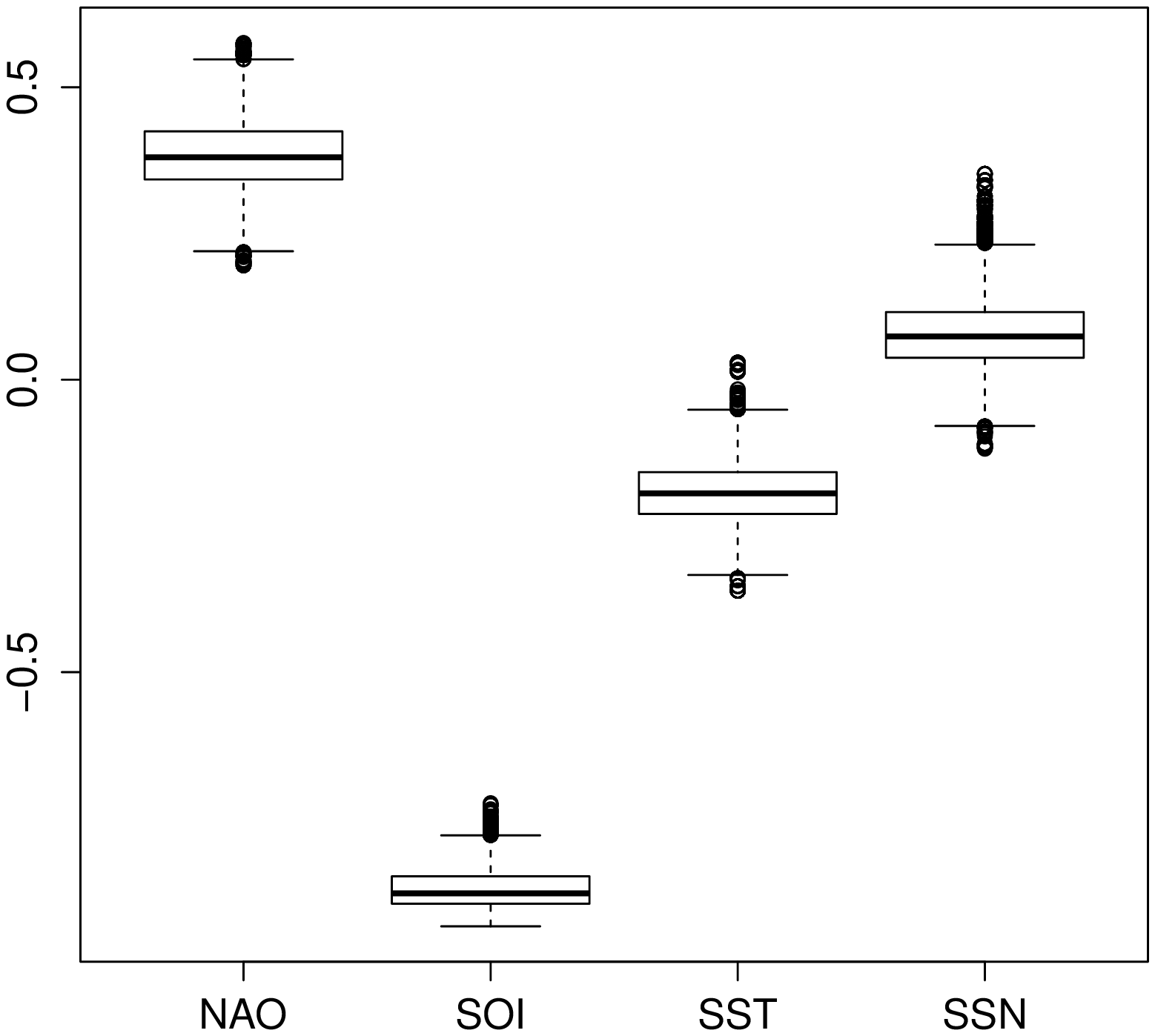}
\end{minipage}
\begin{minipage}{2in}
\includegraphics[width=1.7in,trim=0 30 0 50,clip=TRUE]{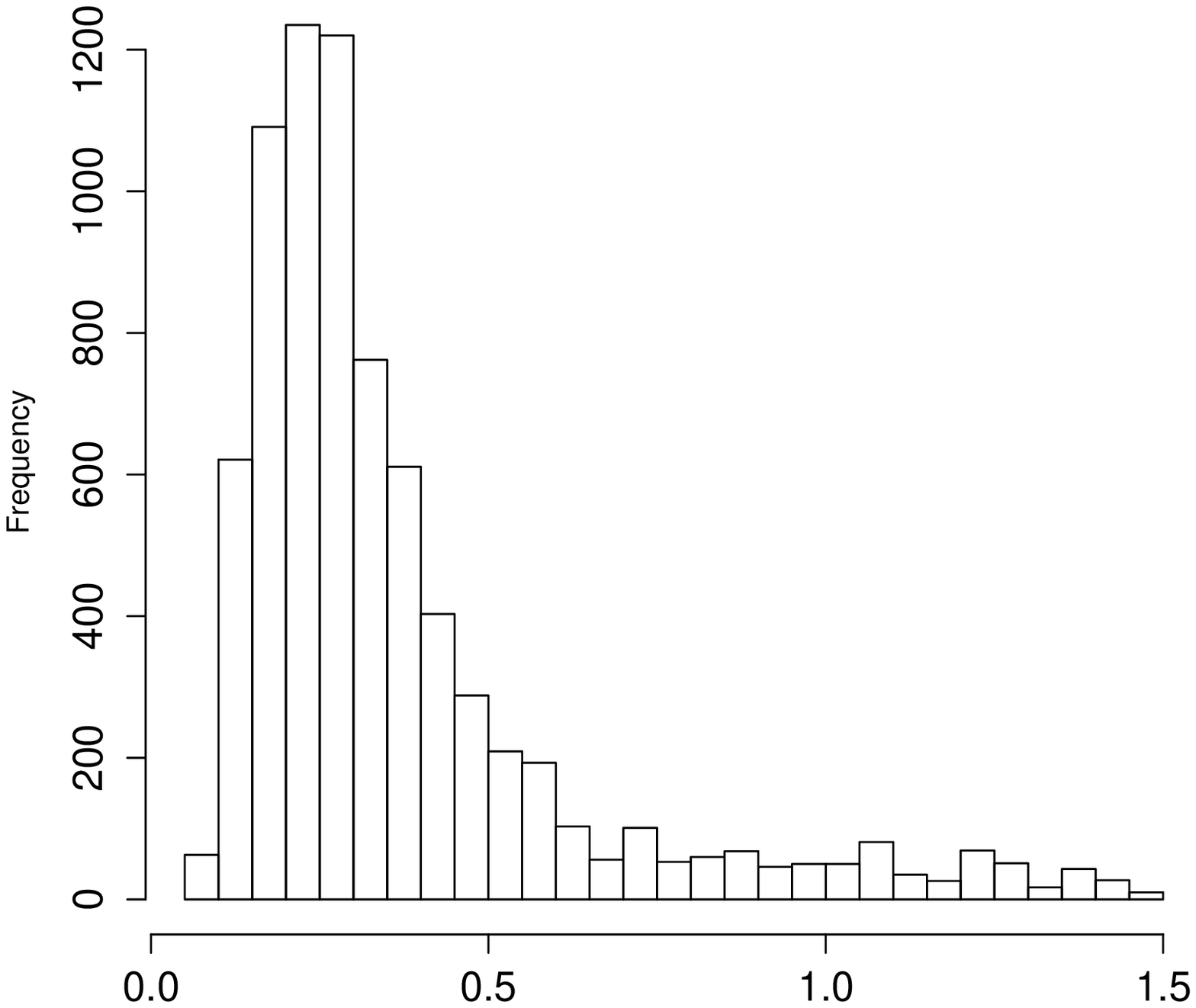}
\end{minipage}

\begin{minipage}{0.75in}
$\tau=0.9$
\end{minipage}
\begin{minipage}{2in}
\includegraphics[width=1.7in,trim=0 30 0 50,clip=TRUE]{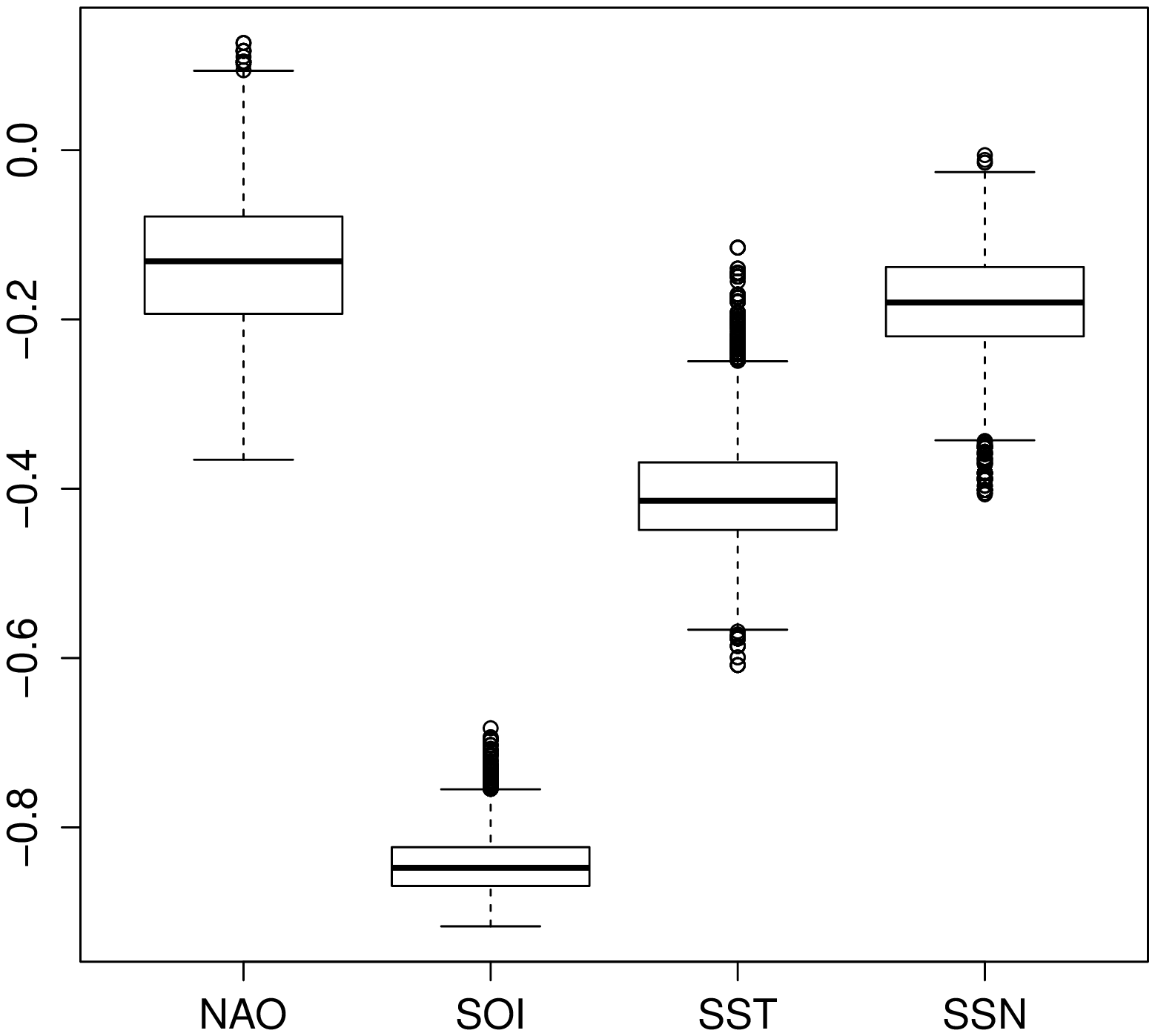}
\end{minipage}
\begin{minipage}{2in}
\includegraphics[width=1.7in,trim=0 30 0 50,clip=TRUE]{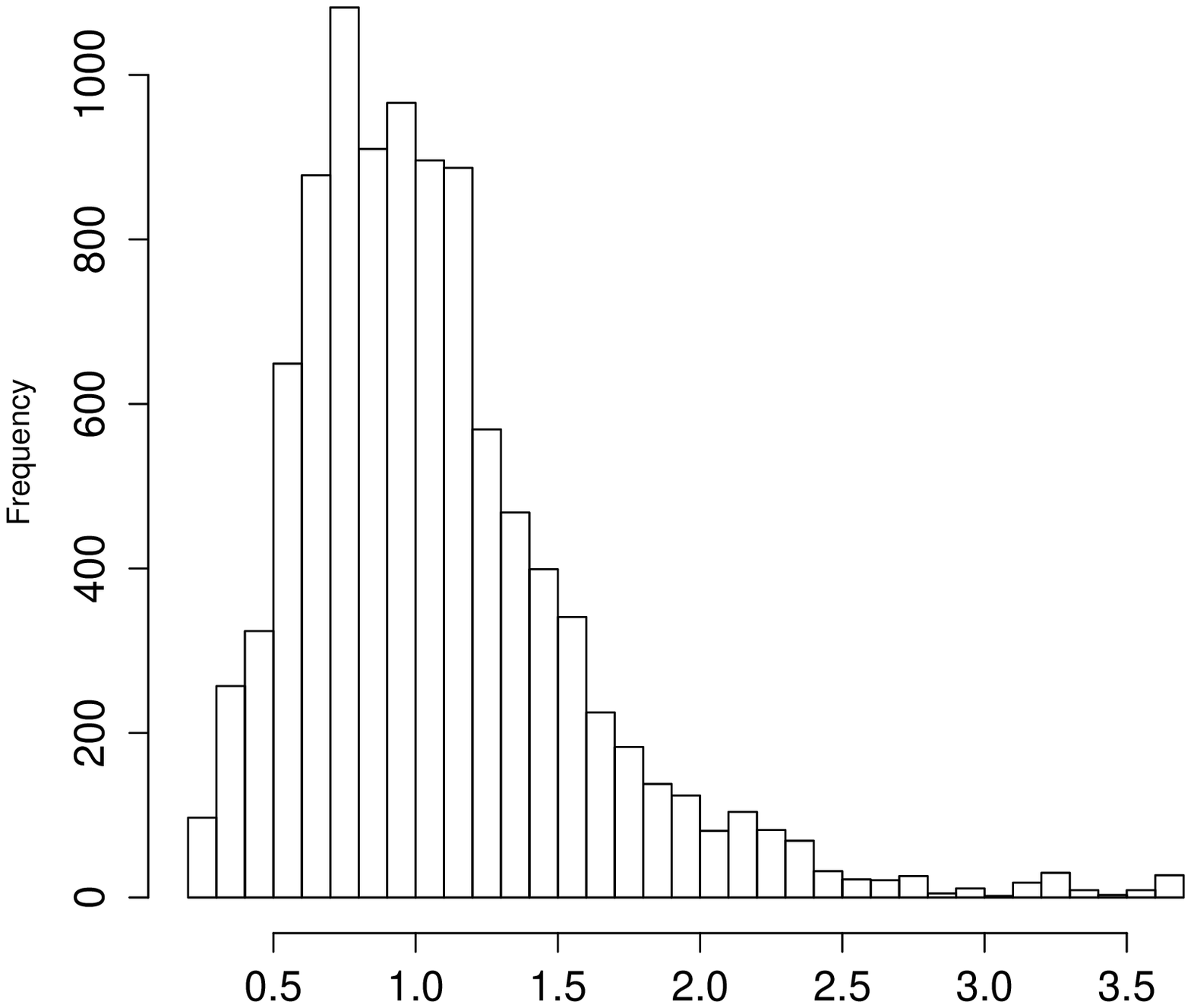}
\end{minipage}

\begin{minipage}{0.75in}
$\tau=0.95$
\end{minipage}
\begin{minipage}{2in}
\includegraphics[width=1.7in,trim=0 30 0 50,clip=TRUE]{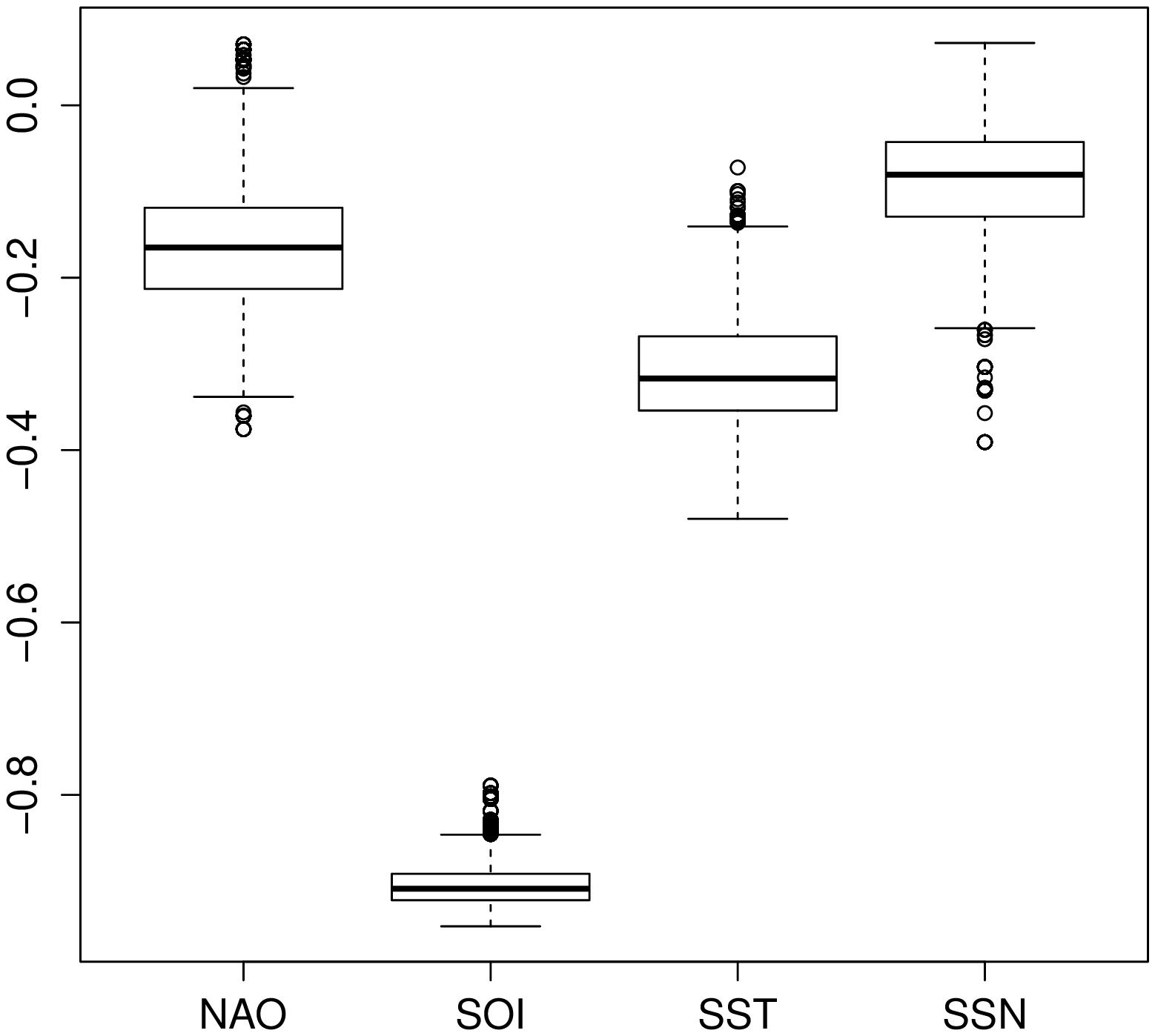}
\end{minipage}
\begin{minipage}{2in}
\includegraphics[width=1.7in,trim=0 30 0 50,clip=TRUE]{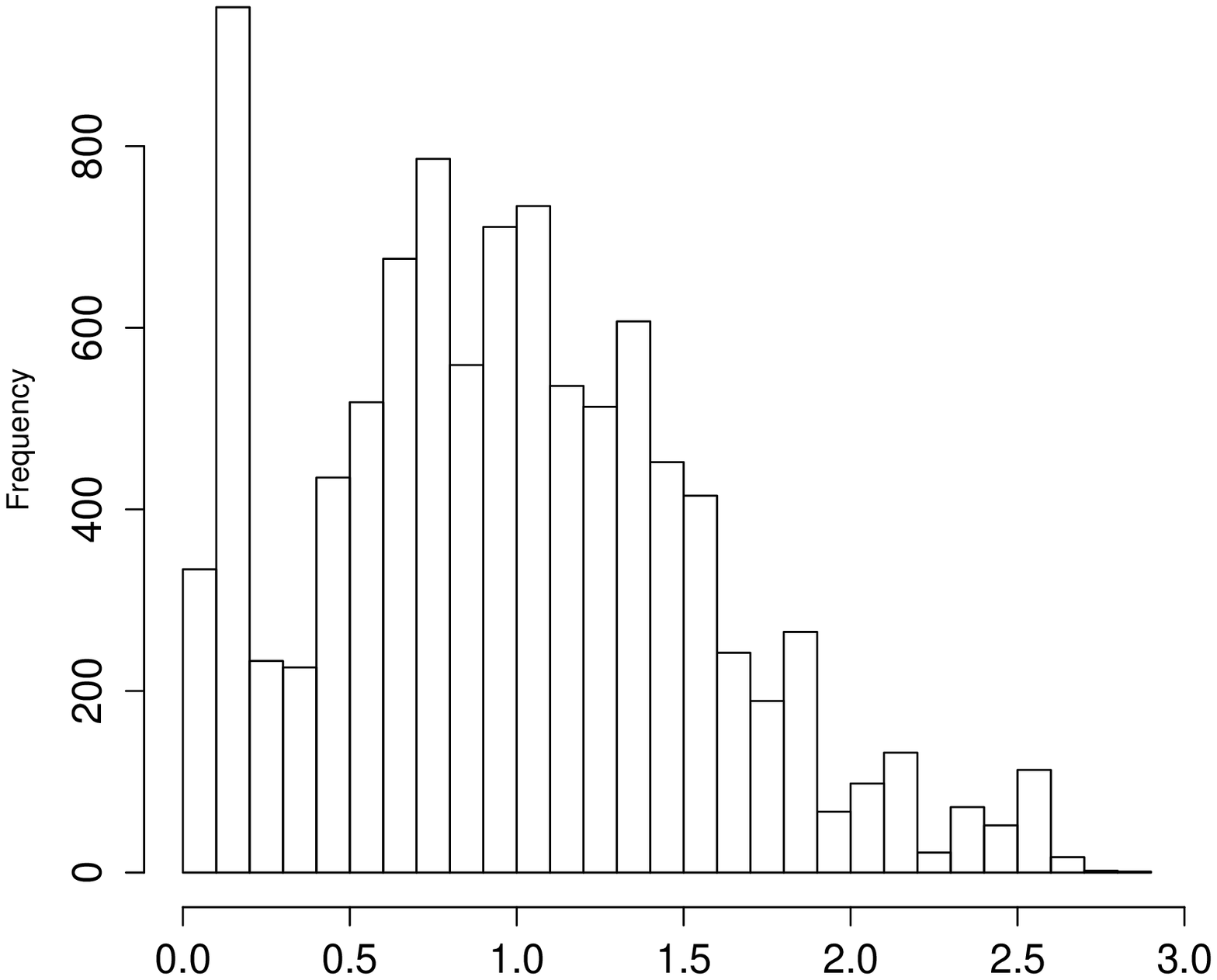}
\end{minipage}

\begin{minipage}{0.75in}
$\tau=0.99$
\end{minipage}
\begin{minipage}{2in}
\includegraphics[width=1.7in,trim=0 30 0 50,clip=TRUE]{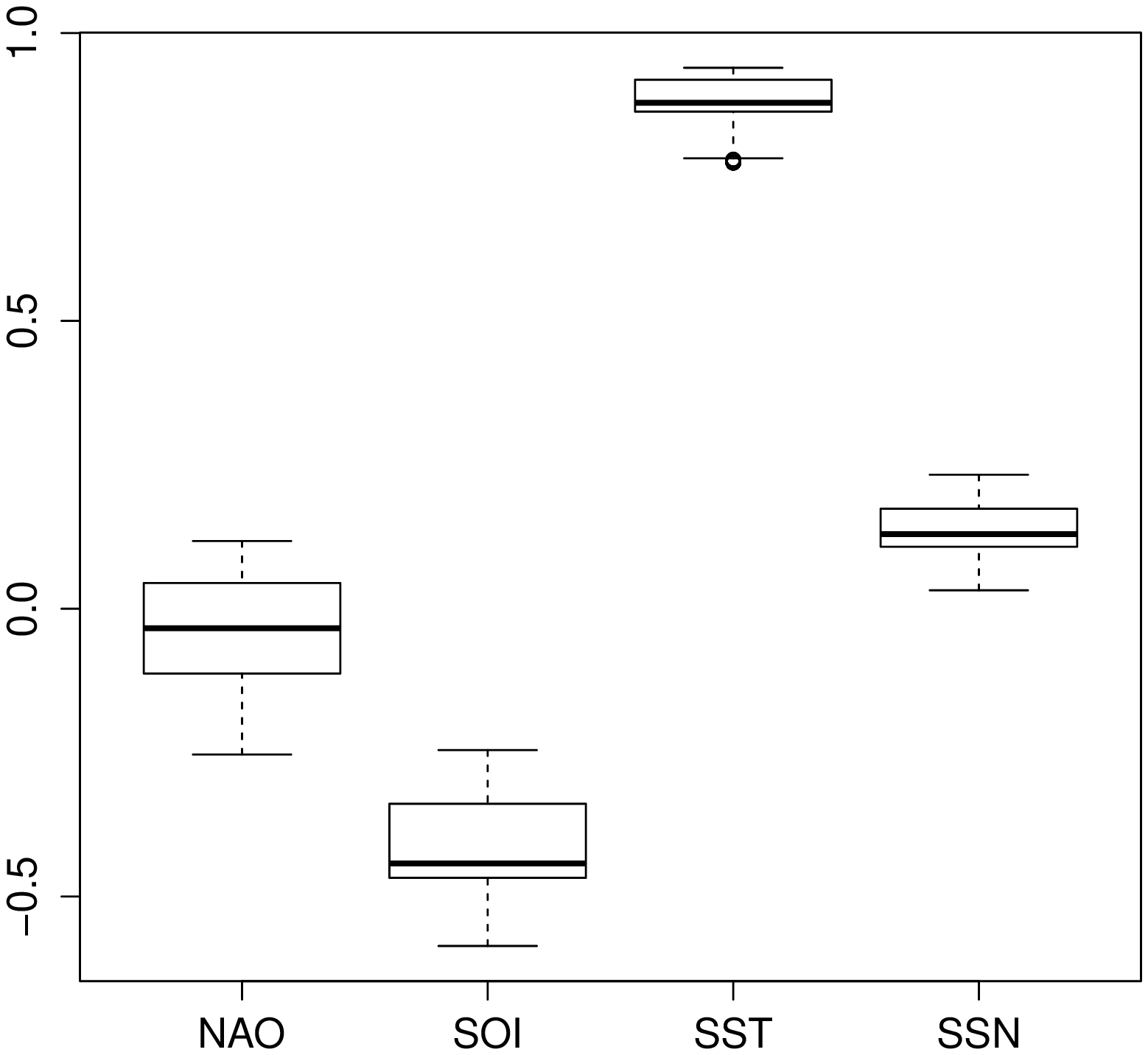}
\end{minipage}
\begin{minipage}{2in}
\includegraphics[width=1.7in,trim=0 30 0 50,clip=TRUE]{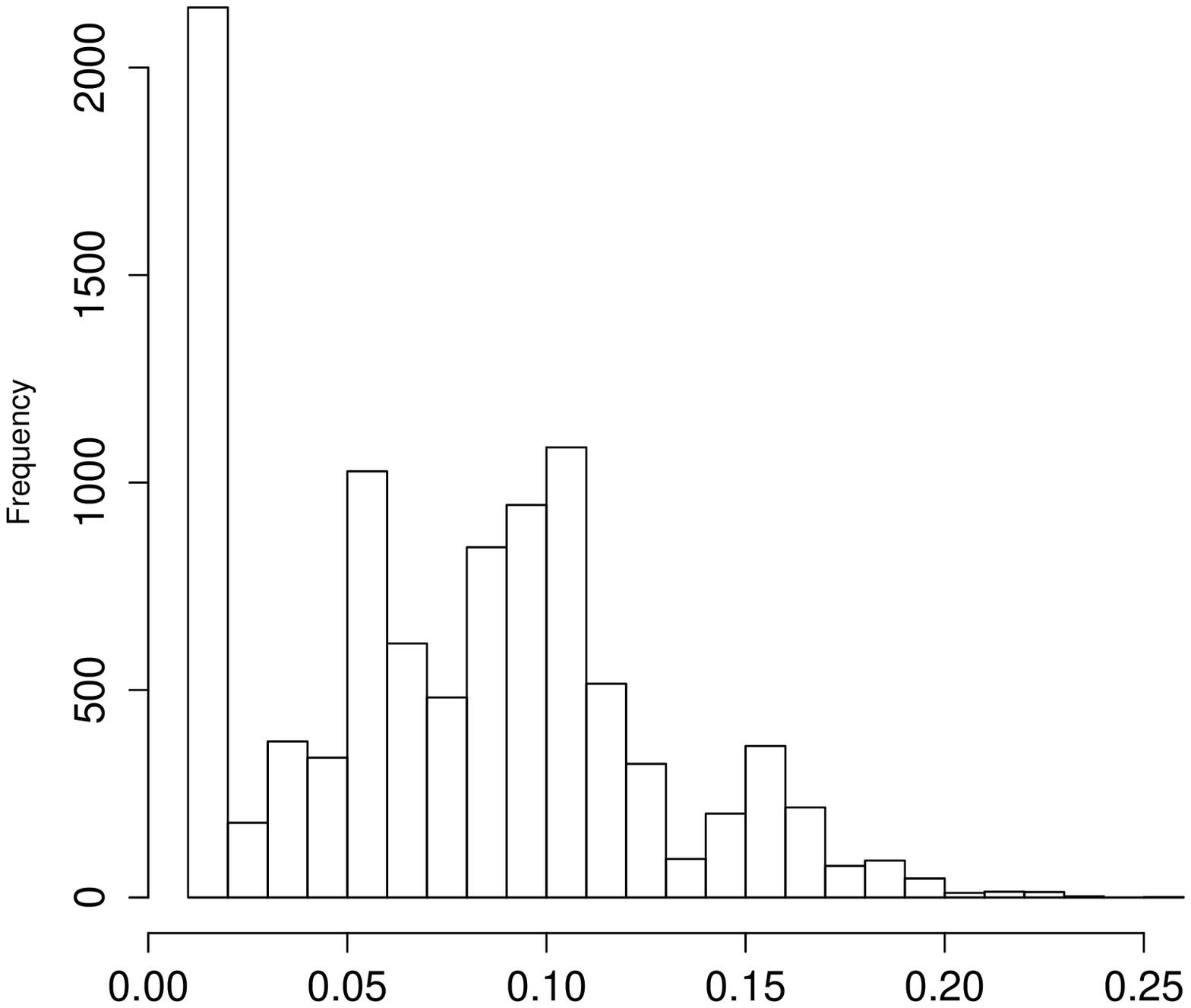}
\end{minipage}
\end{center}

\caption{The boxplots on the left shows the posterior distribution of $\boldsymbol{\beta}$ (after normalization to unit norm). The histograms on the right shows the posterior distribution of $d=1/\|\boldsymbol{\beta}\|^2$.} 
\label{betad_tc}
\end{figure}

Table \ref{result2_TC} compares the obtained index vectors estimated by BQSIM and QSIM 
and Figure \ref{quantileplot_tc} shows the estimated quantile curves of TC intensity at different levels.  From the table, we can see that TC intensity heavily depends on SOI. The index vectors obtained by BQSIM and QSIM are qualitatively similar at lower quantiles, with more obvious deviations at levels above 0.9. This suggests that the estimates are not reliable for high quantile levels, especially for $\tau=0.95$ and $\tau=0.99$. The boxplots in Figure \ref{betad_tc} (left columns) show the samples collected from the posterior distribution of $\boldsymbol{\beta}$ which are normalized to have unit norm. The histograms in Figure \ref{betad_tc} show the implied distribution of $d$ (see equation (\ref{eqn:d})). For Gaussian processes, larger values of $d$ correspond to smoother functions. While it is generally hard to compare the performance of BQSIM and QSIM for real data, our previous simulations suggest that BQSIM is more trustworthy. We also performed model fitting on bootstrapped data and observed that the BQSIM estimates are more stable across bootstrap samples, except for $\tau=0.99$ where estimates obtain from both BQSIM and SIM are quite unstable.

\begin{table}[ht!]
\footnotesize
\caption{Estimates from BQSIM and QSIM for the TCs data.}
\vspace{0.1in}
\label{result2_TC}
\centering 
\begin{tabular}{llccc}
\toprule
\multicolumn{2}{c}{} &{BQSIM} &{QSIM}\\

\midrule
\multirow{4}{*} {$\tau$=0.50}  
& $\rm{NAO(\beta_{1})}$&0.46300&0.49633\\
&$\rm{SOI(\beta_{2})}$&-0.58446&-0.59011\\
&$\rm{SST(\beta_{3})}$&0.01665&-0.06932\\
&$\rm{SSN(\beta_{4})}$&0.51756&0.47965\\ 
\midrule
\multirow{4}{*} {$\tau$=0.75}  
& $\rm{NAO(\beta_{1})}$&0.37840&0.51400\\
&$\rm{SOI(\beta_{2})}$&-0.87369&-0.73418\\
&$\rm{SST(\beta_{3})}$&-0.19373&-0.20965\\
&$\rm{SSN(\beta_{4})}$&0.07462&0.39093\\  
\midrule
\multirow{4}{*} {$\tau$=0.90}  
& $\rm{NAO(\beta_{1})}$&-0.13451&0.36110\\
&$\rm{SOI(\beta_{2})}$&-0.83952&-0.62212\\
&$\rm{SST(\beta_{3})}$&-0.40714&-0.58222\\
&$\rm{SSN(\beta_{4})}$&-0.18082&-0.37594\\  
\midrule
\multirow{4}{*} {$\tau$=0.95}  
& $\rm{NAO(\beta_{1})}$&-0.16689&0.10274\\
&$\rm{SOI(\beta_{2})}$&-0.90598&-0.11676\\
&$\rm{SST(\beta_{3})}$&-0.30985&-0.83128\\
&$\rm{SSN(\beta_{4})}$&-0.08797&-0.53364\\  
\midrule
\multirow{4}{*} {$\tau$=0.99}  
& $\rm{NAO(\beta_{1})}$&-0.03822&0.10860\\
&$\rm{SOI(\beta_{2})}$&-0.41609&-0.12104\\
&$\rm{SST(\beta_{3})}$&0.88491&-0.83293\\
&$\rm{SSN(\beta_{4})}$&0.13593&-0.52894\\ 
                          \bottomrule
\end{tabular}
\end{table}


%
%

\section{Discussion}\label{sec:dis}
In this article, we have proposed a Bayesian quantile regression method for single-index models based on a Gaussian process prior for the unknown link function. As detailed in the Appendix, we designed an efficient MCMC algorithm for posterior inference and demonstrated the superiority of the Bayesian approach to a modern non-Bayesian one. We carefully considered the possibility of marginalizing over the link function in some of the sampling steps, which leads to a partially collapsed sampler that balances sampling efficiency and implementation expediency. The performance of the proposed approach in our simulations is quite encouraging. 

We used zero-mean Gaussian process with squared exponential kernel. It is also possible to explicitly incorporate a mean function in the Gaussian process. This could add some flexibility to the model. For example, if a linear mean function is used, under independent zero-mean normal prior on the linear coefficients, the resulting process would be equivalent to a zero-mean Gaussian process with an additional quadratic term in the kernel function, as shown by \cite{mackay1998introduction}. Thus the consideration on whether to use a non-zero mean is similar to deciding what kind of kernel to use. We take the view that a zero-mean Gaussian process with the quadratic exponential kernel function is already flexible enough to model a variety of curves and thus do not consider these additional possibilities in modelling. 

It is worth noting that we mainly regard ALD as a tool for estimating the conditional quantile, much like in the frequentist approach. By assuming the errors follow ALD, the posterior can give spurious confidence if the underlying data come from a different model than ALD. This is like using a quasi-likelihood to replace the true likelihood in frequentist estimation. In particular, the error distribution may not be accurately estimated by our approach.  As we demonstrated, the method does accurately estimate the index vector and the link function. Finally we remind the readers that ALD can lead to incoherent inferences in the sense that quantile curves are permitted to intersect each other. This is because the inference for distinct quantiles would proceed separately/independently and there is nothing to prevent them from overlapping. 

In terms of computational speed, on an ordinary PC, fitting the BQSIM on a single generated dataset under our simulation setup would take about 10-20 minutes. This is less of a problem for our real data but is quite a burden for simulations. Due to the relatively slow computational speed which is a common problem that plagues MCMC algorithms, some approximation methods such as variational Bayes might be desirable, but this is outside the scope of the current paper. 

As an extension of the current study, one can consider multiple-index models in quantile regression. However, sampling the index matrix poses some serious challenges and is outside the scope of the current investigation.


\section*{Acknowledgements} We thank the Associate Editor and three anonymous referees for their helpful comments that have led to a significant improvement of the manuscript. The research of Heng Lian is supported by Singapore Ministry of Education Tier 1 Grant.

\section*{Appendix: MCMC algorithm details}

The posterior distribution for all of the unknown parameters and latent variables is proportional to the joint distribution, given by 

\begin{eqnarray*}
&&\pi(\boldsymbol{\beta},\boldsymbol{\eta}_n,\boldsymbol{e}_{n},\sigma, \lambda, \gamma|\boldsymbol{y})\\
&\propto&\exp\left\{-\frac{(\boldsymbol{y}-\boldsymbol{\eta}_{n}-k_{1}\boldsymbol{e_{n}})^{T}\boldsymbol{E}^{-1}(\boldsymbol{y}-\boldsymbol{\eta}_{n}-k_{1}\boldsymbol{e}_{n})}{2}\right\}\\
&&\times (\det[\boldsymbol{E}])^{-1/2} \det[\boldsymbol{C}_{n}]^{-1/2}\exp\left\{-\frac{\boldsymbol{\eta}_{n}^{T}\boldsymbol{C}_{n}^{-1}\boldsymbol{\eta}_{n}}{2}\right\}\\
&& \times\mathop{\prod}\limits_{j=1}\limits^{p}\frac{\lambda}{2\sigma}e^{-\lambda|\beta_{j}|/\sigma}\mathop{\prod}\limits_{i=1}\limits^{n}\frac{1}{\sigma}\exp\left\{-\frac{e_{i}}{\sigma}\right\}\\
&&\times\left(\frac{1}{\sigma}\right)^{a_{\sigma}+1}\exp\left\{-\frac{b_{\sigma}}{\sigma}\right\} \left(\frac{1}{\gamma}\right)^{a_{\gamma}+1}\exp\left\{-\frac{b_{\gamma}}{\gamma}\right\}\lambda^{a_{\lambda}-1}\exp\{-b_{\lambda}\lambda\}.
\end{eqnarray*}

The Metropolis-within-Gibbs algorithm may be used to sample from the posterior distribution. Mathematically speaking, it is possible to integrate out $\boldsymbol{\eta}_n$ before sampling, and it is well-known that marginalization can improve the mixing of the chain \citep{liu2008monte}. However, if $\boldsymbol{\eta}_n$ is not sampled, then the full conditional distributions for $e_i$ and $\sigma$ are no longer well-known distributions which leads to extra difficulty in sampling. On the other hand, we note the full conditional distribution  of $\boldsymbol{\beta}$ is 
\begin{equation*}
\begin{split}
\pi(\boldsymbol{\beta}|\boldsymbol{\eta}_n,\lambda,\sigma,\boldsymbol{y})&\propto \det[\boldsymbol{C}_{n}]^{-1/2}\exp\left\{-\frac{\boldsymbol{\eta}_{n}^{T}\boldsymbol{C}_{n}^{-1}\boldsymbol{\eta}_{n}}{2}\right\} \times\mathop{\prod}\limits_{j=1}\limits^{p}\frac{\lambda}{2\sigma}e^{-\lambda|\beta_{j}|/\sigma},
\end{split}
\end{equation*}
and thus the evaluation of the density involves the inverse of $\boldsymbol{C}_n$. Unfortunately, since $\boldsymbol{C}_n$ is a kernel matrix, in many simulations we found it is nearly singular. When $\boldsymbol{\eta}_n$ is integrated out, this singularity problem is avoided since we only have to compute the inverse of the matrix $\boldsymbol{C}_n+\boldsymbol{E}$, where $\boldsymbol{E}$ is a diagonal matrix (see (\ref{BetaPosterior}) below). 

The conditional posterior densities of all the parameters and variables, except for $\boldsymbol{\beta}$ and $\gamma$, are common distributions. The conditional distributions used in the sampling are presented below. $\boldsymbol{\eta}_n$ is marginalized out when considering the posterior conditional distribution of $\boldsymbol{\beta}$ and $\gamma$. 

\begin{equation}
\label{BetaPosterior}
\begin{split}
\pi(\boldsymbol{\beta}|\boldsymbol{e}_n,\sigma, \lambda, \gamma, \boldsymbol{y})&\propto\int\pi(\boldsymbol{y}|\boldsymbol{e}_n,\sigma,\boldsymbol{\eta}_n)\pi(\boldsymbol{\eta}_{n}|\boldsymbol{\beta},\gamma)d\boldsymbol{\eta}_{n}\times\pi(\boldsymbol{\beta}|\sigma,\lambda)\\
&\propto\exp\left\{-\frac{(\boldsymbol{y}-k_{1}\boldsymbol{e}_{n})^{T}(\boldsymbol{E}+\boldsymbol{C}_{n})^{-1}(\boldsymbol{y}-k_{1}\boldsymbol{e}_{n})}{2}\right\}\\
& \times(\det[\boldsymbol{C}_{n}+\boldsymbol{E}])^{-1/2}\times \mathop{\prod}\limits_{j=1}\limits^{p}e^{-\lambda|\beta_{j}|/\sigma},\\
\end{split}
\end{equation}

\begin{equation}
\label{GammaPosterior}
\begin{split}
\pi(\gamma|\boldsymbol{\beta},\boldsymbol{e}_{n},\sigma, \lambda,  \boldsymbol{y})&=\pi(\gamma|\boldsymbol{\beta},\boldsymbol{e}_{n},\sigma,\boldsymbol{y})\\
&\propto\int\pi(\boldsymbol{y}|\boldsymbol{e}_{n},\sigma,\boldsymbol{\eta_{n}})\pi(\boldsymbol{\eta_{n}}|\boldsymbol{\beta}, \gamma)d\boldsymbol{\eta_{n}}\times\pi(\gamma)\\
&\propto\exp\left\{-\frac{(\boldsymbol{y}-k_{1}\boldsymbol{e}_{n})^{T}(\boldsymbol{E}+\boldsymbol{C}_{n})^{-1}(\boldsymbol{y}-k_{1}\boldsymbol{e}_{n})}{2}\right\}\\
&\times (\det[\boldsymbol{C}_{n}+\boldsymbol{E}])^{-1/2} \left(\frac{1}{\gamma}\right)^{a_{\gamma}+1}\exp\left\{-\frac{b_{\gamma}}{\gamma}\right\},\\
\end{split}
\end{equation}\\

\begin{equation}
\label{EtaPosterior}
\begin{split}
\pi(\boldsymbol{\eta_{n}}|\boldsymbol{\beta},\boldsymbol{e}_{n},\sigma, \lambda, \gamma,\boldsymbol{y})&=\pi(\boldsymbol{\eta_{n}}|\boldsymbol{\beta},\boldsymbol{e}_{n},\sigma,\gamma,\boldsymbol{y})   \sim N(\boldsymbol{\mu_{n}},\boldsymbol{\Sigma_{n}}),\\
\boldsymbol{\mu_{n}}&=\boldsymbol{C}_{n}(\boldsymbol{C}_{n}+\boldsymbol{E})^{-1}(\boldsymbol{y}-k_{1}\boldsymbol{e}_{n}),\\
\boldsymbol{\Sigma_{n}}&=\boldsymbol{C}_{n}(\boldsymbol{C}_{n}+\boldsymbol{E})^{-1}\boldsymbol{E},
\end{split}
\end{equation}

\begin{equation*}
\label{SigmaPosterior}
\begin{split}
\pi(\sigma|\boldsymbol{\beta},\boldsymbol{\eta}_n,\boldsymbol{e}_{n}, \lambda, \gamma,\boldsymbol{y})
&=\pi(\sigma| \boldsymbol{\eta_{n}}, \boldsymbol{\beta}_n,\boldsymbol{e}_n, \lambda, \boldsymbol{y}) \sim IG(\alpha_{\sigma},\nu_{\sigma} ),\\
\alpha_{\sigma}&=\frac{3n}{2}+p+a_{\sigma},\\
\nu_{\sigma}&=\mathop{\sum}\limits_{i=1}\limits^{n}\left(\frac{(y_{i}-\eta(\boldsymbol{x_{i}}^{T}\boldsymbol{\beta})-k_{1}e_{i})^{2}}{2k_{2}e_{i}}+e_{i}\right)+\mathop{\sum}\limits_{j=1}\limits^{p}\lambda|\beta_{j}|+b_{\sigma},
\end{split}
\end{equation*}

\begin{equation*}
\label{LambdaPosterior}
\begin{split}
\pi(\lambda|\sigma,\boldsymbol{\beta}, \boldsymbol{\eta}_n,\boldsymbol{e}_{n},\gamma,\boldsymbol{y})=\pi(\lambda|\sigma,\boldsymbol{\beta})\sim Ga(a_{\gamma}+p, b_{\gamma}+\mathop{\sum}\limits_{j=1}\limits^{p}|\beta_{j}|/\sigma^{2}) .\\
\end{split}
\end{equation*}

The full conditional distribution for $e_{i}$ is a generalized inverse Gaussian distribution $(GIG)$,
\begin{equation*}
\label{EPosterior}
\begin{split}
\pi(e_{i}|\sigma,\boldsymbol{\beta}, \boldsymbol{\eta}_n,\lambda, \gamma,\boldsymbol{y})=\pi(e_{i}|\sigma, \boldsymbol{\eta}_n,\boldsymbol{y})\sim GIG\left(\frac{1}{2}, \sqrt{\frac{(y_{i}-\eta(\boldsymbol{x_{i}}^{T}\boldsymbol{\beta}))^{2}}{k_{2}\sigma}},\sqrt{\frac{k_{1}^{2}}{k_{2}\sigma}+\frac{2}{\sigma}}\right),
\end{split}
\end{equation*}
where the probability density function of $GIG(\rho,m,n)$ is 
\begin{equation*}
\label{GIG}
\begin{split}
f(x|\rho,m,n)&=\frac{(n/m)^{\rho}}{2K_{\rho}(mn)}x^{\rho-1}\exp\left\{-\frac{1}{2}(m^{2}x^{-1}+n^{2}x)\right\},\\
&x>0,-\infty<\rho<\infty,m\geq 0,n\geq 0,
\end{split}
\end{equation*}
and $K_{\rho}$ is the modified Bessel function of the third kind \citep{barndorff2001non}.

We use a superscript $(.)^{(t)}$ to denote the sampled values of different quantities at iteration $t$. The variables $ \boldsymbol{\eta}_n^{(t)},\lambda^{(t)}, \gamma^{(t)},\boldsymbol{e}_n^{(t)}$ can be directly generated in R based on the respective full conditional distributions. For $\boldsymbol{\beta}^{(t)}$, we use a Metropolis step with proposal distribution $N(\boldsymbol{\beta}^{(t-1)},\sigma_{\beta}^2I)$, and for $\gamma^{(t)}$, we propose the new value from $\log \gamma^{(t)}\sim N(\log \gamma^{(t-1)},\sigma_\gamma^2)$. In practice, $\sigma_\beta$ and $\sigma_\gamma$ are manually tuned to ensure the acceptance rate to be within $10\%\sim 30\%$. This manual tuning is simplified by transforming all predictors and responses to have mean 0 and variance 1 before running the MCMC algorithm.

 Our sampling strategy is ``partially collapsed" in the sense of \cite{van2008partially}, and in particular is similar to Sampler 7 in that paper. It is easy to see the validity of the constructed sampler (that is, it does not change the stationary distribution). More specifically, to obtain this partially collapsed sampler by modifying the Gibbs sampler, we first marginalize the full conditional distributions for $\boldsymbol{\beta}$ and $\gamma$ to get $\pi(\boldsymbol{\beta},\boldsymbol{\eta}_n|\boldsymbol{e}_n,\sigma, \lambda, \gamma, \boldsymbol{y})$ and $\pi(\gamma,\boldsymbol{\eta}_n|\boldsymbol{\beta},\boldsymbol{e}_{n},\sigma, \lambda,  \boldsymbol{y})$, by moving $\boldsymbol{\eta}_n$ from being conditioned to being sampled. Since $\boldsymbol{\eta}_n$ is sampled again in (\ref{EtaPosterior}) immediately following  $\boldsymbol{\beta}$ and $\gamma$, the two intermediate $\boldsymbol{\eta}_n$'s are redundant and thus can be trimmed, resulting in (\ref{BetaPosterior}) and (\ref{GammaPosterior}) respectively. It was shown in Theorem 1 of \cite{van2008partially} that the marginalization step can only improve the autocorrelation of the chain. As shown in our simulation examples, this improvement is dramatic for our specific problem.

\bibliographystyle{elsart-harv}
\bibliography{reference}

\end{document}